\IfFileExists{elsarticle.cls}{%
  \documentclass[preprint,11pt]{elsarticle}%
  \newcommand{\CMAMEels}{1}%
}{%
  \documentclass[11pt]{article}%
  \newcommand{\CMAMEels}{0}%
}

\usepackage{amsmath}
\usepackage{amssymb}
\usepackage{graphicx}
\usepackage{xcolor}
\usepackage{multirow}
\usepackage{booktabs}       
\usepackage{array}
\usepackage{enumitem}       
\usepackage{caption}
\usepackage{subcaption}
\usepackage{tikz}
\usepackage{pgfplots}
\pgfplotsset{compat=1.18}
\usepgfplotslibrary{groupplots}
\usepackage{hyperref}
\hypersetup{hypertexnames=false,hidelinks}

\ifnum\CMAMEels=1
  \newcommand{\appref}[1]{\ref{#1}}
\else
  \newcommand{\appref}[1]{Appendix~\ref{#1}}
\fi

\ifnum\CMAMEels=0
  \usepackage[numbers,sort&compress]{natbib}
  \usepackage{geometry}
\fi

\ifnum\CMAMEels=1
  \journal{Computer Methods in Applied Mechanics and Engineering}
  \makeatletter
  \@ifundefined{myfooter}{}{%
    \myfooter[C]{Preprint submitted to Computer Methods in Applied Mechanics
      and Engineering}}
  \makeatother
\fi

\newcommand{\CARStitle}{CARS: a centerline-anchored attention surrogate for
coronary blood flow on skeleton-generated lumen geometries}

\newcommand{\CARSabstract}{%
When the parameter of a PDE surrogate is the flow domain itself, the key/value
set of an attention model must encode that domain. We present CARS (Coronary
Artery Rapid Surrogate), a learned surrogate for the finite-volume solution map
of the steady incompressible Navier--Stokes equations on coronary lumens built
from image-derived centerlines. CARS maps a centerline with per-point lumen
radius and a scalar inlet flow rate to pressure and velocity at arbitrary query
points. The attention focal set is the case's own centerline, so geometry fusion
costs $O(N n_f)$ rather than $O(N^2)$ in the number of evaluation points.
Trained per territory on $2{,}700$ OpenFOAM solutions with synthetic focal
lesions, CARS reaches relative $\ell_2$ errors of $0.120/0.149$
(pressure/velocity) on left-coronary and $0.063/0.119$ on right-coronary test
sets, in $87$~ms per case against a solver wall-clock of $572$--$1{,}063$~s. It
outperforms four parameter-matched neural
operators, and accuracy saturates eightfold below the operating focal count.
Fractional flow reserve derived from the predicted pressure matches the CFD
label at Pearson $r = 0.941/0.993$ and reproduces the $\mathrm{FFR}\le 0.8$
decision in $97.7\%/96.3\%$ of held-out cases. The predicted velocity does not
satisfy continuity closely enough for derived quantities, and the weights are
territory-specific.
}

\ifdefined\SPLITAPPENDIX
  \input{appendix_labels}
\fi

\begin{document}

\ifnum\CMAMEels=1
\begin{frontmatter}

\title{\CARStitle}

\author[a]{Reza Akbarian Bafghi\fnref{eqc}}
\ead{reza.akbarianbafghi@colorado.edu}
\author[b]{Sukirt Thakur\fnref{eqc}}
\ead{sukirt@angioinsight.com}
\author[c]{Maziar Raissi\corref{cor1}}
\ead{maziar.raissi@ucr.edu}
\fntext[eqc]{These authors contributed equally.}
\cortext[cor1]{Corresponding author.}
\affiliation[a]{organization={University of Colorado Boulder},
  city={Boulder}, state={Colorado}, country={USA}}
\affiliation[b]{organization={AngioInsight, Inc.},
  city={Minnetonka}, state={Minnesota}, country={USA}}
\affiliation[c]{organization={University of California, Riverside},
  city={Riverside}, state={California}, country={USA}}

\begin{abstract}
\CARSabstract
\end{abstract}


\end{frontmatter}
\else
\title{\CARStitle}
\author{Reza Akbarian Bafghi\thanks{These authors contributed equally.}
  \and Sukirt Thakur\footnotemark[1]
  \and Maziar Raissi\thanks{Corresponding author.}}
\date{}
\maketitle
\begin{abstract}
\CARSabstract
\end{abstract}
\fi

\section{Introduction}

Predicting blood flow in a patient-specific coronary artery is the evaluation
of a parametric solution map whose parameter is the flow domain itself. Each
evaluation solves the steady incompressible Navier--Stokes equations on a lumen
reconstructed from one patient's image, at a prescribed volumetric flow rate,
and returns the pressure and velocity fields. The parameter is
high-dimensional: a coronary tree varies in branching topology, in the number
and caliber of its terminal branches, in curvature and torsion, and in the
position and shape of focal narrowings, and no small set of shape parameters
describes the family. One steady solve is inexpensive by the standards of
computational fluid dynamics and expensive by the standards of a query: the
pipeline that produced the reference solutions used here runs a fixed
$4{,}000$ SIMPLE (Semi-Implicit Method for Pressure-Linked Equations)
iterations per case on a resolved hexahedral mesh
(\appref{app:cfd_pipeline}).\ifdefined\SPLITAPPENDIX\footnote{Appendices A--G
are provided as Supplementary Material.}\fi

\begin{figure}[!t]
    \centering
    \includegraphics{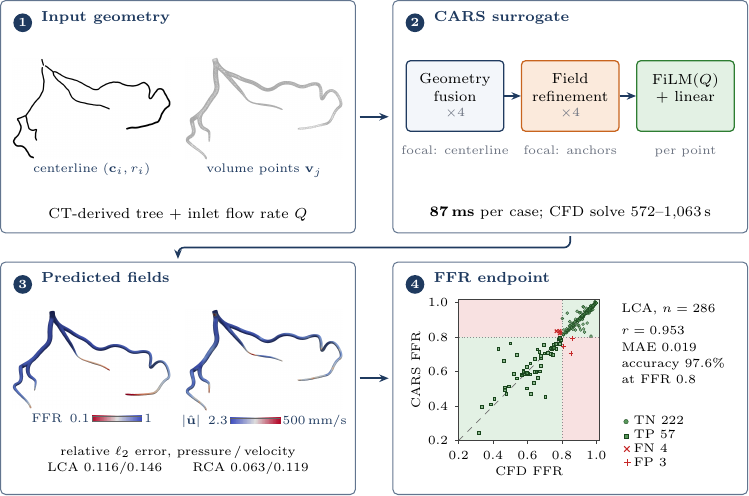}
    \caption{\textbf{CARS at a glance.} From a CT-derived coronary centerline
    (with per-point lumen radius and a set of intra-luminal volume points) and
    a scalar inlet flow rate $Q$, CARS predicts volumetric pressure and
    velocity fields in $87$~ms (fp32) per case. The CFD time in
    panel~2 is the reference solver stage only: a measured median of $572$~s
    (RCA) and $948$--$1{,}063$~s (LCA) on eight MPI ranks. Panel~3 shows one
    predicted case, with pressure shown as the pointwise ratio to the aortic
    pressure $P_{\mathrm{a}}$ and velocity as magnitude in mm/s; the relative
    $\ell_2$ errors beneath it are test-split means for each territory from
    the primary $300$-epoch run.
    Panel~4 is the $286$-case physiologic-range subset of the LCA endpoint
    analysis (\S\ref{sec:clinical_ffr}), classified at the
    $\mathrm{FFR}\le 0.8$ threshold (positive: $\mathrm{FFR}\le 0.8$); red
    quadrants mark disagreement with CFD.}
    \label{fig:hook}
\end{figure}

Two families of methods address parametric PDEs of this kind, and neither
closes this one. Projection-based reduced-order models require a trial space
that transfers across the parameter set; when the parameter is the domain,
that needs a common reference configuration and a family of regular
deformations onto it, which a vascular tree of varying branch count and
connectivity does not admit. Learned surrogates on fixed grids meet the
opposite obstruction, since an intra-luminal volume is neither structured nor
of fixed extent. Attention-based surrogates act on point sets, but attention
over the volume costs $O(N_v^2)$, and a coronary volume needs many points:
the resolution requirement is set by the throat of the tightest stenosis,
while the quantity of interest is a pressure difference across it.

The structure this work exploits is that the domain family, while not
low-dimensional as a set of shapes, is generated by a low-dimensional object: a
vessel is a tube about a curve, and that curve together with a per-point lumen
radius already carries the geometric information on which the solution depends
most strongly. We therefore use the patient-specific centerline as the focal
(key/value) token set of a \emph{tiered} attention mechanism: queries are
formed from all tokens, keys and values only from the centerline, so cost is
$O(N_v n_f)$, linear in the number of evaluation points, and the focal set is
given by the geometry. Two further elements complete the map: the evaluation
points enter as an argument, so one set of weights answers at any evaluation
density in the range tested; and the scalar flow rate enters multiplicatively,
by feature-wise linear modulation, the form consistent with the approximately
linear pressure--flow scaling of this regime. The construction is not specific
to the coronary tree; \S\ref{sec:future} gives the input contract it asks of a
new setting.

The coronary circulation is the validation domain because its reference
solutions are expensive, its geometry is image-derived, and it carries a scalar
functional of the pressure field with an accepted decision threshold, so
surrogate error is interpretable as a decision and not only as a norm. That
functional is fractional flow reserve (FFR), the ratio of distal to proximal
static pressure under hyperemia and the reference standard for deciding whether
a coronary lesion is flow-limiting~\cite{tonino2009ffr} in a disease that
remains a leading cause of death worldwide~\cite{who2025cvd}; obtaining it
non-invasively by solving the Navier--Stokes equations on a lumen segmented
from CT angiography agrees closely with the invasive measurement at the
$\mathrm{FFR}\le 0.8$ threshold~\cite{taylor2013ffrcfd,norgaard2014ctffr}, and
that CT-to-solution pipeline is what a surrogate of this map amortizes.

\begin{figure*}[!t]
    \centering
    \includegraphics{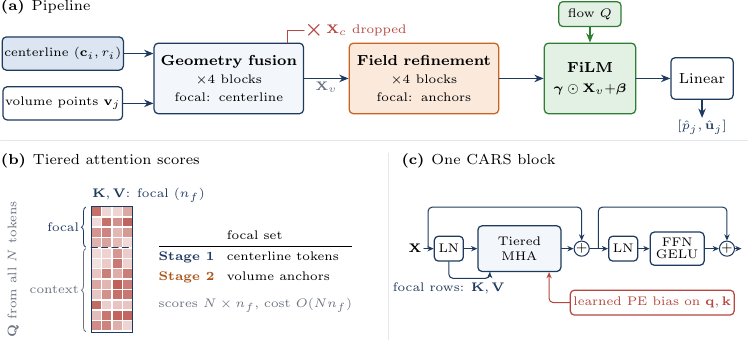}
    \caption{\textbf{CARS architecture.} (a)~Two-stage tiered transformer.
    \textit{Stage 1 (geometry fusion)} uses the centerline tokens as the focal
    set, and the centerline tokens $\mathbf{X}_c$ are discarded after it, so
    only the volume tokens $\mathbf{X}_v$ pass to Stage~2.
    \textit{Stage 2 (field refinement)} uses a uniformly random subsample of the
    volume tokens as \emph{anchors} (the focal set); at the operating point
    $N_{\mathrm{va}} = N_v$, so every point is an anchor. FiLM modulates
    post-Stage-2 volume features by the inlet flow rate $Q$, and a single
    linear layer reads out per-query kinematic pressure $\hat p_j$ and
    three-component velocity $\hat{\mathbf{u}}_j$. (b)~Tiered attention:
    queries come from all $N$ tokens and keys and values from the $n_f$ focal
    tokens only, so the score matrix is $N \times n_f$. Its upper
    $n_f \times n_f$ block is the intra-focal attention, the property that
    distinguishes tiered from plain cross-attention; the remaining $N-n_f$
    context rows read from the focal set but are never read from. At
    $N_{\mathrm{va}}=N_v$ Stage~2 has no context rows and is full
    self-attention. (c)~Each block applies pre-LayerNorm multi-head tiered
    attention, with keys and values taken from the focal rows of the
    normalized sequence and a learned attention-level
    positional bias on the projected queries and keys (the default for all
    CARS results; RoPE3D is the comparison variant), then a GELU feed-forward
    network, each with a residual connection.}
    \label{fig:architecture_overview}
    \vspace{-4mm}
\end{figure*}

We build $6{,}600$ steady finite-volume solutions across the two coronary
territories and evaluate the resulting surrogate, CARS (Figures~\ref{fig:hook}
and~\ref{fig:architecture_overview}), against the $300$-case held-out split of
each: in discrete field norms, in cost, in sensitivity to the geometric
arguments, and through the FFR functional of the predicted pressure field.
\phantomsection\label{sec:intro_claims}%
CARS as specified in \S\ref{sec:methodology} approximates this solution map
more accurately than the four neural-operator baselines and the alternative
focal representations evaluated here, at parameter counts within $3\%$ and on
identical splits, and the FFR functional read off its predicted pressure
reproduces the reference solver's $\mathrm{FFR}\le 0.8$ decision in $97.7\%$
(left coronary artery, LCA) and $96.3\%$ (right coronary artery, RCA) of
held-out cases. Pressure relative $\ell_2$ on the
test split is the primary endpoint; velocity, the endpoint and cost are
secondary. We hypothesize that what matters is that the focal set is
\emph{attached} to the object the domain is generated from, not the number of
focal tokens or the geometry they carry; the ablations are consistent with that
hypothesis but do not isolate it, since every arm which moves the focal set off
the centerline also changes the geometric channels or the positional treatment
(\S\ref{sec:res_coarse}). Table~\ref{tab:scope} states the scope of the
evidence.

The contributions, in the order the experiments take them, are:
\begin{enumerate}
    \item \textbf{A focal set given by the geometry.} An anchored-attention
    surrogate for the parametric steady incompressible Navier--Stokes problem
    on centerline-derived vascular domains, whose attention focal set is the
    centerline with its radius profile, so that geometry fusion is linear in
    the evaluation count.
    \item \textbf{The evaluation set as an argument of the map.} The points at
    which the solution is requested are supplied at inference: pressure error
    varies by at most $0.004$ over the $16$-fold range from $4{,}096$ to
    $65{,}536$ points, inside an equivalence margin of $0.005$
    (Table~\ref{tab:density_precision}).
    \item \textbf{Two architectural findings at matched parameter count.} A
    learned attention-level coordinate embedding lowers error by $27$--$30\%$
    relative to a parameter-free 3D rotary encoding~\cite{su2023rope} at
    $+0.3\%$ parameters, with a frequency-scale confound noted in
    \S\ref{sec:rope3d}; and centerline anchors outperform both learned
    inducing points~\cite{lee2019set} and anchors subsampled from a surface
    discretization~\cite{alkin2024abupt}, while accuracy saturates eightfold
    below the operating focal count.
\end{enumerate}

\paragraph{Relation to a companion paper by the same authors}
A companion paper by the same authors predicts surface pressure and wall shear
stress on coronary geometries through an anisotropic radial-basis
decoder~\cite{bafghi2026centerlines}. The two share the ImageCAS source of
their geometries ($98$ of the $101$ left-coronary and $130$ of the $131$
right-coronary base anatomies recur) and the OpenFOAM pipeline family, and
nothing downstream of it: the tiered-attention construction, the volumetric
prediction task, the input contract of \S\ref{sec:normalization} and every
result reported here are introduced in this paper. The processed datasets are
distinct (no left-coronary case is common to both, and the $794$ right-coronary
case names that recur carry different tensors), and no label field, trained
weights, figure or table is shared.

\begin{table}[!tb]
\centering
\caption{What the evidence in this paper establishes, and what it does not.
Each row on the right names the section that takes it up. G1--G3 are the three
senses of generalization defined in \S\ref{sec:dataset}.}
\label{tab:scope}
\small
\begin{tabular}{p{0.46\linewidth}p{0.46\linewidth}}
\toprule
\textbf{Established here} & \textbf{Not established here} \\
\midrule
Accuracy on unseen radius and flow combinations, on trajectories seen in
training, against baselines matched to $3\%$ on an identical partition (G1)
& Accuracy on unseen anatomies (G2) or unseen acquisition, site or population
(G3); $\sim$$100$ segmentations leave the former unresolved
(\S\ref{sec:limitations})
\\ \addlinespace[3pt]
That the margin is not bought by focal density or parameter count
& That focal-set \emph{provenance} is what produces it, as opposed to the
geometry it carries (\S\ref{sec:res_coarse}) \\ \addlinespace[3pt]
Agreement with the CFD-derived FFR label at its threshold
& Agreement with invasive wire-based FFR; no invasive measurement enters this
work (\S\ref{sec:limitations}) \\ \addlinespace[3pt]
Linear cost of geometry fusion in the evaluation count
& Linear cost of the full model at the operating point, where Stage 2 is full
self-attention (\S\ref{sec:density_precision}) \\ \addlinespace[3pt]
Pressure fields accurate enough to carry the endpoint
& Velocity fields satisfying continuity or momentum balance closely enough for
derived quantities (\S\ref{sec:frame_sensitivity}) \\ \addlinespace[3pt]
Performance on lumens generated as circular sections swept along a centerline
& Transfer to independently segmented, eccentric lumens, where the centerline
reduction stops being lossless (\S\ref{sec:limitations}) \\ \addlinespace[3pt]
Weak sensitivity to evaluation density over a $16$-fold range
& Discretization invariance, and independence of the prediction from the rest
of the query set (\S\ref{sec:frame_sensitivity}) \\ \addlinespace[3pt]
Separate models for the two coronary territories
& Cross-territory transfer, which fails, and invariance to the acquisition
frame, which the construction does not have (\S\ref{sec:cross_territory},
\S\ref{sec:frame_sensitivity}) \\
\bottomrule
\end{tabular}
\vspace{-2mm}
\end{table}

\section{Related Work}

\paragraph{Reduced-order modeling and physics-based ct-FFR}
\label{sec:rom_related}
Solving the Navier--Stokes equations on a lumen segmented from coronary CT
angiography to recover FFR non-invasively was established by Taylor et
al.~\cite{taylor2013ffrcfd} and validated against invasive FFR in the
DISCOVER-FLOW~\cite{koo2011discoverflow}, DeFACTO~\cite{min2012defacto} and
NXT~\cite{norgaard2014ctffr} trials, with per-vessel sensitivity broadly
in the $80$--$90\%$ band and more variable specificity (as low as $54\%$
per patient in DeFACTO) at the
$\mathrm{FFR}\le 0.8$ threshold; the pipeline runs to hours per case even with cloud-based
acceleration~\cite{itu2016machine}. The classical route to a cheap
parametric solver is projection: build a low-dimensional trial space from
solution snapshots and project the governing equations onto
it~\cite{benner2015survey}. Two requirements make this family hard to apply
here. Online efficiency needs affine parameter dependence, or empirical
interpolation to approximate it~\cite{barrault2004eim}, and the quadratic
convective term needs hyper-reduction on top. More binding, when the parameter
is the domain the snapshots live on different meshes, so a common basis exists
only after all domains have been mapped to one reference configuration by a
family of regular parametric deformations~\cite{lassila2010freeform}. Such a
family exists when the geometries deform one topology, the setting of
cardiovascular projection ROM (Ballarin et al.~\cite{ballarin2016cabg}
parametrize fixed-connectivity bypass grafts through their centerlines); it
does not exist for an image-derived population of coronary trees whose branch
count and connectivity differ between patients. One-dimensional models integrate the equations over the cross-section
and solve a hyperbolic system on a network of segments, closed by a tube law and
terminated by lumped-parameter
models~\cite{quarteroni2016cardiovascular,vignon2006outflow}. They are cheap and
topology-agnostic, but the trans-stenotic pressure loss enters through an
empirical closure, and since the FFR endpoint is that loss, the closure sets
their accuracy for the quantity of interest. A third strand parameterizes the
shape by a fixed-length vector of stenosis descriptors or deformation control
points, the assumption behind the ${\sim}12{,}000$ generated stenotic vessels of
Itu et al.~\cite{itu2016machine}. The lesions here are also generated from a
low-dimensional radial deformation, so on that axis the settings are alike; what
a fixed-length vector cannot encode is the image-derived trajectory and the
variable branch count and connectivity. The surrogate developed here keeps the
topology-agnostic character of the reduced models together with the volumetric
output of the reference solve: a change of topology changes the token set, not
the model.

\paragraph{Learned surrogates for cardiovascular flow simulation}
Learning-based surrogates approximate the map from geometry and boundary
conditions to solution fields, amortizing the solver over a population of
cases, and are distinguished mainly by how they represent spatial structure.
Convolutional
networks~\cite{guo2016convolutional,thuerey2020deepflow,jiang2020meshfreeflownet,gharleghi2022transient}
require fixed Cartesian or projection-based discretizations, which restricts
them on irregular patient-specific geometries. Graph neural
networks~\cite{pfaff2021learning,suk2021deep,nannini2025benchmark} accommodate
geometric variability but rely on local message passing, so long-range
dependencies such as the pressure gradient across a stenosis cost depth.
Operator-learning methods, including DeepONet~\cite{lu2019learning} and Fourier
neural operators~\cite{li2020fourier,li2022geofno}, learn maps between function
spaces but in their standard forms assume structured discretizations or an
explicit adaptation to complex geometry. Physics-informed neural
networks~\cite{raissi2019physicsinformed} impose the equations through the
objective but in their basic form require per-case optimization, so their cost
is that of a solve; equation-discovery uses~\cite{chen2021physics} and
metalearned variants~\cite{huang2022metapinn} relax parts of this, the latter
amortizing across a parameter family, which is the property sought here by a
different route.

\paragraph{Attention-based architectures for PDE surrogates}
\label{sec:attn_related}
Attention was introduced for sequence modeling~\cite{vaswani2017attention},
carried to images~\cite{dosovitskiy2020image,liu2021swin} and to unstructured
point sets~\cite{zhao2021pointtransformer}, and transfers to physical fields
because it imposes no grid. Neural operators are constructed so that the
approximation does not depend on the discretization~\cite{kovachki2021neural};
for a model that takes the evaluation set as an argument the corresponding
test is to vary the number of evaluation points at fixed weights
(\S\ref{sec:density_precision}). Global attention over $N$ tokens costs
$O(N^2)$, and essentially every PDE transformer interposes an intermediate set
of size $m \ll N$: a fixed array of
learned latents in Perceiver-IO~\cite{jaegle2022perceiverio} and the Set
Transformer's inducing points~\cite{lee2019set}; orthonormalized neural
eigenfunctions of the kernel integral operator~\cite{xiao2023ono}; learned
physics slices to which tokens are softly assigned~\cite{wu2024transolver};
conditioning tokens in heterogeneous cross-attention with a mixture-of-experts
stage~\cite{hao2023gnot}; or a subsample of the input point set, with surface
points pooled to supernodes~\cite{alkin2024abupt}. In each case the
intermediate set is free parameters, a data-dependent soft partition, or a
subsample of the discretization; in none is it a geometric object supplied by
the problem. The anchors of AB-UPT (anchored-branched universal physics
transformers)~\cite{alkin2024abupt} are a subsample of a surface mesh, so
the anchor count is a sampling hyperparameter and the input contract requires
a surface triangulation; Transolver~\cite{wu2024transolver} recomputes its
slices from token features in every layer, so its coarse space carries no
geometric meaning and its size is chosen a priori. Here the
geometry-conditioning focal set is the centerline itself, so its size is fixed
by the arc length of the vessel; Stage~2 does designate $N_{\text{va}}$ volume
anchors, but they are drawn from the evaluation set and carry no information
the model would not otherwise have, whereas AB-UPT's anchors are the channel
through which surface geometry enters. The baselines of
\S\ref{sec:baseline_comparison} share one input contract, an unstructured
geometric input with a set of evaluation points at matched parameter count.
DeepONet's branch argument is a function on a fixed sensor set shared by no
two patients; Fourier neural operators need a uniform grid;
Geo-FNO~\cite{li2022geofno} and GINO~\cite{li2023gino} map every geometry onto
a fixed latent grid; and mesh-based message passing, of which
MeshGraphNet~\cite{pfaff2021learning} is representative, returns the solution
at the vertices of a mesh it takes as input. These are excluded because they
do not share the contract, not because of accuracy. Attention on point sets also needs coordinates to enter the score:
sinusoidal~\cite{mildenhall2020nerf} and rotary~\cite{su2023rope} encodings
provide that without domain-specific structure, and \S\ref{sec:learned_pe}
replaces the rotary construction with a learned coordinate embedding on the
projected queries and keys.

\paragraph{Machine-learning estimation of the FFR endpoint}
\label{sec:ml_ctffr}
Itu et al.~\cite{itu2016machine} introduced the scalar-feature route to FFR: a
feed-forward network trained on ${\sim}12{,}000$ synthetically generated
stenotic geometries with CFD-derived labels and hand-engineered per-lesion
descriptors, at sub-second inference, subsequently evaluated
clinically~\cite{tesche2018ml}; reduced-order CFD prototypes of on-site ct-FFR
were reported earlier still~\cite{renker2014ffr}. Across that lineage,
including the MACHINE consortium evaluation~\cite{coenen2018ml} and deep models
trained directly on CTA-derived geometry~\cite{kumamaru2020deep}, reported
per-vessel AUC against invasive FFR falls in the ${\sim}0.84$--$0.93$ band,
comparable to the CFD pipelines these models emulate. In that formulation the
output is one scalar per lesion or vessel and the supervision target is itself
a derived scalar, so the volumetric fields are unavailable for any other use;
here the fields are predicted and FFR is extracted from them as a functional.
Per-mesh-vertex prediction of hemodynamic scalars~\cite{nannini2025benchmark}
sits between the two formulations.

\section{Methodology}\label{sec:methodology}

This section states the boundary-value problem being emulated, the map the
surrogate approximates and what its inputs carry, and then the architecture
that realizes the map.

\subsection{Problem formulation}\label{sec:problem}

\paragraph{The boundary-value problem}
Let $g$ denote a coronary geometry and
$\Omega(g)\subset\mathbb{R}^3$ the lumen domain it determines, with boundary
$\partial\Omega(g) = \Gamma_{\text{in}} \cup \Gamma_{\text{wall}} \cup
\bigcup_{i=1}^{N_o}\Gamma^{i}_{\text{out}}$, where $\Gamma_{\text{in}}$ is the
inlet cap, $\Gamma^{i}_{\text{out}}$ the cap of the $i$-th terminal branch, and
$\Gamma_{\text{wall}}$ the vessel wall. Blood is an incompressible Newtonian
fluid in steady flow, so the velocity $\mathbf{u}$ and pressure $p$ satisfy
\begin{align}
    \rho\,(\mathbf{u}\cdot\nabla)\mathbf{u}
      &= -\nabla p + \mu \nabla^{2}\mathbf{u}
      && \text{in } \Omega(g), \label{eq:momentum}\\
    \nabla\cdot\mathbf{u} &= 0
      && \text{in } \Omega(g), \label{eq:continuity}
\end{align}
with dynamic viscosity $\mu$ and density $\rho$. The reference solutions take
$\nu = \mu/\rho = 3.77$~mm$^2$/s and $\rho = 1.06\times 10^{-6}$~kg/mm$^3$ in
the pipeline's self-consistent $(\mathrm{mm},\mathrm{s},\mathrm{kg})$ system,
i.e.\ $\mu = 4.00\times 10^{-3}$~Pa\,s and $\rho = 1{,}060$~kg/m$^3$ in SI.
The boundary conditions are
\begin{align}
    p/\rho &= \bar p_{\text{in}}, \quad \mathbf{u}\ \text{unprescribed}
      && \text{on } \Gamma_{\text{in}}, \label{eq:bc_inlet}\\
    \int_{\Gamma^{i}_{\text{out}}} \mathbf{u}\cdot\mathbf{n}\,\mathrm{d}S
      &= Q_i, \qquad i = 1,\dots,N_o
      && \text{on } \Gamma^{i}_{\text{out}}, \label{eq:bc_outlet}\\
    \mathbf{u} &= \mathbf{0}
      && \text{on } \Gamma_{\text{wall}}. \label{eq:bc_wall}
\end{align}
The inlet carries a fixed kinematic pressure with the velocity left to the
solution. Each terminal branch carries a prescribed volumetric flux, with the
total flow distributed over the outlets in proportion to the square of the
terminal lumen radius,
\begin{equation}
    Q_i \;=\; Q\,\frac{r_i^{2}}{\sum_{k=1}^{N_o} r_k^{2}},
    \label{eq:outlet_split}
\end{equation}
so that the mean velocity is equal across outlets (distinct from the $r^{3}$
scaling of Murray's law~\cite{murray1926physiological}). The wall is rigid and
no-slip.

Two properties of this problem define the parameter set the surrogate is
trained over. First, the outflow distribution is prescribed and not obtained
from a downstream resistance, impedance or Windkessel model; a steady
resistance outlet $p_i = R_i Q_i$ would be admissible in exactly this
formulation and is the more common choice~\cite{vignon2006outflow}. With every
terminal flux prescribed, a lesion cannot divert flow to a parallel branch, so
the simulated pressure drop grows with severity without the flow reduction
that accompanies it physiologically (\S\ref{sec:limitations}), and the total
flow $Q = \sum_{i} Q_i$, not the inlet pressure, parameterizes the operating
state. Second, $Q$ is sampled per case independently of the geometry,
$\mathcal{U}(5{,}000, 8{,}000)$~mm$^3$/s on the left coronary tree and
$\mathcal{U}(2{,}500, 4{,}500)$~mm$^3$/s on the right, so the parameter pair
$(g, Q)$ ranges over a product set and $Q$ carries no mutual information with
the anatomy; the pressure--flow relationship has to be represented explicitly.

\paragraph{Non-dimensionalization}
Let $\bar r$ be the proximal-segment lumen radius and take $L = 2\bar r$ as the
reference length, $U = Q/(\pi \bar r^{2})$ as the reference velocity, and
$\rho U^{2}$ as the reference pressure. In these scales
Eqs.~\eqref{eq:momentum}--\eqref{eq:continuity} carry the single parameter
\begin{equation}
    \mathrm{Re} \;=\; \frac{U L}{\nu} \;=\; \frac{2\,Q}{\pi\,\bar r\,\nu},
    \label{eq:reynolds}
\end{equation}
which over the sampled flow ranges and calibers of this cohort is a few
hundred at the inlet, the regime of resting and hyperemic coronary
flow~\cite{taylor2013ffrcfd} (\appref{app:cfd_pipeline} gives the distribution
and the larger throat values). At that $\mathrm{Re}$ the convective term is
not small, so the approximately linear pressure--flow scaling this paper
relies on has a different origin: away from throats and bifurcations the
segments are slender and the flow nearly unidirectional and nearly fully
developed, and for strictly unidirectional flow
$(\mathbf{u}\cdot\nabla)\mathbf{u}$ vanishes identically, so the axial pressure
gradient balances the viscous wall stress alone and Hagen--Poiseuille holds at
any laminar $\mathrm{Re}$. In a slowly tapering duct the convective
acceleration is largely reversible; the irreversible loss concentrates at
throats and sudden expansions, where it scales quadratically with $Q$. The
pressure drop is therefore linear in $Q$ to leading order away from lesions
and departs from linearity across them, which motivates the multiplicative
flow-rate conditioning of \S\ref{sec:film}.

\paragraph{The supervised pressure variable}
The supervised pressure is the kinematic (static) pressure $p/\rho$ delivered
by the incompressible solver, in mm$^2$/s$^2$; ``pressure'' means this quantity
throughout, and static pressure is the appropriate variable because FFR is a
ratio of static pressures. The solver's pressure is defined up to a constant,
and the gauge used here refers it to the inlet, so $\|p\|$ in
Eq.~\eqref{eq:rel_l2} is the norm of the pressure \emph{drop} through the tree.
This is the conservative choice: referring $p$ to absolute pressure would
inflate the denominator by roughly two orders of magnitude and divide every
relative error reported here by about the same factor, so comparisons against
work reporting on absolute pressure are not like for like.

\paragraph{The solution map}
Eqs.~\eqref{eq:momentum}--\eqref{eq:bc_wall} define a solution map
\begin{equation}
    \mathcal{S} : (g, Q) \;\longmapsto\; (p, \mathbf{u}),
    \qquad
    (p, \mathbf{u}) \in L^{2}(\Omega(g)) \times
    \big[H^{1}(\Omega(g))\big]^{3},
    \label{eq:solution_map}
\end{equation}
assumed well defined and single-valued over the parameter range considered.
The reference solver realizes a finite-volume discretization
$\mathcal{S}_h$ of $\mathcal{S}$, and it is $\mathcal{S}_h$ that supervises
training, so every error reported in this paper is agreement with
$\mathcal{S}_h$. The grid-convergence study of \appref{app:cfd_pipeline}
places $\mathcal{S}_h$ within a median $2.0\%$ (LCA) and $1.4\%$ (RCA) of
$\mathcal{S}$ in a volume-weighted pressure norm and $0.3\%$ in the velocity
norm at production resolution.

\paragraph{The surrogate}
The surrogate approximates $\mathcal{S}$ pointwise. Its arguments are a
centerline $\mathcal{C} = \{(\mathbf{c}_i, r_i)\}_{i=1}^{N_c}$ with
$\mathbf{c}_i \in \mathbb{R}^3$ the position of the $i$-th centerline point and
$r_i > 0$ the local lumen radius; a scalar flow rate $Q\in\mathbb{R}$; and a
finite set of evaluation points
$\mathcal{V} = \{\mathbf{v}_j\}_{j=1}^{N_v} \subset \Omega(g)$. It returns the
pressure and the three velocity components at those points,
\begin{equation}
    \mathcal{F}_\theta : (\mathcal{C}, Q, \mathcal{V})
    \;\longmapsto\; (\mathbf{P}, \mathbf{U})
    = \big\{(\hat p_j, \hat{\mathbf{u}}_j)\big\}_{j=1}^{N_v},
    \label{eq:operator}
\end{equation}
with $\theta$ all trainable parameters and
$\hat{\mathbf{u}}_j = [\hat u_j, \hat v_j, \hat w_j]^\top$. Accuracy is
measured in discrete norms over $\mathcal{V}$; for pressure,
\begin{equation}
    \varepsilon_p(\mathcal{V}) \;=\;
    \frac{\big\|\hat p - p\big\|_{\ell_2(\mathcal{V})}}
         {\big\|p\big\|_{\ell_2(\mathcal{V})}},
    \qquad
    \|x\|_{\ell_2(\mathcal{V})}^{2} = \sum_{j=1}^{N_v} x_j^{2},
    \label{eq:rel_l2}
\end{equation}
and for velocity the same ratio over the full velocity block,
\begin{equation}
    \varepsilon_u(\mathcal{V}) \;=\;
    \frac{\big\|\hat{\mathbf{u}} - \mathbf{u}\big\|_{F}}
         {\big\|\mathbf{u}\big\|_{F}},
    \qquad
    \|\mathbf{u}\|_{F}^{2}
    = \sum_{j=1}^{N_v} \sum_{k \in \{x,y,z\}} u_{j,k}^{2},
    \label{eq:rel_l2_vel}
\end{equation}
which is neither a mean over component-wise ratios nor an error on velocity
magnitude. Both are computed per case on denormalized fields and averaged over
cases. Because
$\mathcal{V}$ is an argument of Eq.~\eqref{eq:operator}, one set of trained
weights answers at any number and placement of evaluation points;
\S\ref{sec:density_precision} measures $\varepsilon$ over a $64$-fold range in
$N_v$. The scope is the \emph{evaluation} set; the centerline is resampled
to a fixed $N_c = 2{,}048$ points throughout.

Two properties Eq.~\eqref{eq:operator} does not have follow from its form.
$\mathcal{F}_\theta$ is a map of the whole set $\mathcal{V}$ and not a
pointwise map applied $N_v$ times: the Stage-2 anchor subsample is drawn from
$\mathcal{V}$ itself, so the predicted value at a fixed point depends on which
other points were requested alongside it, a dependence
\S\ref{sec:frame_sensitivity} measures at $0.019$ on pressure and $0.058$ on
velocity in relative $\ell_2$. And $\mathcal{F}_\theta$ acts on coordinate data
in one fixed frame and is not equivariant under a rigid motion applied jointly
to $\mathcal{C}$ and $\mathcal{V}$, which the boundary-value problem is.

\paragraph{What the centerline argument asserts}
The surrogate does not receive $g$. It receives $\pi(g) = \mathcal{C}$, the
centerline with its radius profile. On the family used here that is not a
restriction: the reference geometries are \emph{generated} from a centerline
and a radius profile by the tubular construction of Eq.~\eqref{eq:tube_map},
discretized by a centerline-based mesher~\cite{decroocq2023mesh} whose
cross-sections are circular by construction, so on this cohort $\pi$ is
injective up to the mesher's deterministic normal-frame roll and
$(\mathbf{c}, r)$ determines $\Omega(g)$ completely; the $N_c = 2{,}048$
centerline points supplied to the surrogate exceed the ${\approx}900$ axial
stations of the reference mesh. Two consequences follow. No claim here rests
on cross-sectional shape being unimportant: what \S\ref{sec:baseline_comparison}
tests is whether, at a comparable geometric input budget, a focal set on the
generating skeleton is more accurate than one subsampled from a surface
discretization of the same lumen, and the surface mesh
AB-UPT~\cite{alkin2024abupt} consumes is a deterministic function of
$(\mathbf{c}, r)$ here that encodes it more coarsely ($8{,}192$ points pooled
to $1{,}024$ supernodes). And injectivity is a property of this cohort, not of
the method, so it governs transfer: applied to lumens with eccentric
cross-sections and asymmetric plaque, $\pi$ ceases to be injective, and since
training minimizes squared error over functions of
$(\pi(g), Q, \mathcal{V})$ under the geometry distribution $\mathcal{D}$, the
minimizer is the conditional expectation over the fiber $\pi^{-1}(\pi(g))$,
\begin{equation}
    \mathcal{F}^{\star}(\pi(g), Q, \mathcal{V}) \;=\;
    \mathbb{E}_{g'\sim\mathcal{D}}
    \Big[\, \mathcal{S}(g', Q)\big|_{\mathcal{V}}
    \;\Big|\; \pi(g') = \pi(g) \Big],
    \label{eq:fiber_expectation}
\end{equation}
whose within-fiber spread becomes an error floor for any model with these
inputs. That floor is identically zero on the present cohort and would have to
be measured on independently segmented lumens.

\paragraph{Why a one-dimensional skeleton is an effective reduced coordinate}
Three properties of Eqs.~\eqref{eq:momentum}--\eqref{eq:bc_wall} make $\pi$
an efficient reduced coordinate for $\mathcal{S}$, and they are properties of
the boundary-value problem and not of the architecture. The solution's
dependence on the domain is strongly anisotropic: by the slenderness argument
above, the axial pressure gradient integrates the radius profile along arc
length, so a change to the trajectory or radius at one station propagates to
the whole downstream subtree, while a change to cross-sectional shape at fixed
area acts locally; a focal set on the centerline spans the direction of
strongest dependence by construction. The axis is an intrinsic coordinate for
this family of domains: away from bifurcations the lumen is a tubular
neighborhood of a curve, and with $\mathbf{c}(s)$ the centerline parameterized
by arc length, $r(s)$ the lumen radius and $(\mathbf{n}_1, \mathbf{n}_2)(s)$ a
normal frame, the map
\begin{equation}
    (s, \boldsymbol{\xi}) \;\longmapsto\;
    \mathbf{c}(s) + r(s)\big(\xi_1 \mathbf{n}_1(s)
    + \xi_2 \mathbf{n}_2(s)\big),
    \qquad |\boldsymbol{\xi}| \le 1,
    \label{eq:tube_map}
\end{equation}
is a diffeomorphism onto the segment whenever the radius stays small relative
to the local radius of curvature and the tube does not self-intersect, so
$\pi$ retains the coordinate system in which axial and cross-sectional
directions separate, and the focal-set size follows the arc length of the
vessel and not the resolution of the discretization. And the alternative
reduction is unavailable: a global reduced basis needs a common reference
domain and a regular deformation family onto it, which varying branch count
and connectivity do not supply, whereas a geometry-attached focal set moves
with the domain. As a numerical scheme, the anchored attention that follows is
a coarse update on the focal set followed by a prolongation to the volume, with
no restriction, because the coarse variables are supplied and need not be
inferred from the fine field; that is a design analogy and not an
approximation bound in $n_f$. It makes one testable prediction, which
\S\ref{sec:error_analysis} measures: if the axis carries the strongest
dependence and is supplied at full resolution, error should not degrade with
operating point, vessel caliber or topological complexity.

\subsection{Inputs and normalization}\label{sec:normalization}

For each case the network is given: an ordered sequence of $N_c$ centerline
points, each carrying the three coordinates $(x, y, z)$ and the lumen radius
$r$; a set of $N_v$ evaluation coordinates $(x, y, z)$; and one scalar, the
inlet flow rate $Q$. Nothing else: no edge list, no branch or segment
identifier, no inlet or outlet flag, no tangent, normal or curvature, no
arc-length coordinate, no per-outlet flux split, no wall distance and no
surface mesh, so branching structure is present only insofar as the
coordinates of the centerline sequence trace it. Every operation on centerline
tokens is either position-wise or attention with the focal set defined by
membership, and the attention-level encoding of \S\ref{sec:pos_encodings}
depends on the token's coordinate alone, so the network is permutation
invariant in the centerline sequence and permutation equivariant in the
evaluation set; we verified this (reversing or shuffling the centerline order
changes the predicted fields by a relative $\ell_2$ of $9\times 10^{-7}$ on
pressure and $2\times 10^{-5}$ on velocity, shuffling the evaluation set by
$1\times 10^{-6}$, all reduction-order noise). The construction has no
built-in invariance to rigid motions of the inputs.

All 3D coordinates, centerline and volume alike, are min--max normalized to
$[0,1]$ using a fixed per-axis bounding box computed over all three splits
(on LCA it coincides exactly with the training-set box),
\begin{equation}
    \tilde{\mathbf{p}} = \frac{\mathbf{p} - \mathbf{p}_{\min}}
    {\mathbf{p}_{\max} - \mathbf{p}_{\min}},
    \label{eq:spatial_norm}
\end{equation}
with $\mathbf{p}_{\min}, \mathbf{p}_{\max} \in \mathbb{R}^3$ held fixed at
inference. Scalar quantities $x \in \{r, Q, p, u, v, w\}$ are z-score
normalized,
\begin{equation}
    \tilde{x} = \frac{x - \mu_x}{\sigma_x},
    \label{eq:zscore}
\end{equation}
with $\mu_x, \sigma_x$ computed once before training and held fixed at
inference, because the pressure
and velocity distributions are heavy-tailed and min--max scaling would compress
the bulk of the distribution. Predictions are denormalized to physical units
before any derived quantity is computed.

\subsection{Spatial position encodings}\label{sec:pos_encodings}

CARS uses two mechanisms: a factored \textbf{sinusoidal embedding}
$\text{PE}_{3D}: \mathbb{R}^3 \to \mathbb{R}^{d_{\text{pos}}}$ used at the
input stage to construct tokens, and an \textbf{attention-level} positional
encoding inside each attention block, which makes the score between two tokens
depend on their coordinates and not only on content similarity. Two
attention-level options are compared in \S\ref{sec:ablations}. The sinusoidal
embedding allocates $d_{\text{pos}}/3$ dimensions per axis (with
$d_{\text{pos}}$ divisible by $6$); for each axis $a$ and pair index
$k = 0, \ldots, d_{\text{pos}}/6 - 1$,
\begin{align}
    \text{PE}_a(p_a, 2k)   &= \sin\!\left(\frac{p_a}{10000^{6k/d_{\text{pos}}}}\right), \\
    \text{PE}_a(p_a, 2k+1) &= \cos\!\left(\frac{p_a}{10000^{6k/d_{\text{pos}}}}\right),
\end{align}
and the per-axis embeddings are concatenated,
\begin{equation}
    \text{PE}_{3D}(\mathbf{p}) = \left[\text{PE}_x(p_x) \,\|\,
    \text{PE}_y(p_y) \,\|\, \text{PE}_z(p_z)\right] \in
    \mathbb{R}^{d_{\text{pos}}}.
\end{equation}
We make no extrapolation claim for the sinusoidal
encoding, because the coordinates are min--max normalized per axis and the one
out-of-range test we run fails outright (\S\ref{sec:cross_territory}).

\subsubsection{3D rotary position embeddings (RoPE3D)}\label{sec:rope3d}

RoPE rotates query and key vectors so that the dot product of the rotated
$\mathbf{q}', \mathbf{k}'$ depends only on the relative position
$\mathbf{p}_q - \mathbf{p}_k$~\cite{su2023rope}. We extend it to 3D by
partitioning each head's $d_{\text{head}}/2$ rotation pairs across the three
axes as evenly as the head width allows and applying a 2D rotation per axis:
with $m_a$ pairs on axis $a$, pair $j$ rotates through
$\theta_a^{(j)} = p_a / 10000^{\,(j-1)/m_a}$, and the rotated query is
$\mathbf{q}' = [\,R_x(\theta_x)\mathbf{q}_x \,|\, R_y(\theta_y)\mathbf{q}_y
\,|\, R_z(\theta_z)\mathbf{q}_z\,]$ with $R_a$ the block-diagonal rotation
over that axis's pairs, and analogously for $\mathbf{k}'$. The score then
depends only on the coordinate displacement, componentwise in each axis, so it
encodes Euclidean proximity directly; RoPE3D adds no learnable parameters, and
a closely related axis-factored rotary encoding is used internally by
AB-UPT~\cite{alkin2024abupt}. The base $10^{4}$ is the language-model default,
where positions are token indices running to $10^{3}$ or beyond; here
coordinates lie in $[0,1]$, so the highest-frequency pair completes at most
$0.16$ of a cycle across the whole domain and nine of the eleven pairs on the
first axis are small-angle everywhere. The learned alternative below can
realize whatever length scale the problem wants, so the margin measured
between them in \S\ref{sec:ablations} confounds the form of the encoding with
the choice of frequency scale; the control, a RoPE3D run with the base
rescaled to the lumen radius, has not been run. The full construction is in
\appref{supp:rope3d}.

\subsubsection{Learned coordinate embedding}\label{sec:learned_pe}

The alternative replaces the rotation by a learned additive bias on the
projected queries and keys. A single two-layer MLP
$f: \mathbb{R}^3 \to \mathbb{R}^{h\,d_{\text{head}}}$, of hidden width $64$
with GELU activation, maps a token's coordinate $\mathbf{p}_i$ to one bias
vector per head,
\begin{equation}
    \mathbf{b}_i = f(\mathbf{p}_i)
    \;\in\; \mathbb{R}^{h \times d_{\text{head}}},
    \label{eq:learned_bias}
\end{equation}
and the head-$h$ component is added to both the query and the key of token $i$
after projection, so with rotation disabled the score between tokens $i$ and
$j$ becomes
\begin{equation}
    \frac{\big(\mathbf{q}_i + \mathbf{b}_{h,i}\big)^{\!\top}
          \big(\mathbf{k}_j + \mathbf{b}_{h,j}\big)}{\sqrt{d_{\text{head}}}}
    \;=\;
    \frac{\mathbf{q}_i^{\!\top}\mathbf{k}_j
        + \mathbf{q}_i^{\!\top}\mathbf{b}_{h,j}
        + \mathbf{b}_{h,i}^{\!\top}\mathbf{k}_j
        + \mathbf{b}_{h,i}^{\!\top}\mathbf{b}_{h,j}}
         {\sqrt{d_{\text{head}}}}.
    \label{eq:learned_bias_score}
\end{equation}
The last summand is a learned function of \emph{both} coordinates and supplies
a position-pair interaction, while the two cross terms let position modulate
content; unlike RoPE3D it is not translation-invariant by construction. The
cost is $12{,}736$ parameters ($+0.3\%$), and the learned variant is the
default for all CARS results in this paper.

\subsection{Tokenization of the vessel and the volume}
\label{sec:token_construction}

Centerline geometry and volume query points are converted into token sequences
of common width $d_{\text{model}}$. Each centerline token combines a
sinusoidal embedding of position with a learned embedding of radius by
additive fusion,
\begin{equation}
    \mathbf{x}_{c,i} = \text{MLP}_{\text{pos}}(\text{PE}_{3D}(\mathbf{c}_i))
    + \text{MLP}_{r}(r_i),
\end{equation}
while volume tokens carry only spatial information,
\begin{equation}
    \mathbf{x}_{v,j} = \text{MLP}_{\text{vol}}(\text{PE}_{3D}(\mathbf{v}_j)).
\end{equation}
The two MLPs share architecture (two-layer GELU) but have independent weights,
and additive fusion is adopted without an ablation isolating it. No a priori
field values are provided at volume tokens, so all hemodynamic information
must be propagated from the centerline tokens through attention.

\subsection{Tiered attention with centerline anchors}
\label{sec:tiered_attention}

With $N_c = 2{,}048$ centerline tokens and $N_v = 16{,}384$ volume tokens
per case at the working point, standard $O(N^2)$ self-attention over
$N = N_c + N_v$ tokens is dominated by volume-to-volume interactions. In the
laminar, pressure-driven, geometry-dominated regime of coronary flow, the
pressure and velocity at an intra-luminal point are determined predominantly
by the upstream centerline trajectory and the local lumen radius, the two
quantities the centerline tokens encode, so the centerline can mediate
long-range hemodynamic information across stenoses and branch points. We
therefore split the token sequence
$\mathbf{X} \in \mathbb{R}^{N \times d_{\text{model}}}$ into a small set of
\textit{focal} tokens $\mathbf{X}_f \in \mathbb{R}^{n_f \times d_{\text{model}}}$
(the centerline) and a large set of \textit{context} tokens (the volume
queries), and constrain attention to use only focal tokens as keys and values.
With projections $\mathbf{Q} = \mathbf{X} W^Q$, $\mathbf{K} = \mathbf{X}_f W^K$,
$\mathbf{V} = \mathbf{X}_f W^V$ and
$W^Q, W^K, W^V \in \mathbb{R}^{d_{\text{model}} \times d_{\text{head}}}$,
\begin{equation}
    \text{TieredAttn}(\mathbf{X}, \mathbf{X}_f) = \mathrm{softmax}\!\left(
    \frac{\mathbf{Q}\mathbf{K}^\top}{\sqrt{d_{\text{head}}}}\right)\mathbf{V}
    \;\in\; \mathbb{R}^{N \times d_{\text{head}}},
    \label{eq:tiered_attn}
\end{equation}
so queries are formed from \emph{all} tokens while keys and values come from
the $n_f$ focal tokens only. The attention matrix has shape $N \times n_f$,
reducing memory from $O(N^2)$ to $O(N \cdot n_f)$ and FLOPs from $O(N^2 d)$ to
$O(N \cdot n_f d)$, roughly a $9\times$ reduction in the geometry-fusion stage
at the settings used throughout ($N \approx 18{,}432$, $n_f = 2{,}048$); the
complete two-stage cost is Eq.~\eqref{eq:complexity}. Because keys and values
are drawn from the focal set alone, information never flows from the context
tokens back onto it: the focal representation is updated by focal-to-focal
attention, and the result is carried to the context tokens as a
softmax-weighted combination of focal values, an interpolation with learned,
query-dependent basis functions forming a partition of unity over the focal
set. The focal tokens are not learned global parameters but are provided by
the case's own centerline.

Tiered attention is applied independently across $h$ heads, each on a
$d_{\text{head}} = d_{\text{model}}/h$ subspace with per-head projections
$W^Q_i, W^K_i, W^V_i$, the head outputs concatenated and projected by
$W^O \in \mathbb{R}^{d_{\text{model}} \times d_{\text{model}}}$,
\begin{equation}
    \text{MultiHead}(\mathbf{X}, \mathbf{X}_f) = [\text{head}_1 \,\|\,
    \cdots \,\|\, \text{head}_h]\, W^O.
\end{equation}
Each CARS block applies multi-head tiered attention followed by a position-wise
feed-forward network, with pre-LayerNorm and residual connections,
\begin{align}
    \mathbf{X}'  &= \mathbf{X} + \text{MultiHead}(\text{LN}(\mathbf{X}),
    \text{LN}(\mathbf{X})_f), \label{eq:block_attn}\\
    \mathbf{X}'' &= \mathbf{X}' + \text{FFN}(\text{LN}(\mathbf{X}')),
    \label{eq:block_ffn}
\end{align}
where $\text{LN}(\cdot)_f$ denotes the first $n_f$ rows of the normalized
sequence and the FFN is a two-layer MLP with GELU activation and hidden width
$4 d_{\text{model}}$; pre-LayerNorm gives more stable training dynamics with
the asymmetric attention pattern~\cite{xiong2020prelayernorm}. The
attention-level positional encoding enters the score inside
Eq.~\eqref{eq:tiered_attn}: RoPE3D yields a locality prior by construction,
whereas the learned embedding can represent one but is not guaranteed to.

\subsection{Two-stage processing}\label{sec:two_stage}

The block stack is split into two stages with different focal/context
partitions. \textit{Stage 1: geometry fusion.} The concatenated sequence
$[\mathbf{X}_c; \mathbf{X}_v]$ is processed by $L_{\text{geo}}{=}4$
tiered-attention blocks with the centerline tokens as focal anchors
($n_f = N_c$). Volume tokens acquire geometry-conditioned representations,
while centerline tokens, which also appear as queries, attend to one another
and propagate information along the vessel tree; the centerline tokens are
then discarded. \textit{Stage 2: field refinement.} The remaining volume tokens
are processed by $L_{\text{refine}}{=}4$ further blocks, where $N_{\text{va}}$
of the tokens act as \textit{volume anchors} (focal) and the remaining queries
attend to them. The anchors are a uniformly random subsample of the evaluation
points, drawn per case with no learned or geometric selection rule, so the
anchor set carries no information beyond the points themselves. The default is
$N_{\text{va}} = 16{,}384$ with no separate query set, equal to the training
evaluation density, so at the operating point of every accuracy number in this
paper every point is an anchor and Stage~2 is full self-attention; the anchor
set becomes a strict subsample only when the model is queried above that
density, the regime \appref{app:complexity} isolates for the linear-cost
measurement. The two stages implement a coarse-to-fine flow: geometry
conditioning establishes large-scale hemodynamic structure, and anchor-based
refinement, in which anchors communicate through the query side of attention
and query points interpolate from refined anchors, improves local consistency.

\subsection{Flow-rate conditioning and decoding}\label{sec:film}

After Stage~2 the volume tokens are modulated by the inlet flow rate $Q$ via
Feature-wise Linear Modulation (FiLM)~\cite{perez2018film}: a small MLP maps
the normalized flow rate $\tilde Q$ to per-feature scale and shift parameters,
applied element-wise to the volume features,
\begin{align}
    [\boldsymbol{\gamma}, \boldsymbol{\beta}] &=
    \text{MLP}_{\text{FiLM}}(\tilde{Q}) \in \mathbb{R}^{2 d_{\text{model}}},
    \label{eq:film_params}\\
    \text{FiLM}(\mathbf{X}_v; Q) &= \boldsymbol{\gamma} \odot
    \mathbf{X}_v + \boldsymbol{\beta},
    \label{eq:film}
\end{align}
with $\boldsymbol{\gamma}, \boldsymbol{\beta}$ broadcast across volume tokens,
and the final layer initialized so that FiLM is the identity at the start of
training. A multiplicative modulation is the physically appropriate form
because pressure scales approximately linearly with $Q$ at fixed geometry over
the sampled range (\S\ref{sec:problem}); with
$\boldsymbol{\gamma} \propto \tilde Q$, Eq.~\eqref{eq:film} reproduces that
$p \propto Q$ scaling up to the affine offsets of the normalization, and the
network need only correct the offset.

A single linear layer then maps each modulated volume token to four output
channels, $[\hat{p}_j, \hat{u}_j, \hat{v}_j, \hat{w}_j] = \mathbf{W}_{\text{out}}
\mathbf{x}_{v,j}'' + \mathbf{b}_{\text{out}}$, so any spatial variation in the
predicted fields is produced upstream of the readout; predictions are
denormalized before any clinical metric is computed.

\section{Experimental Setup}

\subsection{Dataset}\label{sec:dataset}

Coronary geometries are derived from the public ImageCAS
dataset~\cite{zeng2023imagecas} of coronary CT-angiography volumes acquired at
a single Chinese tertiary center; CFD simulations run on these geometries by
the OpenFOAM pipeline of \appref{app:cfd_pipeline} produce the volumetric
pressure and velocity fields used as ground truth. The unit of the dataset is
a point of the parameter set of Eq.~\eqref{eq:solution_map}, not a patient:
each ImageCAS segmentation supplies a centerline trajectory, and cases are
drawn from it by imposing different synthetic focal narrowings on the radius
profile (a radial deformation of the radius profile) and drawing a fresh $Q$
for each. Each $(x, y, z, r, Q)$ tuple is solved once, and the solutions are
partitioned at random into $2{,}700$ training, $300$ validation and $300$ test
cases per territory. The split is over samples of the parameter set, the
standard protocol for a parametric-PDE surrogate, and it is identical for
every model compared in this paper (CARS, all four neural-operator baselines,
every ablation arm, the 1D ROM and the scalar regressors), so all
model-to-model comparisons are like-for-like. The $3{,}300$ LCA cases come
from $101$ base anatomies and the $3{,}300$ RCA cases from $131$, so the
trajectory component of $g$ recurs across splits and cases within a split are
not mutually independent: $39\%$ (LCA) and $57\%$ (RCA) of the variance in
per-case pressure error lies between patients. All confidence intervals on the
FFR endpoint are accordingly computed by resampling patients. A median of
$35$ (LCA) and $28$ (RCA) cases derive from each anatomy, and $34$ (LCA) and
$22$ (RCA) of the $300$ test cases are identical in both trajectory and radius
profile to a training case and differ only in $Q$. Excluding those cases raises
the mean pressure error from $0.117$ to $0.127$ on LCA and from $0.063$ to
$0.065$ on RCA, and the velocity error from $0.147$ to $0.155$ and $0.119$ to
$0.122$ (\appref{sup:dedup}).

\label{sec:generalization_levels}%
A claim that a surrogate ``generalizes'' can mean three things here.
\textbf{G1, parameter-space generalization}: unseen combinations of radius
perturbation and inlet flow rate, on centerline trajectories seen during
training; every field error and endpoint figure in this paper measures this.
\textbf{G2, anatomy generalization}: unseen centerline trajectories, from
segmentations absent from training; the reported numbers do not measure it,
and $101$ and $131$ segmentations leave an anatomy-disjoint partition about a
dozen independent test anatomies, so the experiment was run and is reported in
\appref{supp:anatomy_disjoint}. \textbf{G3, clinical
generalization}: unseen acquisition protocol, scanner vendor, site or
population; not tested here. The cross-territory experiment of
\S\ref{sec:cross_territory} bears on neither G2 nor G3 directly.

Each case contributes a centerline resampled to $2{,}048$ points via
branch-aware arc-length interpolation with per-point lumen radius, $16{,}384$
volume points subsampled per epoch from the simulation domain, and per-point
targets (kinematic pressure and three-component velocity). The volume
subsample is drawn uniformly at random over the \emph{cell centers} of the
reference mesh, not over volume; since the mesher grades cells towards the
wall, with a near-wall cell height of $0.75\%$ of the local radius, this
measure weights the near-wall region more heavily than a volume-weighted norm,
and the field errors of Eqs.~\eqref{eq:rel_l2}--\eqref{eq:rel_l2_vel} are to
be read in that measure. It differs from the volume-weighted norm of the
grid-convergence study, so the two are not directly comparable. Cohort summary
statistics are given in Table~\ref{tab:cohort}.

\begin{table}[!tb]
\centering
\caption{Cohort summary for the LCA and RCA datasets, computed on the
centerlines and CFD outputs used in this work. Patient age and sex are
omitted: ImageCAS~\cite{zeng2023imagecas} reports them for the collection as a
whole and not for the per-territory subsets drawn here, and no analysis is
stratified by them. Lesion severity is Eq.~\eqref{eq:severity}, the fractional
radius deficit relative to the reference shoulder radius; the FFR-positive
rate (CFD-derived $\mathrm{FFR}\le 0.8$ on the filtered $[0.3, 1.0]$ subset)
is reported on the test split.}
\label{tab:cohort}
\begin{tabular}{lcc}
\toprule
 & LCA & RCA \\
\midrule
Cases (train / val / test)        & $2{,}700/300/300$       & $2{,}700/300/300$ \\
Base anatomies (segmentations)    & $101$                   & $131$ \\
Source                            & ImageCAS                & ImageCAS \\
Acquisition                       & Coronary CTA            & Coronary CTA \\
Inlet flow-rate range (mm$^{3}$/s) & $\mathcal{U}(5{,}000, 8{,}000)$ & $\mathcal{U}(2{,}500, 4{,}500)$ \\
\quad realized mean $\pm$ sd      & $6{,}431 \pm 875$       & $3{,}515 \pm 571$ \\
\quad equivalent (mL/min)         & ${\approx}386$          & ${\approx}211$ \\
Centerline length, median (mm)    & $182$                   & $146$ \\
Lesion severity, median           & $0.46$                  & $0.24$ \\
\quad $\ge 0.30$ (test set)       & $91\%$                  & $45\%$ \\
$\mathrm{FFR}\le 0.8$ rate ($n$)  & $21\%$ ($61/286$)       & $32\%$ ($93/293$) \\
\bottomrule
\end{tabular}
\vspace{-2mm}
\end{table}

\subsection{Baselines and comparisons}

CARS is evaluated along two axes. First, against four established
architectures adapted for volumetric hemodynamic prediction:
GNOT~\cite{hao2023gnot}, ONO~\cite{xiao2023ono},
Transolver~\cite{wu2024transolver} and AB-UPT~\cite{alkin2024abupt}, all with
parameter counts within $3\%$ of CARS (range $3.76$--$3.87$M) and trained on
the same dataset with identical preprocessing. Second, through component
ablations that isolate each design choice: positional-encoding mechanism
(RoPE3D vs.\ learned bias), attention structure (tiered vs.\ cross-only vs.\
Perceiver bottleneck), conditioning (FiLM vs.\ concat vs.\ none), anchor design
(centerline vs.\ learned inducing points), and two-stage vs.\ single-stage
architecture.

\subsection{Metrics}\label{sec:metrics}

We report normalized MSE during training (the optimization target), relative
$\ell_2$ error on denormalized pressure and velocity fields, per-case inference
latency on a single A100 80\,GB, and the downstream FFR endpoint (Pearson $r$,
ROC AUC, Bland--Altman bias and limits of agreement, threshold accuracy at
$\mathrm{FFR}\le 0.8$ with $95\%$ CIs) against the CFD-derived label. The
primary endpoint is the mean relative $\ell_2$ error of the \emph{pressure}
field, Eq.~\eqref{eq:rel_l2}, on the held-out test split in the cell-center
measure of \S\ref{sec:dataset}: pressure is the field the clinical functional
depends on and the field on which the reference solver's discretization error
has been quantified. Velocity error, the FFR statistics and cost are secondary
endpoints, and no claim rests on a velocity or endpoint margin alone. This
hierarchy was not registered in advance, so it does not license the
multiplicity control a pre-specified analysis plan would.

\subsection{Training}\label{sec:training}

The model is trained to minimize normalized mean squared error on pressure and
velocity components at every volume query point; relative $\ell_2$ on
denormalized predictions is the evaluation metric, so the numbers are
scale-free across vessels of different size. Optimization uses
AdamW~\cite{loshchilov2019adamw} with learning rate $5\times 10^{-5}$, weight
decay $0.05$, cosine annealing with $5\%$ linear warm-up and an end learning
rate of $10^{-6}$, gradient clipping at global norm $1.0$, dropout $0.1$, and
a parameter EMA with decay $0.9999$ for evaluation. We train at batch size $1$
(larger batches destabilized the AdamW optimization in our setup). Default
architecture: $d_\text{model}{=}192$, $h{=}3$,
$L_\text{geo}{=}L_\text{refine}{=}4$, $d_\text{pos}{=}192$. Two training
budgets are reported: a $100$-epoch budget for matched comparison against the
neural-operator baselines, and a $300$-epoch budget adopted as the primary
configuration after the hyperparameter sweep of \S\ref{sec:hp_sweep} showed
that the model had not saturated at $100$ epochs. Full hyperparameters and
per-baseline configurations are in \appref{app:training} and
\appref{app:baselines}.

\section{Results}

\subsection{It reproduces the reference fields}\label{sec:res_fields}
\label{sec:headline}

\begin{figure}[!t]
    \centering
    \includegraphics[width=0.85\linewidth]{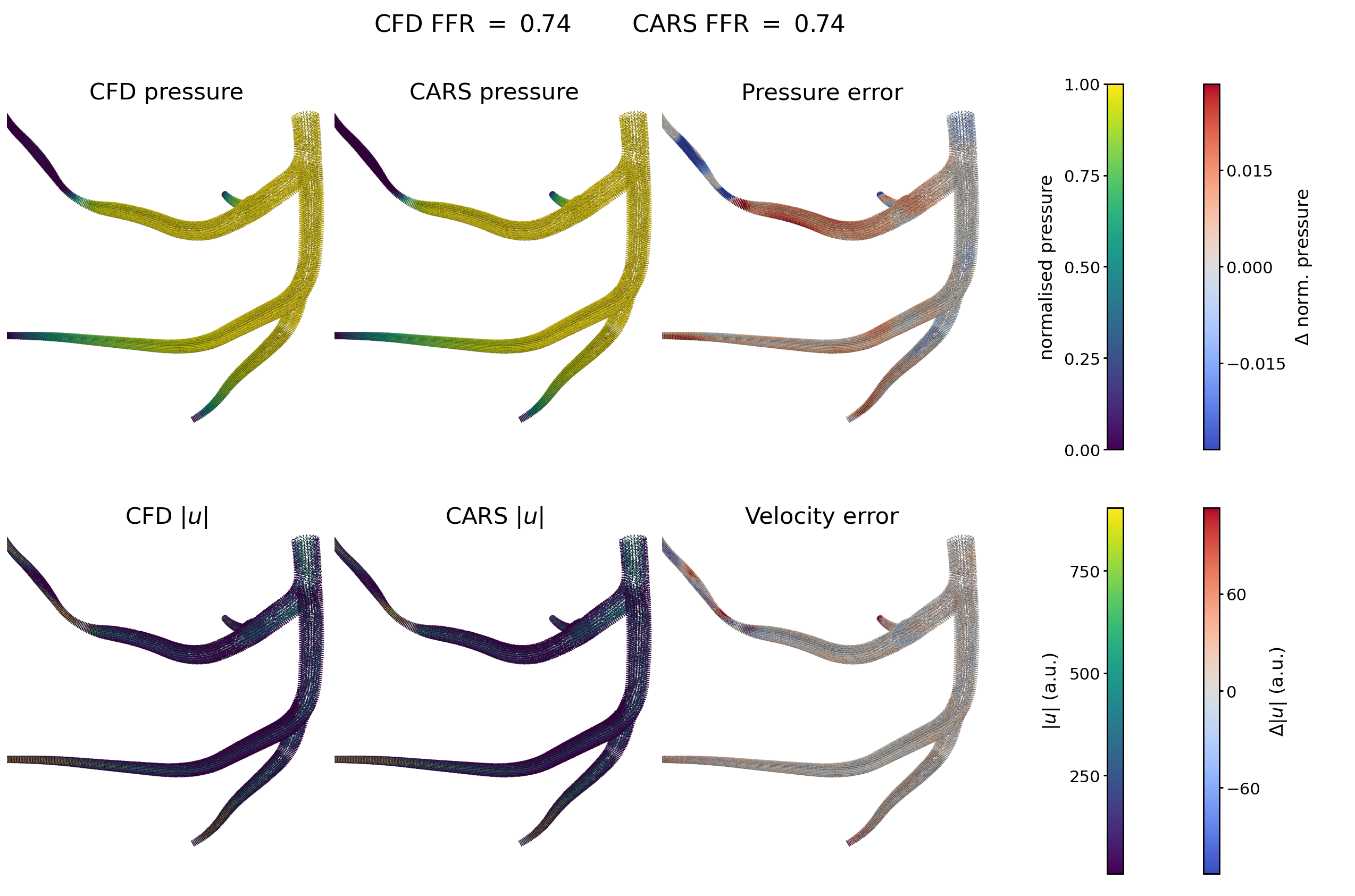}
    \caption{\textbf{Volumetric field prediction versus CFD} on a
    flow-limiting LCA case (CFD $\mathrm{FFR}=0.74$, CARS $0.74$; relative
    $\ell_2$ pressure error $0.060$, just below the cohort median).
    \textit{Top:} normalized kinematic pressure: CFD (left), CARS (center),
    signed error (right). \textit{Bottom:} velocity magnitude, same order. The
    CFD and CARS columns share a color scale; error panels use a symmetric
    diverging scale at the $95$th percentile of $|\text{error}|$. The viewing
    direction projects the tree with least self-overlap and is shared by all
    six panels.}
    \label{fig:field_compare}
\end{figure}

CARS in its primary $300$-epoch configuration reaches a mean test relative
$\ell_2$ error of $0.116$/$0.146$ (pressure / velocity) on LCA and
$0.063$/$0.119$ on RCA against the CFD reference in a single run; retraining
the LCA configuration from three seeds, identical in every respect but the
initialization, gives $0.120 \pm 0.003$ / $0.149 \pm 0.005$ (mean $\pm$ s.d.,
$n = 3$; Table~\ref{tab:baselines}), the figures quoted as the headline, while
the RCA figures are one run with no seed spread measured. The typical case is
considerably more accurate than the mean: on LCA the pressure error has median
$0.065$, roughly half the mean, with ninetieth and ninety-fifth percentiles of
$0.288$ and $0.432$; on RCA the median is $0.041$ with percentiles $0.121$ and
$0.188$; velocity is the more uniform field, its mean and median differing by
$0.017$ (LCA) and $0.014$ (RCA) against $0.053$ and $0.023$ for pressure. The
means are what every
comparison in this paper is made on, for comparability with the baseline
literature, and they are set by a right tail: the worst LCA case reaches
$0.950$ and $3.7\%$ exceed $0.5$, against a worst RCA case of $0.494$ and none
above $0.5$ (Table~\ref{tab:error_tails}); \S\ref{sec:error_analysis} identifies
what the tail cases have in common. Per-case latency is $87$~ms in fp32 on a
single A100. Figure~\ref{fig:field_compare} shows a flow-limiting LCA case
chosen for being typical (pressure error $0.060$, just below the cohort
median; CFD-derived $\mathrm{FFR}$ of $0.74$ reproduced exactly): the
predicted pressure reproduces the along-vessel gradient and the caliber-driven
drop into the distal branches, the predicted velocity magnitude reproduces the
low-velocity bulk and the localized stenosis jet, and the signed residual is
low in amplitude, spatially smooth, of either sign and largest along the
proximal segment, not a localized defect at the throat.

These errors are quoted against a reference whose own accuracy is measured. A
three-level grid-convergence study on the upstream solver
(\appref{app:cfd_pipeline}) puts the discretization error of the reference
fields at a median $2.0\%$ (LCA, interquartile range $1.14$--$5.86\%$) and
$1.36\%$ (RCA, $0.73$--$2.37\%$) in a volume-weighted pressure norm and
$0.3\%$ in the velocity norm at production resolution, with a median observed
order of $2.51$ on pressure. The reference error is smaller than the surrogate
error on the median case by about a factor of three, though in a different
norm. At the $20$~mm endpoint station the reference labels sit a median
$|\Delta \mathrm{FFR}|$ of $0.0010$ from their grid-extrapolated values,
against a surrogate MAE of $0.008$ on RCA and $0.019$ on LCA, roughly eight
and nineteen times the label bias; iteration error is smaller still, at
$5.6\times 10^{-6}$ relative $\ell_2$ on pressure and $7.8\times 10^{-7}$ on
FFR.

\subsection{How much the focal set matters}\label{sec:res_coarse}

Three measurements bear on the generating-object hypothesis of
\S\ref{sec:intro_claims} without identifying it: what happens when the focal
set is replaced, how accuracy depends on its size, and how the complete
construction compares with architectures that reach the same geometry by other
means. The identifying experiment would hold the focal token count, the
geometric information supplied to the focal tokens and the positional
treatment fixed, and vary only where the focal points sit: on the centerline
in one arm, and at the same number of interior points of the same lumen in the
other, each carrying the same per-point radius channel. Neither arm below is
that experiment: the learned-anchor arm removes the radius profile and the
Stage-1 positional encoding along with the geometric attachment, and AB-UPT
differs in encoder, positional treatment and conditioning as well as in anchor
provenance. What the measurements do establish is a bound: the focal set is
not doing the work through sheer density, because accuracy saturates eightfold
below the operating focal count, and not through a parameter advantage,
because every arm is matched to within $3\%$.

\subsubsection{Component ablations}\label{sec:ablations}

CARS makes five architectural choices that depart from generic transformer
surrogates: an attention-level positional encoding, FiLM flow-rate
conditioning, tiered attention with geometry-defined focal tokens,
centerline-derived anchors, and a two-stage architecture. Each is ablated at
parameter-matched budget on the LCA cohort under the $100$-epoch protocol,
reporting \emph{validation} relative $\ell_2$ error (Table~\ref{tab:ablation}).
The component ablations use the RoPE3D variant as reference, and the two rows
whose RoPE3D deltas were small or sign-ambiguous were also run on the
stronger learned-PE backbone;
what each arm receives, and the run-to-run bounds for the deltas, are in
\appref{app:reablation}.

\begin{table}[!tb]
\centering
\caption{Component ablations on the LCA dataset at the matched $100$-epoch
budget, validation relative $\ell_2$ error, all configurations
parameter-matched to within ${\sim}3\%$ ($d_\text{model}=168$ for the
Perceiver variant; ``Remove FiLM'' deletes the conditioning pathway). The
component rows use the RoPE3D
reference ($0.205/0.287$); the paired positional-encoding head-to-head was a
separate re-run whose RoPE3D figure ($0.199/0.269$) differs from it by the
run-to-run spread (\appref{app:reablation}). ``noRoPE'' removes the rotary
encoding inside attention; ``cross-only'' additionally removes self-attention
among focal tokens. Learned-PE-backbone rows are compared against the
learned-PE reference.}
\label{tab:ablation}
\footnotesize
\setlength{\tabcolsep}{3pt}
\begin{tabular}{llrcc}
\toprule
Group & Variant & Params & Pressure & Velocity \\
\midrule
\textit{Reference}  & CARS, RoPE3D (component reference)   & 3{,}844{,}548 & 0.205 & 0.287 \\
\addlinespace[2pt]
\textit{Positional} & RoPE3D (paired re-run)               & 3{,}844{,}548 & 0.199 & 0.269 \\
\textit{encoding}   & Learned PE (default)                 & 3{,}857{,}284 & \textbf{0.140} & \textbf{0.196} \\
\addlinespace[2pt]
\textit{Attention}  & Tiered noRoPE                        & 3{,}844{,}548 & 0.221 & 0.275 \\
                    & Cross-only (noRoPE)                  & 3{,}844{,}548 & 0.262 & 0.638 \\
                    & Tiered $\to$ Perceiver               & 3{,}951{,}372 & 0.354 & 0.559 \\
\addlinespace[2pt]
\textit{Conditioning} & FiLM $\to$ Concat                  & 3{,}844{,}356 & 0.224 & 0.240 \\
                    & Remove FiLM                          & 3{,}745{,}348 & 0.329 & 0.487 \\
                    & FiLM $\to$ Concat, learned-PE backbone & 3{,}857{,}092 & 0.164 & 0.220 \\
\addlinespace[2pt]
\textit{Architecture} & CL anchors $\to$ Learned ($512$)   & 3{,}831{,}300 & 0.385 & 0.434 \\
                    & Two-stage $\to$ Single-stage         & 3{,}844{,}548 & 0.176 & 0.386 \\
                    & Two-stage $\to$ Single, learned-PE backbone & 3{,}857{,}284 & 0.150 & 0.261 \\
\bottomrule
\end{tabular}
\vspace{-2mm}
\end{table}

\paragraph{Positional encoding}
A small learned coordinate embedding ($+0.3\%$ parameters) reduces pressure
error by $30\%$ and velocity error by $27\%$ relative to the parameter-free 3D
rotary construction ($0.199/0.269 \to 0.140/0.196$), and is the default in all
subsequent CARS configurations. The margin is a comparison of implementations:
the rotary arm inherits the $10^{4}$ wavelength base from language modeling
while coordinates here live in $[0,1]$ (\S\ref{sec:rope3d}), so part of the
margin may be the frequency scale and the comparison at a tuned base is a
control we have not run.

\paragraph{Focal self-attention and the focal representation}
Removing self-attention among focal tokens (cross-only) more than doubles
velocity error; that variant also drops the rotary encoding, so the controlled
comparison is against the noRoPE row, $0.275 \to 0.638$. Replacing tiered
attention with a Perceiver-style latent bottleneck degrades both fields
($0.354/0.559$). Replacing the geometry-provided centerline anchors with $512$
learned inducing points nearly doubles validation pressure error
($0.205 \to 0.385$) and gives $0.434$ on velocity; that arm removes the
centerline coordinates, the radius profile and the Stage-1 positional encoding
along with the geometric attachment (\appref{app:reablation}), so it bounds
the value of the whole geometry-attached focal representation and not of
geometric attachment alone.

\paragraph{Conditioning and staging}
Removing FiLM, so that $Q$ does not reach the network at all, degrades both
fields substantially ($0.329/0.487$). Replacing FiLM with concatenation is
intermediate on the RoPE3D backbone but resolves uniformly in FiLM's favor on
the learned-PE backbone ($0.164/0.220$ against $0.140/0.196$). Collapsing the
two stages into one shows a pressure--velocity trade-off on RoPE3D
($0.176/0.386$) and resolves uniformly in favor of two-stage on the learned-PE
backbone ($0.150/0.261$ against $0.140/0.196$).

\paragraph{Accuracy against the focal count}
Since cost is nearly flat in $n_f$ (\appref{app:complexity}), what accuracy
does over the same range is the question that matters for the coarse-space
reading. Thinning the $2{,}048$-point centerline by uniform arc-length stride
at fixed trained weights costs $0.0024$ on pressure and $0.0021$ on velocity
for an eightfold reduction to $256$ points, comparable to the seed-to-seed
standard deviation and an order of magnitude below the error itself; only at
$128$ points does the error rise materially, by $0.017$ on pressure
(Table~\ref{tab:nf_accuracy}). The knee sits near $256$--$512$, where the arc-length
spacing on a median $182$~mm LCA centerline is $0.7$--$0.36$~mm, comparable to
the median lumen radius: the focal set has to resolve the radius profile at
the scale of the lumen, and refining it beyond that buys nothing. The default
$n_f = 2{,}048$ is therefore well past the point of diminishing returns, which
removes one explanation of the margins above and in
\S\ref{sec:baseline_comparison}, namely that CARS wins by attending to more
geometry than the alternatives; it does not by itself show that the
\emph{provenance} of the focal set is what matters.

\paragraph{Run-to-run bounds}
Every ablation is a single seed. Retraining the $100$-epoch learned-PE
configuration from three seeds gives a seed-to-seed standard deviation of
$0.0026$ on pressure and $0.0058$ on velocity, and the two RoPE3D runs above
give a more conservative two-point spread of $0.006$ and $0.018$; against
either, the architectural effects the paper draws conclusions from are larger
by $4$--$25\times$ on the relevant field (\appref{app:reablation}).

\subsubsection{Comparison with neural-operator baselines}
\label{sec:baseline_comparison}

Table~\ref{tab:baselines} compares CARS against GNOT~\cite{hao2023gnot},
ONO~\cite{xiao2023ono}, Transolver~\cite{wu2024transolver} and an AB-UPT
baseline~\cite{alkin2024abupt} from the same architectural family, all matched
in parameter count ($3.76$--$3.87$M) and trained on the same data with
identical preprocessing. Two properties bound the comparison
(\appref{app:baselines}): the baselines were trained at CARS's optimizer
settings and not at settings selected for each, and the flow-rate conditioning
pathway is not matched across arms, so a quantified share of the gap over the
three non-FiLM baselines is attributable to conditioning. The first three
baselines are conditioned on the centerline only; AB-UPT consumes the surface
mesh, which on this cohort encodes the same geometry the centerline determines
(\S\ref{sec:problem}), so that comparison isolates the encoding. Under the
matched $100$-epoch budget, CARS test error is $0.150/0.199$ against
$0.444/0.686$ for the strongest non-AB-UPT baseline (GNOT), a $66\%/71\%$
reduction; the val-to-test gap is small (${\sim}7\%$ pressure, ${<}2\%$
velocity), and the parameter-free RoPE3D variant of CARS ($0.201/0.272$) still
beats every baseline on pressure and all but AB-UPT on velocity, so the gain
is not carried by the positional encoding alone. AB-UPT is the strongest
baseline and shares anchor attention, rotary encoding and FiLM with CARS, so
the residual gap localizes CARS's three additions: tiered centerline-focal
attention, the learned coordinate embedding, and the two-stage separation.
Retraining both arms from the same three random initializations gives
$0.153 \pm 0.003$ / $0.205 \pm 0.006$ for CARS against $0.205 \pm 0.003$ /
$0.231 \pm 0.008$ for AB-UPT at $100$ epochs, paired differences of
$0.052 \pm 0.004$ and $0.026 \pm 0.012$ ($t = 23.2$, $p = 0.002$ and
$t = 3.8$, $p = 0.062$ on two degrees of freedom). At $300$ epochs, the budget
the headline rests on, the same three seeds give a margin of $0.039$ on
pressure and $0.023$ on velocity ($0.039 \pm 0.014$ and $0.023 \pm 0.012$;
$t = 4.6$, $p = 0.044$ and $t = 3.3$, $p = 0.079$), with CARS ahead on both
fields at every seed. At $n = 3$ a rank test cannot go below $p = 0.25$ and
the velocity margin misses $\alpha = 0.05$; what the design supports is that
the difference is consistent in direction across every pair and several times
the within-arm spread. The margin narrows by only a quarter between budgets,
so the advantage is not primarily one of optimization speed; AB-UPT's seed
spread is four times CARS's; and the single AB-UPT run in the table is that
arm's best of four. Neither arm is at its asymptote, so this is a
fixed-budget comparison. In latency, the pure-transformer baselines are faster
($9$--$23$~ms) because they avoid the concatenated focal--context attention
over ${\sim}18{,}432$ tokens, and AB-UPT is slower ($166$~ms) because of its
branched surface+volume decoder; the one model in the same accuracy band as
CARS is ${\sim}2\times$ slower and every faster model is at least $2\times$
less accurate (Figure~\ref{fig:pareto}). A dimensionally consistent
one-dimensional Poiseuille--Bernoulli reduced-order model driven by the same
$r^2$ outflow split reaches threshold agreement of only $49.0\%/77.1\%$ on the
FFR endpoint (LCA/RCA), below the majority-class rate on LCA, because it does
not resolve the trans-stenotic loss that sets the endpoint (\appref{supp:rom}).

\begin{table}[!tb]
\centering
\caption{Baseline comparison on the LCA dataset at parameter counts within
$3\%$ of CARS. Relative $\ell_2$ error on pressure / velocity in fp32; latency
(ms) is the median per-case fp32 forward pass on a single A100 80\,GB (batch $1$,
$16{,}384$ volume points, $2{,}048$ centerline points). $^{\dagger}$ marks
models that consume the surface mesh in place of the centerline. The $300$-epoch single
runs are each arm's most favorable seed, so the single-run gap is not the
margin; matched-seed rows retrain both arms from the same three
initializations (mean $\pm$ s.d.), and the $300$-epoch matched row is the
headline. The reference solver's median wall-clock is $572$~s (RCA) and
$948$--$1{,}063$~s (LCA) on eight MPI ranks of an Azure F16s\_v2 node
(\S\ref{sec:cost}). $^{\ast}$bfloat16. n/a: latency and validation error
are reported for the single runs only.}
\label{tab:baselines}
\setlength{\tabcolsep}{2pt}
\scriptsize
\begin{tabular}{lrrcccc}
\toprule
Model & Params & ms & Val.\ $p$ & Val.\ $\mathbf{u}$ & Test $p$ & Test $\mathbf{u}$ \\
\midrule
\multicolumn{7}{l}{\emph{$100$ epochs, single run}} \\
CARS (learned PE)   & 3.86M & $87$  & \textbf{0.140} & \textbf{0.196} & \textbf{0.150} & \textbf{0.199} \\
CARS (RoPE3D)       & 3.84M & $87$  & 0.199 & 0.269 & 0.201 & 0.272 \\
AB-UPT$^{\dagger}$  & 3.76M & $166$ & 0.197 & 0.252 & 0.211 & 0.254 \\
Transolver          & 3.81M & $8.7$ & 0.518 & 0.714 & 0.514 & 0.718 \\
GNOT                & 3.87M & $21.4$ & 0.430 & 0.688 & 0.444 & 0.686 \\
ONO                 & 3.84M & $22.5$ & 0.835 & 0.970 & 0.804 & 0.957 \\
\addlinespace[3pt]
\multicolumn{7}{l}{\emph{$100$ epochs, three matched initializations per arm}} \\
CARS (learned PE)   & 3.86M & n/a & n/a & n/a & $\mathbf{0.153 \pm 0.003}$ & $\mathbf{0.205 \pm 0.006}$ \\
AB-UPT$^{\dagger}$  & 3.76M & n/a & n/a & n/a & $0.205 \pm 0.003$ & $0.231 \pm 0.008$ \\
\addlinespace[3pt]
\multicolumn{7}{l}{\emph{$300$ epochs, single run (CARS seed $42$, AB-UPT seed $1234$)}} \\
CARS (learned PE)   & 3.86M & $87$ / $15^{\ast}$ & \textbf{0.108} & \textbf{0.141} & \textbf{0.116} & \textbf{0.146} \\
AB-UPT$^{\dagger}$  & 3.76M & $166$ & 0.133 & 0.157 & 0.130 & 0.162 \\
\addlinespace[3pt]
\multicolumn{7}{l}{\emph{$300$ epochs, three matched initializations per arm (headline)}} \\
CARS (learned PE)   & 3.86M & n/a & n/a & n/a & $\mathbf{0.1198 \pm 0.0032}$ & $\mathbf{0.1491 \pm 0.0047}$ \\
AB-UPT$^{\dagger}$  & 3.76M & n/a & n/a & n/a & $0.1584 \pm 0.0138$ & $0.1723 \pm 0.0106$ \\
\bottomrule
\end{tabular}
\vspace{-2mm}
\end{table}

\subsection{What it costs}\label{sec:res_cost}
\label{sec:hp_sweep}

Cost enters this construction twice: in how accuracy scales with the training
budget and model size, and in how inference cost scales with the number of
points at which the field is requested. A three-axis sweep over training
length, depth and width (Table~\ref{tab:hp_sweep}) shows training length is the
most effective lever per unit compute: $100\!\to\!300$ epochs at the reference
size reduces validation pressure error by $23\%$ and velocity error by $28\%$
at no parameter cost, doubling depth to $8{+}8$ blocks at $100$ epochs matches
what $200$ epochs achieves at ${\sim}2\times$ the parameters and
${\sim}1.5\times$ the wall-clock per epoch, and widening to
$d_\text{model}{=}256$ does not pay off within the $100$-epoch budget. The
$300$-epoch run is therefore the primary configuration, with validation
pressure still improving at that budget.

\label{sec:density_precision}%
Three properties of the query-based formulation are measured on the trained
checkpoint with no retraining (Table~\ref{tab:density_precision}). First,
evaluation density. Both fields converge from above: pressure is degraded only
at the sparsest setting ($0.137$ LCA, $0.080$ RCA at $1{,}024$ points) and
then moves by at most $0.004$ over the $16$-fold range from $4{,}096$ to
$65{,}536$. Taking $0.005$ as the equivalence margin, an order of magnitude above the
standard deviation from resampling evaluation points
(\appref{app:reablation}) and comparable to the seed-to-seed spread, the
whole interval lies
inside the margin, so pressure error is statistically indistinguishable across
a sixteenfold change in evaluation density; it does not follow that the map is
discretization invariant in the sense a neural operator establishes by
construction, and the result says nothing outside the tested range. Velocity
is slower, reaching within $5\%$ of its converged value at the trained density
of $16{,}384$ and within $2\%$ by $32{,}768$. Second, the linear-cost regime.
Cost is linear in the evaluation count only when the Stage-2 anchor set is
held fixed as that count grows (Eq.~\eqref{eq:complexity}), and every accuracy
number so far is measured at $N_{va} = N_v$, where it is not. Holding the
evaluation budget at $16{,}384$ points and reducing the anchor count, with the
remainder supplied as pure query points, an eightfold reduction of the anchor
set from $16{,}384$ to $2{,}048$ costs $0.008$ on pressure error ($7\%$ in
relative terms, while removing $87\%$ of the quadratic term) and $0.073$ on
velocity, a $50\%$ relative increase; pressure error at $N_{va} = 1{,}024$
with $15{,}360$ further query points ($0.138$) agrees to $0.001$ with the error
when only $1{,}024$ points are requested ($0.137$), so accuracy is governed by
the anchor count and query points beyond the anchor set are interpolated from
it. A deployment that wants the FFR endpoint, a functional of pressure alone,
at high evaluation density can therefore take the linear regime cheaply; one
that wants resolved velocity fields cannot. Training directly in the linear
regime, at $N_{va} = 2{,}048$ and $22.5$~ms per case, is worse on pressure
($0.141$ against $0.125$ for the flagship thinned to the same anchor count)
and better on velocity ($0.202$ against $0.220$); the full fixed-anchor sweep
and the measured scaling slopes are in \appref{app:complexity}. Third,
precision. bfloat16 raises pressure error by $21\%$ on LCA
($0.117 \to 0.141$) and $37\%$ on RCA ($0.064 \to 0.088$) and degrades
velocity by a factor of ${\approx}3.6$; the FFR endpoint is correspondingly
robust, threshold accuracy moving from $97.6\%$ to $96.9\%$ on LCA and from
$96.6\%$ to $96.9\%$ on RCA (nearest-point protocol, physiologic-range
subset). All field errors and endpoint metrics in this paper are therefore
reported in fp32 at $87$~ms per case, and the $15$~ms bfloat16 figure is an
operating point for the FFR endpoint specifically.

\begin{table}[!tb]
\centering
\caption{Evaluation-density and precision sensitivity, mean relative $\ell_2$
over the full $300$-case test split of each territory, trained weights
unchanged; every requested point is a Stage-2 anchor. The fp32 precision row
is the $16{,}384$ density row on the identical evaluation points. The
trained-density row reproduces the single-run headline to within $0.001$.}
\label{tab:density_precision}
\small
\begin{tabular}{llcccc}
\toprule
 & & \multicolumn{2}{c}{LCA} & \multicolumn{2}{c}{RCA} \\
\cmidrule(lr){3-4}\cmidrule(lr){5-6}
 & Setting & Pressure & Velocity & Pressure & Velocity \\
\midrule
\multirow{5}{*}{Density (fp32)}
 & $1{,}024$          & 0.137 & 0.302 & 0.080 & 0.267 \\
 & $4{,}096$          & 0.120 & 0.175 & 0.065 & 0.140 \\
 & $16{,}384$ (trained) & 0.117 & 0.147 & 0.064 & 0.118 \\
 & $32{,}768$         & 0.116 & 0.143 & 0.063 & 0.116 \\
 & $65{,}536$         & 0.116 & 0.141 & 0.063 & 0.115 \\
\addlinespace[2pt]
\multirow{2}{*}{Precision}
 & fp32               & 0.117 & 0.147 & 0.064 & 0.118 \\
 & bfloat16           & 0.141 & 0.532 & 0.088 & 0.426 \\
\bottomrule
\end{tabular}
\vspace{-2mm}
\end{table}

\subsection{The derived endpoint}\label{sec:res_endpoint}

\subsubsection{Fractional flow reserve endpoint}\label{sec:clinical_ffr}

Field norms average over the domain. The FFR endpoint does the opposite: it is
a scalar functional of the predicted pressure field evaluated at a single
point, so it tests the field where a downstream decision is taken. Given a
predicted pressure field $\hat p$ on $\Omega(g)$, the endpoint is
\begin{equation}
    \mathrm{FFR}[\hat p] \;=\;
    \frac{P_{\mathrm{a}} + c_\rho\,\hat p\big(\mathbf{x}_{\mathrm{m}}(\mathcal{C})\big)}
         {P_{\mathrm{a}}},
    \qquad
    c_\rho = 1.06\times 10^{-3}~\mathrm{Pa}\,/\,(\mathrm{mm}^2/\mathrm{s}^2),
    \label{eq:ffr_functional}
\end{equation}
with $P_{\mathrm{a}}$ the aortic reference pressure and
$\mathbf{x}_{\mathrm{m}}(\mathcal{C}) \in \Omega(g)$ a measurement point
determined by the centerline alone: the lesion is the eligible centerline
point of maximum radius deficit (Eq.~\eqref{eq:severity}) and
$\mathbf{x}_{\mathrm{m}}$ lies $20$~mm of arc length distal to it within the
same branch (\appref{app:ffr_protocol}). Because $\mathbf{x}_{\mathrm{m}}$
depends only on $\mathcal{C}$, the identical functional is applied to the
predicted and to the reference field, and any difference in $\mathrm{FFR}$ is
a difference in the pressure field at that point. Because the evaluation set
is an argument of the surrogate, the predicted pressure is read directly at
$\mathbf{x}_{\mathrm{m}}$ by appending it to the standard $16{,}384$-point
query set, which removes the sampling variability of a nearest-neighbor
lookup (a median $0.31$~mm (LCA) and $0.35$~mm (RCA) away, contributing about
$0.01$ of run-to-run variation in $r$); the reference is taken at the nearest
production cell center, a median $0.12$~mm (LCA) and $0.19$~mm (RCA) away, and
inverse-distance weighting over the eight nearest centers changes every
statistic by less than $0.001$. Exact querying changes no threshold decision
on LCA and one on RCA, and improves the full-cohort LCA correlation from
$0.921$ to $0.941$ (Table~\ref{tab:ffr_protocol}). Every endpoint figure in this
subsection and in Table~\ref{tab:ffr_ci} uses it; the sensitivity analyses of
\S\ref{sec:robustness_uq} and the precision comparison of
\S\ref{sec:density_precision} retain the nearest-point protocol on the
physiologic-range subset, because each compares rows against a baseline
computed the same way, so their baselines read $97.6$--$97.9\%$ (LCA) and $96.2$--$96.6\%$ (RCA), and
no trend
within them is affected.

\begin{figure}[!t]
    \centering
    \begin{subfigure}[t]{0.48\linewidth}
        \centering
        \includegraphics[width=\linewidth]{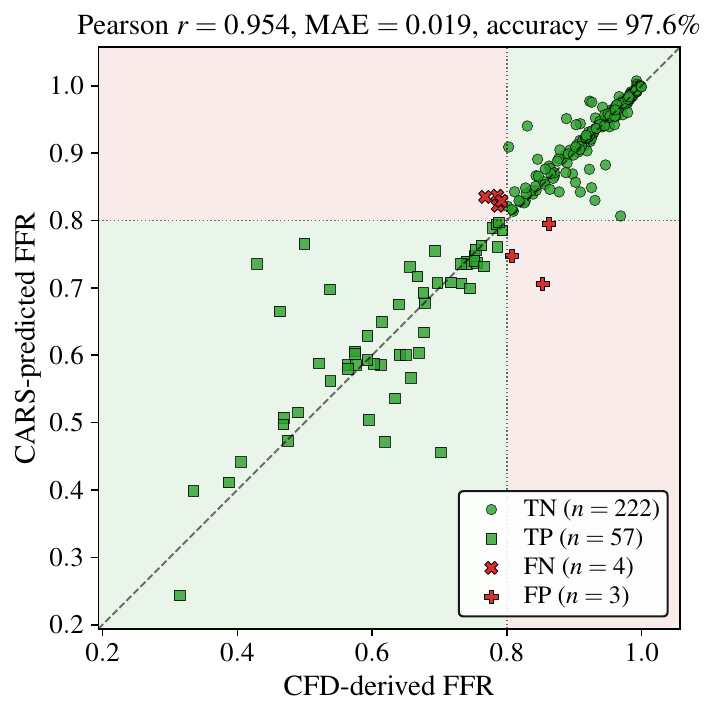}
        \caption{LCA}
    \end{subfigure}\hfill
    \begin{subfigure}[t]{0.48\linewidth}
        \centering
        \includegraphics[width=\linewidth]{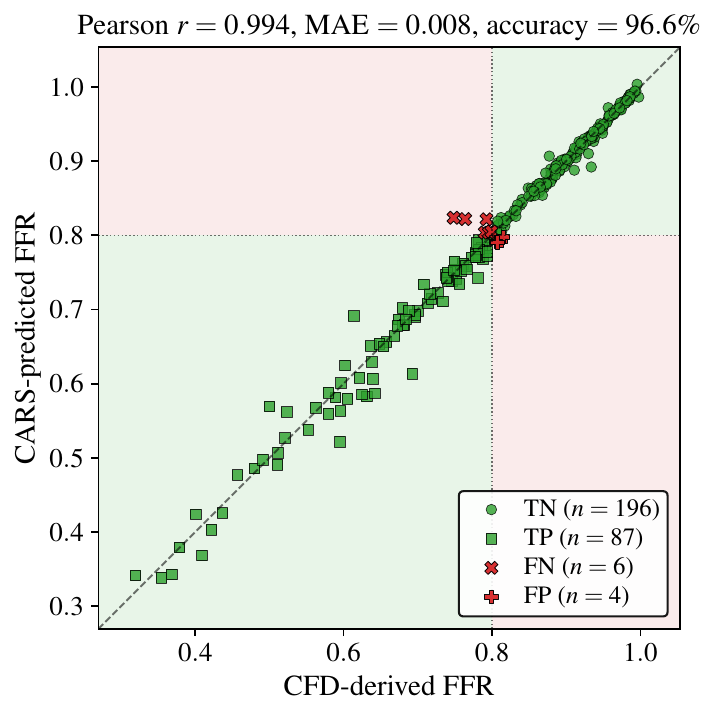}
        \caption{RCA}
    \end{subfigure}
    \caption{Predicted vs.\ CFD-derived FFR on the physiologic-range subset of
    each territory, under the nearest-point lookup
    ($r = 0.954$/$0.994$); the primary exact-coordinate protocol of
    \S\ref{sec:clinical_ffr} gives $r = 0.953$/$0.994$ on the same cases.
    Dotted lines mark the clinical $\mathrm{FFR}=0.8$ threshold; dashed line
    $y=x$.}
    \label{fig:ffr_correlation}
\end{figure}

On the full $300$-case test split of each territory, with no exclusion of any
kind, predicted and reference FFR agree at Pearson $r = 0.941$/$0.993$ on
LCA/RCA with MAE $0.026$/$0.010$ and agreement at the $\mathrm{FFR}\le 0.8$
threshold of $97.7\%$/$96.3\%$ ($3$/$5$ false positives and $4$/$6$ false
negatives; Figure~\ref{fig:ffr_correlation}, Table~\ref{tab:ffr_ci}). This
full cohort is the primary analysis and the figure to compare against external
reports. Restricting to the $286$/$293$ cases whose reference FFR lies in the
physiologic range $[0.3, 1.0]$, the subset every sensitivity analysis uses so
that its rows are mutually comparable, gives $r = 0.953$/$0.994$, MAE
$0.019$/$0.008$ and $97.6\%$/$96.2\%$; the excluded tail is dominated by total
occlusion and severe diffuse disease and changes the decision statistic by at
most one case (\appref{app:ffr_agreement}). Bland--Altman bias is negligible
($+0.003$ LCA, $-0.001$ RCA) with $95\%$ limits of agreement
$[-0.082,+0.088]$ and $[-0.030,+0.029]$, and ROC AUC for the threshold
decision is $0.995$/$0.995$ on that subset and $0.996$/$0.996$ on the full
cohort (\appref{app:ffr_agreement}, Table~\ref{tab:ffr_ci}). An Itu-style scalar
regressor~\cite{itu2016machine} and an XGBoost upper bound trained on the same
splits and labels from $15$ hand-engineered per-vessel geometric features reach
AUC $0.92/0.88$ and $0.96/0.93$ on LCA/RCA against CARS's $0.995/0.995$ on the
same subset, with non-overlapping bootstrap $95\%$ CIs (Table~\ref{tab:ffr_baselines}), so the volumetric formulation contributes information for the
endpoint that scalar feature regression does not extract; these AUCs are not
comparable to the ${\sim}0.84$--$0.93$ band of the prior
literature~\cite{coenen2018ml,tesche2018ml,kumamaru2020deep}, which is
measured against \emph{invasive} FFR on clinical cohorts. Evaluating
Eq.~\eqref{eq:ffr_functional} adds millisecond-scale post-processing to the
forward pass.

\begin{table}[!tb]
\centering
\caption{Threshold-level operating point at $\mathrm{FFR}\le 0.8$ under the
exact-coordinate protocol, on the full test split (primary) and on the
physiologic-range $\mathrm{FFR}_{\mathrm{gt}}\in[0.3, 1.0]$ subset. Intervals
are patient-clustered nonparametric bootstrap $95\%$ CIs
($n_{\mathrm{boot}}=10{,}000$): patients are resampled with replacement and
all cases of a drawn patient enter together, so the $300$ cases are treated as
$96$ (LCA) and $111$ (RCA) independent units, and the subsets as $94$ and
$110$.}
\label{tab:ffr_ci}
\setlength{\tabcolsep}{3.5pt}
\scriptsize
\begin{tabular}{lllll}
\toprule
 & \multicolumn{2}{c}{LCA} & \multicolumn{2}{c}{RCA} \\
\cmidrule(lr){2-3}\cmidrule(lr){4-5}
Metric & full ($300$) & subset ($286$) & full ($300$) & subset ($293$) \\
\midrule
Pearson $r$  & $0.941\;[0.921, 0.979]$ & $0.953\;[0.923, 0.974]$ & $0.993\;[0.987, 0.996]$ & $0.994\;[0.991, 0.996]$ \\
MAE          & $0.026\;[0.017, 0.038]$ & $0.019\;[0.015, 0.024]$ & $0.010\;[0.008, 0.013]$ & $0.008\;[0.007, 0.010]$ \\
ROC AUC      & $0.996\;[0.990, 0.999]$ & $0.995\;[0.989, 0.999]$ & $0.996\;[0.991, 0.999]$ & $0.995\;[0.990, 0.999]$ \\
Sensitivity  & $0.941\;[0.875, 0.987]$ & $0.934\;[0.861, 0.986]$ & $0.940\;[0.894, 0.980]$ & $0.935\;[0.888, 0.978]$ \\
Specificity  & $0.987\;[0.971, 1.000]$ & $0.987\;[0.970, 1.000]$ & $0.975\;[0.948, 0.995]$ & $0.975\;[0.948, 0.995]$ \\
Accuracy     & $0.977\;[0.959, 0.991]$ & $0.976\;[0.957, 0.990]$ & $0.963\;[0.942, 0.983]$ & $0.962\;[0.939, 0.983]$ \\
\bottomrule
\end{tabular}
\vspace{-2mm}
\end{table}

\subsubsection{Parametric sensitivity to the geometric inputs}
\label{sec:robustness_uq}

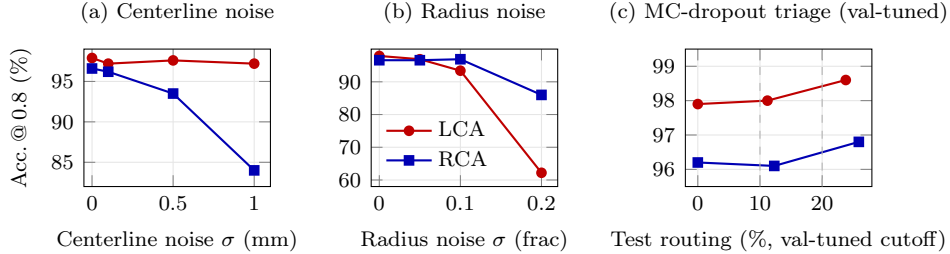
\begin{figure}[!t]
\centering
\begin{tikzpicture}
\begin{axis}[
    name=plotA,
    width=0.32\linewidth, height=3.4cm,
    xlabel={Centerline noise $\sigma$ (mm)},
    ylabel={Acc.\,@\,$0.8$ (\%)},
    xmin=-0.05, xmax=1.1, ymin=82, ymax=99,
    grid=major, grid style={gray!20},
    label style={font=\scriptsize}, tick label style={font=\scriptsize},
    title={(a) Centerline noise}, title style={font=\scriptsize},
    every axis plot/.append style={mark size=1.6pt, thick},
]
\addplot[mark=*, color=red!75!black] coordinates
    {(0,97.9) (0.1,97.2) (0.5,97.6) (1.0,97.2)};
\addplot[mark=square*, color=blue!70!black] coordinates
    {(0,96.6) (0.1,96.2) (0.5,93.5) (1.0,84.0)};
\end{axis}
\begin{axis}[
    name=plotB,
    at={(plotA.right of north east)}, anchor=left of north west,
    xshift=8pt,
    width=0.32\linewidth, height=3.4cm,
    xlabel={Radius noise $\sigma$ (frac)},
    xmin=-0.01, xmax=0.22, ymin=58, ymax=100,
    ytick={60,70,80,90},
    legend style={font=\scriptsize, draw=none, fill=none,
                  at={(0.03,0.04)}, anchor=south west, cells={anchor=west}},
    grid=major, grid style={gray!20},
    label style={font=\scriptsize}, tick label style={font=\scriptsize},
    title={(b) Radius noise}, title style={font=\scriptsize},
    every axis plot/.append style={mark size=1.6pt, thick},
]
\addplot[mark=*, color=red!75!black] coordinates
    {(0,97.9) (0.05,96.9) (0.10,93.4) (0.20,62.2)};
\addlegendentry{LCA}
\addplot[mark=square*, color=blue!70!black] coordinates
    {(0,96.6) (0.05,96.6) (0.10,96.9) (0.20,86.0)};
\addlegendentry{RCA}
\end{axis}
\begin{axis}[
    at={(plotB.right of north east)}, anchor=left of north west,
    xshift=8pt,
    width=0.32\linewidth, height=3.4cm,
    xlabel={Test routing (\%, val-tuned cutoff)},
    xmin=-2, xmax=28, ymin=95.5, ymax=99.5,
    grid=major, grid style={gray!20},
    label style={font=\scriptsize}, tick label style={font=\scriptsize},
    title={(c) MC-dropout triage (val-tuned)}, title style={font=\scriptsize},
    every axis plot/.append style={mark size=1.6pt, thick},
]
\addplot[mark=*, color=red!75!black] coordinates
    {(0,97.9) (11.2,98.0) (23.8,98.6)};
\addplot[mark=square*, color=blue!70!black] coordinates
    {(0,96.2) (12.3,96.1) (25.9,96.8)};
\draw[dashed, gray!60] (axis cs:10,95.5) -- (axis cs:10,99.5);
\draw[dashed, gray!60] (axis cs:20,95.5) -- (axis cs:20,99.5);
\end{axis}
\end{tikzpicture}
\caption{Robustness and uncertainty triage for the FFR endpoint (threshold
accuracy at $\mathrm{FFR}\le 0.8$, nearest-point protocol, physiologic-range
subset). \textbf{(a)} Centerline-position noise (additive isotropic Gaussian,
mm). \textbf{(b)} Radius noise (multiplicative fractional Gaussian).
\textbf{(c)} MC-dropout triage under a \emph{val-tuned} protocol: the
$\sigma$-percentile cutoff is set on the val split at the $(1-f)$-th
percentile for target routing fraction $f\in\{10\%, 20\%\}$ (dashed
verticals) and applied unchanged to test; the plotted $x$-coordinate is the
realized test routing fraction, and the $0\%$ point is the MC-mean baseline.}
\label{fig:robustness_triage}
\end{figure}

The geometric argument of Eq.~\eqref{eq:operator} is recovered from an image,
so the practical question is how the map propagates a perturbation of it. We
perturb the centerline positions $\mathbf{c}_i$ and radius profile $r_i$ of
each test case and re-evaluate the trained map, reading sensitivity through
the endpoint functional, the most demanding readout of a local perturbation;
both perturbation families leave the branch decomposition intact, so what is
measured is sensitivity to the geometric parameters and not to topology
(Figure~\ref{fig:robustness_triage}; tables in \appref{supp:sensitivity}).
Under independent per-point noise, LCA is unchanged by position perturbations
up to $\sigma = 1$~mm, consistent with independent per-point displacements
averaging out across the token set, while RCA, whose trees are shorter and
carry a single dominant trunk, loses three points of threshold accuracy at
$0.5$~mm ($96.6\%$ to $93.5\%$) and more at $1$~mm. Radius perturbations
degrade the endpoint at a rate consistent with the $r^{-4}$ sensitivity of
the viscous pressure drop: $5\%$ relative noise leaves both territories
intact, $10\%$ produces moderate degradation on LCA while RCA is unaffected,
and $20\%$ degrades both substantially, so the radius argument is the binding
one and the two territories are complementary. A spatially correlated
perturbation, one bias applied uniformly over a random $10$~mm arc-length
window, leaves both territories at their unperturbed operating point for
position biases up to $2$~mm, and at a $20\%$ window-level radius bias LCA
remains within $0.4$~pp while RCA loses $1.3$~pp, because only a $10$~mm
region is biased and the measurement point usually lies outside it. Finally,
$K=20$ MC-dropout forward passes give a predicted standard deviation $\sigma$
on the endpoint that correlates with its absolute error at $\rho = 0.45/0.40$
but is overconfident as an interval ($1\sigma$ coverage $52\%$/$44\%$ against
a calibrated $68\%$), so it is used only as an ordering over cases: a
$\sigma$-percentile cutoff set on the validation split and applied unchanged
to test withholds the highest-$\sigma$ cases and removes $31$--$61\%$ of the
summed endpoint error for $11$--$26\%$ of cases, though on RCA at the $10\%$
target the withheld cases carried error without carrying threshold errors
($96.1\%$ against $96.2\%$ with no routing).

\subsection{Error structure, invariances and transfer}\label{sec:res_limits}

\subsubsection{Error versus geometric and flow complexity}
\label{sec:error_analysis}

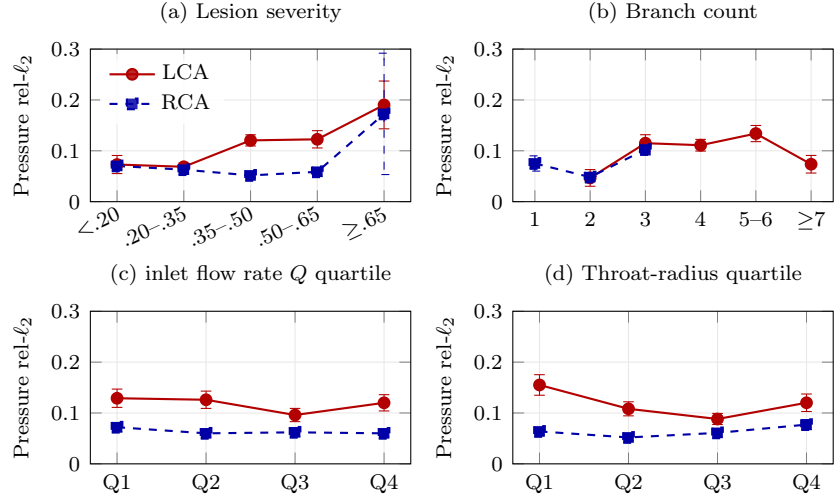
\begin{figure}[!t]
\centering
\begin{tikzpicture}
\begin{groupplot}[
    group style={group size=2 by 2, horizontal sep=1.35cm,
                 vertical sep=1.45cm},
    width=0.46\linewidth, height=3.6cm,
    ymin=0, ymax=0.30,
    ylabel={Pressure rel-$\ell_2$},
    label style={font=\scriptsize},
    tick label style={font=\scriptsize},
    title style={font=\scriptsize},
    grid=major, grid style={gray!18},
    error bars/error bar style={line width=0.4pt},
]
\nextgroupplot[title={(a) Lesion severity},
    xtick={1,2,3,4,5}, xmin=0.6, xmax=5.4,
    xticklabels={$<$.20, .20--.35, .35--.50, .50--.65, $\ge$.65},
    x tick label style={rotate=25, anchor=north east, inner sep=1pt,
                        font=\scriptsize},
    legend style={font=\scriptsize, at={(0.03,0.97)}, anchor=north west,
                  draw=none, fill=white, fill opacity=0.85, text opacity=1},
]
\addplot[mark=*, red!70!black, thick, error bars/.cd, y dir=both, y explicit]
  coordinates {(1,0.0730) +- (0,0.0178) (2,0.0686) +- (0,0.0074)
               (3,0.1205) +- (0,0.0110) (4,0.1226) +- (0,0.0170)
               (5,0.1902) +- (0,0.0470)};
\addlegendentry{LCA}
\addplot[mark=square*, blue!65!black, thick, dashed,
         error bars/.cd, y dir=both, y explicit]
  coordinates {(1,0.0702) +- (0,0.0072) (2,0.0629) +- (0,0.0089)
               (3,0.0516) +- (0,0.0044) (4,0.0587) +- (0,0.0086)
               (5,0.1726) +- (0,0.1193)};
\addlegendentry{RCA}

\nextgroupplot[title={(b) Branch count},
    xtick={1,2,3,4,5,6}, xmin=0.6, xmax=6.4,
    xticklabels={1, 2, 3, 4, 5--6, $\ge$7},
    x tick label style={font=\scriptsize},
]
\addplot[mark=*, red!70!black, thick, error bars/.cd, y dir=both, y explicit]
  coordinates {(2,0.0467) +- (0,0.0164) (3,0.1150) +- (0,0.0166)
               (4,0.1110) +- (0,0.0115) (5,0.1339) +- (0,0.0159)
               (6,0.0736) +- (0,0.0173)};
\addplot[mark=square*, blue!65!black, thick, dashed,
         error bars/.cd, y dir=both, y explicit]
  coordinates {(1,0.0751) +- (0,0.0150) (2,0.0483) +- (0,0.0033)
               (3,0.1025) +- (0,0.0107)};

\nextgroupplot[title={(c) inlet flow rate $Q$ quartile},
    xtick={1,2,3,4}, xmin=0.7, xmax=4.3,
    xticklabels={Q1, Q2, Q3, Q4},
    x tick label style={font=\scriptsize},
]
\addplot[mark=*, red!70!black, thick, error bars/.cd, y dir=both, y explicit]
  coordinates {(1,0.1290) +- (0,0.0180) (2,0.1260) +- (0,0.0170)
               (3,0.0960) +- (0,0.0130) (4,0.1200) +- (0,0.0160)};
\addplot[mark=square*, blue!65!black, thick, dashed,
         error bars/.cd, y dir=both, y explicit]
  coordinates {(1,0.0720) +- (0,0.0090) (2,0.0600) +- (0,0.0070)
               (3,0.0620) +- (0,0.0080) (4,0.0600) +- (0,0.0070)};

\nextgroupplot[title={(d) Throat-radius quartile},
    xtick={1,2,3,4}, xmin=0.7, xmax=4.3,
    xticklabels={Q1, Q2, Q3, Q4},
    x tick label style={font=\scriptsize},
]
\addplot[mark=*, red!70!black, thick, error bars/.cd, y dir=both, y explicit]
  coordinates {(1,0.1551) +- (0,0.0202) (2,0.1082) +- (0,0.0138)
               (3,0.0882) +- (0,0.0109) (4,0.1201) +- (0,0.0172)};
\addplot[mark=square*, blue!65!black, thick, dashed,
         error bars/.cd, y dir=both, y explicit]
  coordinates {(1,0.0637) +- (0,0.0064) (2,0.0519) +- (0,0.0063)
               (3,0.0609) +- (0,0.0081) (4,0.0771) +- (0,0.0099)};
\end{groupplot}
\end{tikzpicture}
\caption{Mean per-case pressure relative $\ell_2$ error by bin, with $\pm 1$
standard error of the mean, over the $300$-case test split of each territory
(bin definitions and velocity columns in Table~\ref{tab:error_covariates}; all four
panels share the vertical scale). Severity in (a) carries the largest trend of
the four covariates, and only on LCA. Quartiles in (c) and (d) are
per-territory. Bins with fewer than $10$ cases carry wide intervals: severity
$\ge 0.65$ on RCA ($n = 2$), two branches ($n = 3$) and $\ge 7$ branches
($n = 7$) on LCA.}
\label{fig:error_covariates}
\end{figure}

Because the parameter pair $(g, Q)$ is summarized by a small number of
scalars, the per-case error of the primary configuration can be resolved
against each of them (Figure~\ref{fig:error_covariates}; values in
Table~\ref{tab:error_covariates}), with rank correlations tested by a patient-clustered
permutation and corrected for multiplicity across the twenty tests by
Benjamini--Hochberg, since same-patient cases share the centerline exactly. No
covariate produces a large or reliable dependence. Lesion severity shows the
largest binned trend, a factor of $2.8$ on LCA pressure from $0.069$ to
$0.190$ across bands ($\rho = +0.166$), but the per-case $p$-values of $0.004$
and $0.005$ become $q = 0.11$ and $0.15$ after clustering and multiplicity
correction; severity and throat radius are strongly anti-correlated
($\rho = -0.635$) and cannot be separated on this sample, and RCA returns
nothing ($\rho = -0.048$). Pressure error is insensitive to inlet flow rate
($|\rho| \le 0.090$; RCA velocity error rises weakly, $\rho = +0.124$), also within a patient where $Q$ is essentially the only
covariate that varies (within-anatomy $\rho = +0.04$ LCA, $+0.09$ RCA), which
is the behavior expected if the flow-rate
dependence is carried by the multiplicative conditioning; inlet caliber is
flat ($|\rho| \le 0.091$), and lesion throat radius, the only caliber measure
with a nominal signal, fails correction ($\rho = -0.117$, $p = 0.042$ per
case, $0.126$ clustered, on LCA). Branch count is the one covariate whose association
survives correction, on three of the four territory--field pairs
($q = 0.034$, $0.002$, $0.006$), but the dependence is weak, not monotone in
the bins, and partly a proxy for tree extent ($\rho = 0.632$ with arc length
on LCA). Set against per-case errors spanning a factor of $52$ on LCA ($0.018$
to $0.950$) and $38$ on RCA, with the top decile carrying $40.5\%$ (LCA) and
$36.3\%$ (RCA) of the summed pressure error, none of these locates where the
surrogate is inaccurate: for the $44$ LCA cases above $0.2$, median severity,
branch count and throat radius are $0.488$, $4$ and $0.668$~mm against cohort
medians of $0.463$, $4$ and $0.721$~mm. The regimes in which the operator
might have been expected to become harder to approximate (higher flow rates,
smaller vessels, more branched trees) are not where this surrogate is less
accurate, which is what one expects if the axis carries the strongest
dependence and is supplied at full resolution; the residual that sets the mean
is case-specific. It is a property of the cases and not of the optimization:
across the three $300$-epoch seeds, per-case pressure errors correlate at
Pearson $0.89$--$0.93$, and of the $44$ worst cases per seed, $33$ are in the
worst $44$ of all three (pairwise overlaps $36$--$38$ against an expectation of
$0.9$ under independence), so ensembling over seeds would not remove the tail.
Nor is it an artifact of the globally $z$-scored objective against a per-case
relative metric: relative error correlates with per-case $\|p\|$ only weakly
(Spearman $-0.137$) while absolute error rises with it ($+0.574$), and exactly
$11$ of the $44$ worst cases fall in the smallest-$\|p\|$ quartile, the chance
expectation. The tail is approximation error on specific geometries,
reproducible and therefore available to be diagnosed.

\subsubsection{What the construction does not enforce}
\label{sec:frame_sensitivity}
\label{sec:pde_consistency}

The boundary-value problem is invariant under a rigid motion applied jointly
to the domain and the points at which the solution is read, is pointwise in
the sense that the solution at a location does not depend on which other
locations are asked about, and satisfies Eqs.~\eqref{eq:momentum} and
\eqref{eq:continuity}. CARS has none of these properties by construction, so
how much each costs is a measurement, made on the trained LCA checkpoint with
no retraining (\appref{supp:limits}). Rotating the centerline
and evaluation points jointly about the case centroid, scored against the
unrotated reference with the predicted velocity counter-rotated, two degrees
of rotation or one millimeter of translation cost less than the seed-to-seed
standard deviation; five degrees or two millimeters cost $15\%$ and $8\%$ of
the pressure error; twenty degrees or ten millimeters raise it by a factor of
three to three and a half; and a rotation drawn uniformly from $SO(3)$ leaves
a median relative pressure change of $1.02$, so the output is then unrelated
to the correct one. Every geometry is expressed in one acquisition frame and
the positional encodings act on absolute coordinates in it, so the tolerance
measured here covers the residual misalignment of a consistently oriented
cardiac CT series and not a different patient-positioning convention. Holding
$2{,}048$ points fixed and predicting them twice, each time embedded in a
different set of $14{,}336$ further points from the same case, the two
requests disagree by a mean relative $\ell_2$ of $0.019$ on pressure (median
$0.015$) and $0.058$ on velocity, a sixth and two-fifths of the respective
errors: the reported errors carry this much irreducible variation from the
evaluation sampling alone, and fixing the anchor set to a deterministic
function of the geometry would remove it at the accuracy cost bounded in
\appref{app:complexity}. Consistency with the governing equations, evaluated
by automatic differentiation of the network on fifty test cases per territory
with $4{,}096$ interior points each and made dimensionless by the per-case
mean speed $U$ and median lumen diameter $D$, is loose: the median
$|\nabla\!\cdot\!\mathbf{u}|\,D/U$ is $0.26$ on LCA and $0.30$ on RCA, so the
local dilatation is of order a quarter of the characteristic velocity
gradient, and the momentum residual scaled by $U^2/D$ has medians of $2.6$ and
$2.9$, falling to $1.57$ once points whose finite-difference Laplacian stencil
leaves the lumen are removed, still of the order of the convective term. The
inlet flux recovered from the predicted velocity by disc quadrature differs
from the prescribed $Q$, an \emph{input} delivered through a dedicated
conditioning pathway, by a median of $12.2\%$ on LCA and $5.5\%$ on RCA. CARS
therefore reproduces the reference fields in the supervised norms, and the
derived pressure functional to within a few per cent, without preserving the
governing equations pointwise: it approximates the solution map realized by a
particular finite-volume discretization and does not inherit its conservation
properties. Quantities that integrate the pressure field (pressure drops, the
endpoint) are insensitive to residual error that averages out; quantities that
differentiate the velocity field (wall shear stress, vorticity, dissipation)
are not, and should not be read off this surrogate without physics-constrained
training terms or a projection step enforcing $\nabla\!\cdot\!\mathbf{u}=0$
after prediction.

\subsubsection{Cross-territory generalization: RCA}\label{sec:cross_territory}
\label{sec:rca}

Right coronary artery geometries differ from LCA in topology (typically a
single dominant trunk), inlet flow rate (mean $3{,}515\pm 571$~mm$^3$/s,
${\sim}55\%$ of the LCA mean) and spatial distribution. Zero-shot
LCA$\to$RCA inference of the LCA-trained checkpoint reaches relative $\ell_2$
of order $8$ on both fields even after re-normalizing spatial coordinates by
the RCA-set bounding box, two orders of magnitude worse than the
within-territory result: the flow-rate distributions are non-overlapping (a
gap of ${\sim}3$ standard deviations), so the FiLM pathway operates out of
distribution on every RCA case, and the centerline geometry distribution is
shifted, so the learned positional bias and the centerline-anchor tokens are
both extrapolating. The failure is not one of calibration, and per-territory
conditioning does not repair it (Table~\ref{tab:xterritory}): normalizing
positions and radius with RCA statistics roughly halves the pooled pressure
error, from $15.7$ to $7.0$, normalizing $Q$ with RCA statistics as well makes
it \emph{worse}, to $22.5$, and in every variant the single global scale
factor minimizing the error is indistinguishable from zero, so the pooled
correlation between predicted and reference pressure is $+0.02$ to $+0.06$
and the zero-shot output is uncorrelated with the right field, a
representational failure under extrapolation. Training in the non-dimensional
variables of Eq.~\eqref{eq:reynolds} remains untested and is the one remaining
candidate we consider promising. A fresh CARS instance trained from scratch on
the RCA split with identical configuration reaches $0.063/0.119$ and, on the
FFR endpoint, Pearson $r = 0.994$, MAE $0.008$ and threshold accuracy $96.2\%$
(sensitivity $93.5\%$, specificity $97.5\%$) on the physiologic-range subset;
field errors, $r$ and MAE improve on the LCA result under matched
architecture, which we attribute to the simpler topology of the right
coronary tree, while threshold accuracy and specificity do not ($96.2\%$
against $97.6\%$, $97.5\%$ against $98.7\%$), and inference cost is
indistinguishable between territories. The RCA cohort therefore
evidences within-distribution generalization and not cross-territory
transfer; the centerline carries no territory label, so a single checkpoint
cannot separate the two regimes without cross-territory training or an
explicit territory token.

\section{Discussion}

\subsection{Structure of the endpoint error}\label{sec:endpoint_error}

\begin{figure}[!t]
    \centering
    \includegraphics[width=0.85\linewidth]{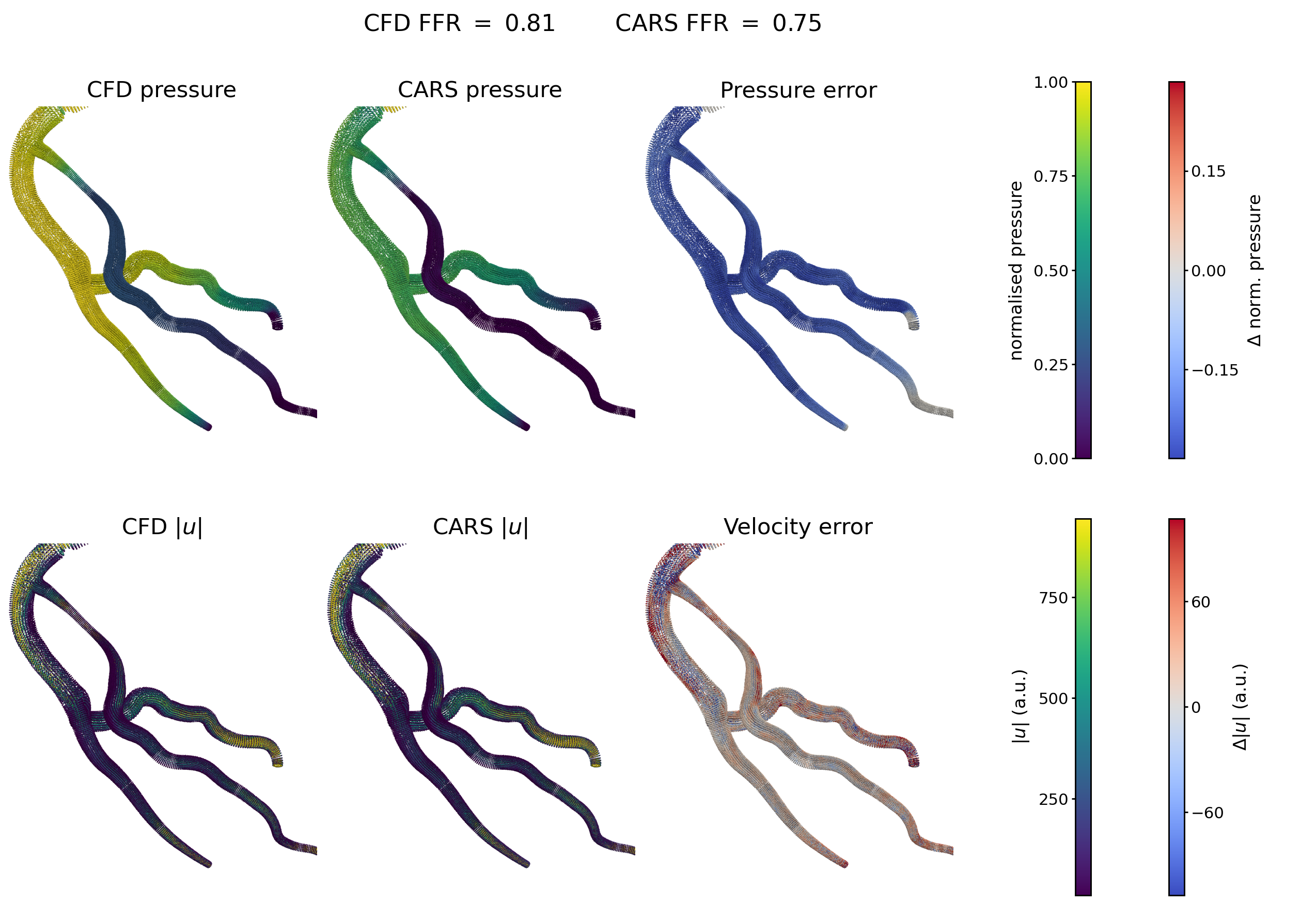}
    \caption{\textbf{Field-level structure of a discordant case}: an LCA
    false positive (CFD $\mathrm{FFR}=0.808$, CARS $0.75$, case pressure error
    $0.20$), whose reference value sits $0.008$ from the decision threshold.
    Panels and color scales as in Figure~\ref{fig:field_compare}. The whole
    predicted pressure field is displaced to a lower level, visible as a
    uniformly green CARS panel against a uniformly yellow CFD panel and an
    error panel that is blue almost everywhere; the shape of the field is
    right and its level is offset, so the ratio of
    Eq.~\eqref{eq:ffr_functional} crosses the threshold without any localized
    geometric miss.}
    \label{fig:field_discordant}
\end{figure}

Where the endpoint disagrees with the reference, the disagreement is a sign
change in the threshold comparison and not a large field error. CARS produces
$7/286$ discordant cases on LCA (3~FP, 4~FN) and $11/293$ on RCA (5~FP, 6~FN)
on the physiologic-range subset, and the majority lie within $\pm 0.05$ of
the $\mathrm{FFR}=0.8$ level, where the functional of
Eq.~\eqref{eq:ffr_functional} is most sensitive to a small error in the
pressure at the measurement point. On an LCA false positive
(Figure~\ref{fig:field_discordant}) the predicted pressure is too low along
the entire vessel by a spatially near-uniform amount: a global offset in the
operating point, not a focal geometric miss, which the ratio converts into a
threshold crossing, and the error mode the $\sigma$ ordering of
\S\ref{sec:robustness_uq} is best placed to flag. The construction separates
by field: pressure is accurate, stable under thinning of the focal and anchor
sets, insensitive to evaluation density and good enough to carry the derived
decision, whereas velocity degrades far faster under anchor thinning ($+50\%$
against $+7\%$ relative at eightfold thinning),
an order of magnitude faster under reduced precision, and does not satisfy
continuity closely enough to differentiate. That asymmetry determines which
quantities the surrogate can be used for.

\subsection{Position of the surrogate in the end-to-end cost}\label{sec:cost}

The arguments of Eq.~\eqref{eq:operator} are themselves the output of an
upstream segmentation and centerline-extraction step, and evaluating
Eq.~\eqref{eq:ffr_functional} additionally requires the geometric
lesion-localization step of \appref{app:ffr_protocol}. Those steps cost
seconds to minutes per case, against $87$~ms for the forward pass, so the
end-to-end time from a new image is dominated by the geometry pipeline; what
Table~\ref{tab:baselines} reports is a speed-up of the flow solution, the
stage this work replaces, by a factor of about $6{,}600$ to $12{,}000$ in
wall-clock on different hardware classes. The evaluation set used for every
number here is the set of cell centers of the reference mesh, so obtaining the
points at which the field is reported still requires a volume mesh; points
sampled from $(\mathbf{c}, r)$ through Eq.~\eqref{eq:tube_map} would serve,
but that changes the inference-time point distribution from the one the model
was trained on, and we have not measured the effect. CARS replaces the solve
on a given mesh, not the meshing. The ${\sim}1$~hour figure in circulation is
a literature estimate for a full CT-to-FFR pipeline including segmentation and
meshing, so dividing it by a forward pass compares two different quantities
and inflates the ratio by roughly an order of magnitude.

\subsection{Limitations}\label{sec:limitations}

Table~\ref{tab:scope} states what the evidence does and does not establish;
the limitations below name what would remove each, and further ones (the
convergence filter preceding the splits, developing flow over much of the
tree, the geometric component of the reference error, non-physiological
corners of the product set, the outcome-adjacent endpoint subset, a single
acquisition source, RoPE3D under length extrapolation, and smooth-only
perturbations) are tabulated with their remedies in
\appref{supp:limitations}.

\paragraph{The target is the solver, not the patient}
Every error measures agreement with a steady, laminar, rigid-wall, Newtonian
OpenFOAM pipeline with prescribed $r^{2}$ flow-split outlets
(\appref{app:cfd_pipeline}). What the surrogate captures is therefore the
bulk laminar pressure drop along the vessel, the focal loss across stenoses
driven by lumen narrowing, the post-stenotic pressure recovery, and the
$p \propto Q$ scaling built into the conditioning; pulsatile and transient
effects, transitional jets distal to severe stenoses, compliance and
microvascular resistance are outside the supervision, and wall shear
stress~\cite{stone2018wss} is computable post hoc but is not a supervised
output.
The AUC of any emulator of this pipeline against \emph{invasive} FFR is
bounded by the pipeline's own, which the prior ML-CT-FFR literature places in
the ${\sim}0.84$--$0.93$
band~\cite{itu2016machine,coenen2018ml,tesche2018ml,kumamaru2020deep};
closing that gap is a calibration step against a paired invasive cohort.

\paragraph{The reported errors are parameter-space results}
The partition is over samples of $(x, y, z, r, Q)$, applied identically to
every model, so the comparisons are unaffected; what it bounds is the
absolute figures, which quantify interpolation over the sampled parameter set
(G1) and not transfer to unseen anatomies (G2) or acquisition settings (G3).
Trajectories recur across splits, and a minority of test cases coincide with
a training case in both trajectory and radius profile, differing only in $Q$
(\S\ref{sec:dataset}). An anatomy-disjoint re-partition of the same LCA cases
(\appref{supp:anatomy_disjoint}) degrades both CARS and AB-UPT sharply; CARS is the
more accurate on both fields and on the large majority of the eleven test
anatomies, and the patient-clustered interval spans a factor of two, so
resolving the absolute level is a data-collection problem.

\paragraph{The lumens are tubular by construction}
Because $\pi$ is injective on this cohort (\S\ref{sec:problem}), the
experiments compare \emph{encodings} of a fixed geometric content and cannot
say how much geometry a surrogate of this map needs; on independently
segmented anatomy the fiber spread of Eq.~\eqref{eq:fiber_expectation}
becomes a genuine error floor, unmeasured here. The same construction fixes
the lesion family: every stenosis is a radial deformation of the radius
profile, so trans-lesional loss is driven by caliber and not by plaque
morphology.

\paragraph{The frame is part of the input, and terminal fluxes are prescribed}
Nothing in the architecture is invariant under a rigid motion of the inputs
(\S\ref{sec:frame_sensitivity}), so inputs must arrive in approximately the
frame of the training cohort, a constraint invisible within a single-center
cohort; building the equivariance in is the direction a deployment would
take. Because every outlet flux is fixed by Eq.~\eqref{eq:outlet_split}, a
lesion cannot divert flow into a parallel branch, so in these labels the
pressure drop grows with severity without the flow reduction FFR exists to
probe, a severity-dependent bias not confined to a tail at a median LCA
severity of $0.46$, and the closed-form split makes the map easier than the
physical one; a resistance-outlet re-solve on a severity-spanning subset would
quantify the label shift.

\paragraph{The baseline comparison is not fully matched}
Five asymmetries, recorded in \appref{app:baselines}, limit it: the
baselines were trained at CARS's optimizer settings; the conditioning pathway
differs across arms, priced at $0.024$ on pressure by this paper's own
ablation; AB-UPT's parameter match is nominal because its surface heads carry
zero loss weight; AB-UPT encodes the same geometry more coarsely; and AB-UPT
trains in mixed precision. Initialization is not among them.

\subsection{Extensions}\label{sec:future}

At the level of the construction, what is presented is an \emph{input
contract} for PDE surrogates on branching or tubular domains carrying a
natural low-dimensional skeleton: a skeleton with a caliber profile
extractable by a deterministic pre-processing step, and a solution whose
dependence on the geometry is dominated by transport along it. That is scope
and not demonstrated reach, since the one transfer experiment reported here
is negative: airway trees, duct networks, fractured media and heat-exchanger
manifolds meet the contract, and each would need its own reference data and
training run. Four experiments follow directly from what this paper measures.
\emph{Objective alignment}: the loss weights velocity three to one against
the primary pressure endpoint (\appref{app:training}), and a sweep over that
weight would establish whether the reported pressure accuracy is limited by
the architecture or by the objective. \emph{Positional-encoding control}: a
scale-matched RoPE3D run would separate the encoding choice from the
frequency-scale choice (\S\ref{sec:rope3d}). \emph{Focal-set identification}:
the experiment of \S\ref{sec:res_coarse}, which varies only where the focal
points sit, would establish whether provenance itself explains the margin.
\emph{External validation}: an independent multi-site CTA and CFD cohort with
an anatomy-disjoint partition at a scale that resolves it, together with
post-hoc calibration against invasive wire-based FFR, is what would carry the
method towards use; both require only centerlines with radius profiles and a
flow rate. Extensions to transient and pulsatile supervision, compliant walls,
non-Newtonian rheology and physics-constrained training each require new
reference data and not a change to the construction.

\section{Conclusion}\label{sec:conclusion_claims}

CARS is a two-stage anchored-attention surrogate for the parametric steady
incompressible Navier--Stokes problem on centerline-derived coronary lumens.
It takes a centerline with a per-point lumen radius and a scalar inlet flow
rate, accepts the evaluation points as an argument, and returns pressure and
velocity in $87$~ms per case. On identical splits and at parameter counts
matched to within $3\%$ it is the most accurate of the configurations
compared here, ahead of four neural-operator baselines and of the alternative
focal representations, and the FFR functional read off its predicted pressure
field reproduces the reference solver's $\mathrm{FFR}\le 0.8$ decision in
$97.7\%$ (LCA) and $96.3\%$ (RCA) of held-out cases. The results support a
geometry-defined focal set for this class of surrogate: the margin is neither
a density effect nor a parameter-count effect, though the ablations do not
isolate the provenance of the focal set from the geometric information its
tokens carry. The same building blocks, anchor attention over a
geometry-defined focal set, attention-level geometric positional encoding,
and multiplicative conditioning on a scalar boundary variable, apply to any
problem meeting the input contract of \S\ref{sec:future}, of which airway
trees, branching pipe networks, channelized porous media and tube bundles are
candidate instances; coronary flow is the only one evaluated here, and the
cross-territory result shows that meeting the contract does not guarantee
transfer across cohorts. Two extensions carry the surrogate toward
clinical use: post-hoc calibration of the CFD reference against invasive
wire-based FFR, and expansion of the geometry distribution across scanner
vendors, acquisition protocols and patient subgroups beyond a single
institution.

\section*{CRediT authorship contribution statement}

\textbf{Reza Akbarian Bafghi}: Conceptualization, Methodology, Software,
Formal analysis, Investigation, Validation, Writing -- original draft,
Visualization.
\textbf{Sukirt Thakur}: Conceptualization, Methodology, Software,
Data curation, Formal analysis, Investigation, Validation, Writing --
original draft, Visualization.
\textbf{Maziar Raissi}: Conceptualization, Methodology, Writing -- review
\& editing, Supervision, Project administration.

Reza Akbarian Bafghi and Sukirt Thakur contributed equally.

\section*{Declaration of competing interest}

S.T. is employed by AngioInsight, Inc., a company active in coronary
hemodynamics. The authors declare no other competing financial interests or
personal relationships that could have appeared to influence the work reported
in this paper. No invasive measurement, patient-identifiable data or
proprietary clinical pipeline enters this work.

\section*{Funding}

This research did not receive any specific grant from funding agencies in the
public, commercial, or not-for-profit sectors.

\section*{Data availability}

The coronary CT-angiography volumes underlying this work are publicly
available as part of the ImageCAS release~\cite{zeng2023imagecas}. The
CFD-derived fields used as supervisory targets were generated from them by the
OpenFOAM pipeline of \appref{app:cfd_pipeline}, whose mesher draws each
stenosis without a recorded seed, so the cases cannot be regenerated exactly
from the source volumes (\appref{app:gci}). The derived dataset, trained
models and code are not publicly available.

\section*{Declaration of generative AI and AI-assisted technologies in the manuscript preparation process}

During the preparation of this work the authors used AI to
assist with editing the text and the code. After using this tool, the authors
reviewed and edited the content as needed and take full responsibility for the
content of the published article.

\clearpage

\ifnum\CMAMEels=1
  \bibliographystyle{elsarticle-num}
\else
  \bibliographystyle{unsrtnat}
\fi

\ifdefined\SPLITAPPENDIX
  \input{appendix_refs}
\fi
\bibliography{references}

\ifdefined\SPLITAPPENDIX\else
\clearpage

\appendix
\setcounter{table}{0}
\setcounter{figure}{0}
\makeatletter
\@addtoreset{table}{section}
\@addtoreset{figure}{section}
\makeatother

\ifdefined\SUPPBUILD
\setcounter{page}{1}
\renewcommand{\thepage}{S\arabic{page}}
\begin{center}
{\Large Supplementary Material for\\[2pt]
``\CARStitle''}\\[8pt]
Reza Akbarian Bafghi, Sukirt Thakur, Maziar Raissi
\end{center}
\vspace{6pt}
\fi


\section{Training and architecture details}\label{app:training}\label{app:math}

\subsection{Architecture hyperparameters}

Table~\ref{tab:hyperparams} lists the defaults used throughout.

\begin{table}[h]
\centering
\caption{Default CARS model hyperparameters used in all reported experiments
unless otherwise stated.}
\label{tab:hyperparams}
\begin{tabular}{ll}
\toprule
Parameter & Value \\
\midrule
$d_\text{model}$ & 192 \\
Attention heads $h$ & 3 \\
FFN expansion ratio & 4 \\
Geometry-fusion blocks $L_\text{geo}$ & 4 \\
Field-refinement blocks $L_\text{refine}$ & 4 \\
Sinusoidal position dim $d_\text{pos}$ & 192 \\
Centerline tokens $N_c$ & 2{,}048 \\
Volume tokens $N_v$ & 16{,}384 \\
Stage-2 volume anchors $N_{\text{va}}$ & 16{,}384 \\
Stage-2 pure query points $N_{\text{vq}}$ & 0 \\
Dropout & 0.1 \\
Output channels & 4 (1 pressure + 3 velocity) \\
\bottomrule
\end{tabular}
\end{table}

\subsection{Loss and optimization}

The training loss is the mean squared error between predicted and
ground-truth pressure and velocity components in normalized space,
$\mathcal{L} = \mathrm{MSE}(\hat p, p) + \sum_{k\in\{x,y,z\}}
\mathrm{MSE}(\hat u_k, u_k)$. Pressure contributes one term and velocity
three, and all four targets are z-scored, so the effective weighting is
$1{:}3$ in favor of velocity although pressure is the primary endpoint. The
pressure errors reported here are therefore what the architecture achieves
\emph{despite} an objective tilted towards velocity, and the
pressure--velocity trade-offs visible in the fixed-anchor sweep and the
bfloat16 comparison may be partly an artifact of this weighting; a sweep over
the relative weight needs retraining and has not been run.

Optimization uses AdamW~\cite{loshchilov2019adamw} with learning rate
$\alpha = 5\times 10^{-5}$, $\beta = (0.9, 0.999)$, weight decay
$\lambda = 0.05$, and $\epsilon = 10^{-8}$; cosine annealing with linear
warm-up over the first $5\%$ of steps, decaying to $10^{-6}$; gradient
clipping at global norm $1.0$; dropout $0.1$; and an exponential moving
average of parameters with decay $\gamma = 0.9999$ for validation and
inference. Training runs at batch size $1$ for $100$ epochs in ablation studies
and $300$ epochs in full runs, because larger batches destabilized
optimization in this AdamW setup. We record batch size $1$ as an unresolved
optimization pathology: the learning rate was tuned at batch size $1$ and not
jointly with it, so the instability may be an artifact of holding $\alpha$
fixed while the effective noise scale changed, and the finding that training
length is the most effective lever per unit compute may partly reflect a slow
optimizer setting. Every CARS run trains in fp32, and inference, including the
reported $87$~ms latency, is fp32 except for the $15$~ms bfloat16 operating
point quoted as such. Checkpoints are saved every $10$ epochs.

\subsection{Computational complexity, analytic and measured}
\label{app:complexity}

For $N_c$ centerline tokens, $N_v$ evaluation points and $N_{va}$ Stage-2
anchors, the forward cost of the two stages of \S\ref{sec:two_stage} is
\begin{equation}
    O\big(\underbrace{(N_v + N_c)\,N_c\,d}_{\text{Stage 1}}
        \;+\; \underbrace{N_v\,N_{va}\,d}_{\text{Stage 2}}
        \;+\; \underbrace{(N_v + N_c)\,d^{2}}_{\text{feed-forward}}\big) ,
    \label{eq:complexity}
\end{equation}
where $N_v$ counts \emph{all} evaluation points, the anchors included, whereas
the $N_c$ centerline tokens of Stage~1 are disjoint from the evaluation
tokens. The expression is linear in $N_v$ only when both $N_c$ and $N_{va}$
are held fixed as $N_v$ grows. At the configuration behind every accuracy
number in this paper $N_{va} = N_v = 16{,}384$, so the middle term is
$O(N_v^{2} d)$ and Stage~2 is full self-attention; the accuracy cost of
entering the linear regime is Table~\ref{tab:fixed_anchor}. Stage~1 is
quadratic in the centerline count, which is harmless because $N_c = 2{,}048$
is small and well past the point where accuracy stops improving.

Wall-clock is the median of $15$ timed forward passes after $5$ warmup passes
on an idle A100 80GB PCIe in fp32; peak memory is
\texttt{max\_memory\_allocated} around a single pass. Table~\ref{tab:complexity}
reports fitted log--log slopes over three regimes. Holding the evaluation set
at $16{,}384$ points and varying the focal count over $256$--$4{,}096$ changes
cost by less than $25\%$ (slope $0.07$ in time), so within this range the
position-wise feed-forward work $O(N d^{2})$ dominates. Holding the Stage-2
anchor set at $2{,}048$ and adding query points, cost grows close to linearly
(overall slope $0.66$, local slope $0.90$ over $32{,}768 \to 65{,}536$), which
is the regime the complexity claim describes. If instead every evaluation
point is also an anchor, the cost approaches quadratic (slope $1.27$, rising
to $1.70$ over $16{,}384 \to 32{,}768$, against $1.81$ for a standard
self-attention layer of the same width over the same token count); the density
sweep of Table~\ref{tab:density_precision} traverses this regime. Peak-memory
slopes are less informative ($0.64$ tiered, $0.72$ reference) because the
reference uses a memory-efficient attention implementation. No approximation
bound in $n_f$ is derived here; an interpolation-error estimate in terms of
focal arc-length spacing, with the straight-tube Poiseuille case as the
tractable instance, is the missing analysis.

\begin{table}[!tb]
\centering
\caption{Measured scaling on an idle A100 80GB PCIe, fp32. Slopes are fitted
to $\log$(cost) against $\log$(size) over the stated range. The claim is a
slope of $1$ for the tiered model and $2$ for global self-attention. n/a: not
measured.}
\label{tab:complexity}
\setlength{\tabcolsep}{3.5pt}
\small
\begin{tabular}{llcc}
\toprule
Swept quantity & Range & Time slope & Memory slope \\
\midrule
Focal (centerline) tokens, $N$ fixed  & $256$--$4{,}096$    & $0.07$ & $0.05$ \\
Evaluation points, anchors fixed      & $2{,}048$--$65{,}536$ & $0.66$ & n/a \\
\quad upper half of that range        & $32{,}768$--$65{,}536$ & $0.90$ & n/a \\
Evaluation points, all as anchors     & $2{,}048$--$32{,}768$ & $1.27$ & $0.64$ \\
\quad upper half of that range        & $16{,}384$--$32{,}768$ & $1.70$ & n/a \\
\addlinespace[2pt]
Global self-attention (reference)     & $3{,}072$--$67{,}584$ & $1.81$ & $0.72$ \\
\bottomrule
\end{tabular}
\vspace{-2mm}
\end{table}

\begin{table}[!tb]
\centering
\caption{Accuracy in the fixed-anchor regime, LCA, $300$-case test split,
fp32, mean relative $\ell_2$ over cases. Trained weights unchanged; the total
evaluation budget is held at $16{,}384$ points, the anchor count $N_{va}$ is
reduced and the remaining points are supplied as pure Stage-2 query points,
with the error scored on those query points only. The Stage-2 attention
matrix is $16{,}384 \times N_{va}$, so its cost relative to the operating point
is $N_{va}/16{,}384$. The last row is a separate model trained from scratch
with the anchor set fixed at $2{,}048$; both it and the thinned flagship run
at a measured median $22.5$~ms against the operating point's $87$~ms.}
\label{tab:fixed_anchor}
\small
\begin{tabular}{lcccc}
\toprule
$N_{va}$ & Query points & Stage-2 cost & Pressure & Velocity \\
\midrule
$1{,}024$   & $15{,}360$ & $0.063\times$ & $0.138$ & $0.311$ \\
$2{,}048$   & $14{,}336$ & $0.125\times$ & $0.125$ & $0.220$ \\
$4{,}096$   & $12{,}288$ & $0.25\times$  & $0.121$ & $0.174$ \\
$8{,}192$   & $8{,}192$  & $0.50\times$  & $0.117$ & $0.154$ \\
\addlinespace[2pt]
$16{,}384$ (operating point) & $0$ & $1\times$ & $0.117$ & $0.147$ \\
\addlinespace[2pt]
\multicolumn{5}{l}{\emph{trained in the linear regime, $300$ epochs, same architecture and data}} \\
$2{,}048$   & $14{,}336$ & $0.125\times$ & $0.141$ & $0.202$ \\
\bottomrule
\end{tabular}
\vspace{-2mm}
\end{table}

\subsection{3D rotary position embeddings}
\label{supp:rope3d}

RoPE rotates query and key vectors so that, for the rotated
$\mathbf{q}', \mathbf{k}'$, the dot product $\mathbf{q}'^\top \mathbf{k}'$
depends only on the relative position
$\mathbf{p}_q - \mathbf{p}_k$~\cite{su2023rope}. We extend this to 3D by
partitioning each attention head's $d_{\text{head}}/2$ rotation pairs across
the three axes as evenly as the head width allows, and applying a 2D rotation
per axis. Writing $m_a$ for the number of pairs assigned to axis $a$, so that
$\sum_a m_a = d_{\text{head}}/2$, define for each axis $a \in \{x, y, z\}$ and
pair index $j = 1, \ldots, m_a$ the rotation angle
$\theta_a^{(j)} = p_a / 10000^{\,(j-1)/m_a}$ and the block-diagonal rotation
matrix
\begin{equation}
    R_a(\theta_a) = \mathrm{diag}\!\left(
    \begin{pmatrix} \cos\theta_a^{(1)} & -\sin\theta_a^{(1)} \\
    \sin\theta_a^{(1)} & \cos\theta_a^{(1)} \end{pmatrix},
    \ldots,
    \begin{pmatrix} \cos\theta_a^{(m_a)} & -\sin\theta_a^{(m_a)} \\
    \sin\theta_a^{(m_a)} & \cos\theta_a^{(m_a)} \end{pmatrix}
    \right).
\end{equation}
Writing the head as its three axis blocks,
$\mathbf{q} = [\mathbf{q}_x \,|\, \mathbf{q}_y \,|\, \mathbf{q}_z]$ and
analogously for $\mathbf{k}$, the rotated query is
\begin{equation}
    \mathbf{q}' = \big[\,R_x(\theta_x)\mathbf{q}_x \;\big|\;
                        R_y(\theta_y)\mathbf{q}_y \;\big|\;
                        R_z(\theta_z)\mathbf{q}_z \,\big],
\end{equation}
and analogously for $\mathbf{k}'$. The resulting attention score depends only
on the coordinate displacement $\mathbf{p}_q - \mathbf{p}_k$, componentwise in
each axis, and so encodes 3D Euclidean proximity directly. RoPE3D adds no
learnable parameters and negligible computation.

\paragraph{The frequency ladder is inherited, and it is not adapted to this
coordinate range}
The base $10^{4}$ in $\theta_a^{(j)}$ is the language-model default, where
positions are token indices running to $10^{3}$ or beyond. Here coordinates
are normalized to $[0,1]$ (\S\ref{sec:normalization}), so the angle in the
highest-frequency pair reaches at most $1$~radian across the entire domain,
which is $0.16$ of a cycle, and the ladder falls from there: with
$d_{\text{head}} = 64$ and eleven pairs on the first axis, the per-pair
maximum angles are $1$, $0.43$, $0.19$, $0.08$, $0.035$, and below $0.016$ for
the remaining six. Nine of the eleven rotations are therefore small-angle over
the whole coronary tree, and RoPE3D as configured is close to the identity in
most of its channels. The same applies to the input-stage sinusoidal
embedding, where $75\%$ of channels traverse less than $0.1$~radian and $47\%$
less than $0.01$ across the domain, so those channels supply a scaled linear
ramp and a constant rather than an oscillation. The learned alternative of
\S\ref{sec:learned_pe} is an unconstrained MLP of the coordinate and can
realize whatever length scale the problem wants, while the rotary arm is
pinned to a wavelength ladder chosen for a different kind of input. The margin
measured between them therefore confounds the form of the encoding with the
choice of frequency scale: a learned encoding beats \emph{this configuration}
of RoPE3D rather than rotary encoding as such. Rescaling the base so that the
shortest wavelength matches the lumen radius, or non-dimensionalizing the
coordinates by a characteristic length, is the control that would separate the
two; it needs retraining and has not been run.

\subsection{Hyperparameter sweep}

Table~\ref{tab:hp_sweep} varies training length, depth and width one axis at
a time from the $100$-epoch learned-PE reference.

\begin{table}[!htb]
\centering
\caption{Single-axis hyperparameter sweep on the learned-PE configuration. The
reference row is the $100$-epoch learned-PE run of Table~\ref{tab:ablation}.
Bold rows mark the best pressure and velocity across the sweep. All errors are
\emph{validation} relative $\ell_2$; the corresponding test errors for the
$300$-epoch row are $0.116$ / $0.146$ on pressure / velocity
(\S\ref{sec:headline}).}
\label{tab:hp_sweep}
\small
\begin{tabular}{llrcc}
\toprule
Axis & Configuration & Params & Val.\ press. & Val.\ vel. \\
\midrule
\textit{Reference} & $100$ ep, $d{=}192$, $4{+}4$ blocks & 3{,}857{,}284 & 0.140 & 0.196 \\
\addlinespace[2pt]
\textit{Epochs}    & $200$ epochs                       & 3{,}857{,}284 & 0.117 & 0.161 \\
                   & $300$ epochs                       & 3{,}857{,}284 & \textbf{0.108} & \textbf{0.141} \\
\addlinespace[2pt]
\textit{Depth}     & $6{+}6$ blocks                     & 5{,}636{,}740 & 0.126 & 0.173 \\
                   & $8{+}8$ blocks                     & 7{,}416{,}196 & 0.115 & 0.154 \\
\addlinespace[2pt]
\textit{Width}     & $d{=}256$, $h{=}4$                 & 6{,}797{,}572 & 0.140 & 0.187 \\
\bottomrule
\end{tabular}
\end{table}

\section{Field accuracy: ablation contracts, seed variance and error distribution}\label{app:reablation}

\subsection{What each ablation arm receives}
\label{app:ablation_contracts}

The ablations of Table~\ref{tab:ablation} are described by the component they
replace, but two arms change more than that description implies.
Table~\ref{tab:ablation_contracts} states the input contract of every arm. All
arms receive the volume evaluation points $\mathcal{V}$ and the scalar flow
rate $Q$, except where FiLM is removed. The learned-anchor arm replaces the
centerline token stream by $512$ free parameters and so receives neither the
centerline coordinates nor the radius profile, and because the anchors carry
no spatial positions the attention-level positional encoding is inactive in
Stage~1; that row prices three simultaneous changes, and the $0.205 \to 0.385$
pressure delta bounds the joint effect from above and attributes none of it.
Isolating the focal-set contribution requires an arm that keeps the centerline
and radius as inputs while drawing the focal set from learned inducing points,
which has not been run. ``Remove FiLM'' prices the availability of the flow
rate; FiLM$\to$Concat prices the form of the conditioning.

\begin{table}[!tb]
\centering
\small
\caption{Input contract of each ablation arm. ``PE (S1)'' is the
attention-level positional encoding in Stage~1; it is active in Stage~2 in
every arm.}
\label{tab:ablation_contracts}
\footnotesize
\setlength{\tabcolsep}{4pt}
\begin{tabular}{@{}lccccp{0.30\linewidth}@{}}
\toprule
Arm & CL $(x,y,z)$ & $r$ & $Q$ & PE (S1) & What the row prices \\
\midrule
CARS (default)                 & yes & yes & yes & yes & (reference) \\
Tiered $\to$ Cross-only        & yes & yes & yes & yes & focal self-attention \\
Tiered $\to$ Perceiver         & yes & yes & yes & yes & attention structure \\
FiLM $\to$ Concat              & yes & yes & yes & yes & form of conditioning \\
Two-stage $\to$ Single-stage   & yes & yes & yes & yes & staging \\
Remove FiLM                    & yes & yes & no  & yes & availability of $Q$ \\
CL anchors $\to$ Learned       & no  & no  & yes & no  & focal set, geometric input and Stage-1 PE jointly \\
\bottomrule
\end{tabular}
\end{table}

\subsection{Reference configuration}

The component ablations use RoPE3D as the reference ($0.205/0.287$ at
$100$ epochs). The positional-encoding head-to-head is a separate, paired run
of CARS-RoPE3D and CARS-learned-PE (identical hyperparameters, fixed seed, a
different random-initialization state); its RoPE3D figure is $0.199/0.269$.
The two RoPE3D numbers index the same configuration in two runs, and the difference is consistent with the run-to-run variability at this
batch size. We use $0.205/0.287$ for deltas within the component rows and
$0.199/0.269$ when quoting CARS-RoPE3D against learned PE and in
Table~\ref{tab:baselines}; the positional-encoding margin is the paired one,
$0.199/0.269 \to 0.140/0.196$ ($30\%/27\%$). Three rows with large RoPE3D
deltas (Perceiver, learned anchors, remove FiLM) were not run on the
learned-PE backbone; the two small or sign-ambiguous rows (FiLM$\to$Concat,
single-stage) were, and both resolve in favor of the CARS defaults
(Table~\ref{tab:ablation}).

\subsection{Run-to-run variance}

Tables report point estimates from one seed per configuration. Re-evaluating
the primary $300$-epoch model over eight independent draws of the full test
set gives a standard deviation of $0.0003$ on pressure and $0.0002$ on
velocity (mean $0.117$/$0.147$), so reported field errors are reproducible to
a few parts in $10^{4}$ and the corresponding band on the threshold-agreement
figure is $0.26$ percentage points. Retraining the $100$-epoch learned-PE
configuration from three seeds ($42$, $43$, $44$) gives test pressure
$0.1497$, $0.1538$, $0.1544$ and velocity $0.1993$, $0.2108$, $0.2060$, i.e.\
$0.1526 \pm 0.0026$ and $0.2054 \pm 0.0058$; seed $42$ rounds to the
$0.150/0.199$ of Table~\ref{tab:baselines}. The two CARS-RoPE3D runs above
give a more conservative two-point spread of $\Delta_p \approx 0.006$ and
$\Delta_v \approx 0.018$. At $300$ epochs the three seeds of the primary
configuration (Table~\ref{tab:seeds300}) give $0.1198 \pm 0.0032$ and $0.1491 \pm 0.0047$, so the
two-run velocity bound is conservative by roughly a factor of four at that
budget. Seed spread is not a constant of the problem: AB-UPT at the same
budget spreads by $0.0138$ on pressure, four times CARS's figure. Against the
conservative bound the architectural effects the paper draws conclusions
from are larger by $4$--$10\times$ on the relevant field, and against the
measured three-seed spread by $10$--$25\times$.

\begin{table}[!htb]
\centering
\caption{The primary $300$-epoch LCA configuration and the AB-UPT baseline,
each retrained from the same three seeds and evaluated under one fp32 test
protocol on the $300$-case test split (best-validation checkpoint, EMA weights
for CARS). Seed $42$ is the CARS run reported as the single-run figure of
Table~\ref{tab:baselines}; seed $1234$ is the AB-UPT single run and is listed for reference
but excluded from the matched statistics, since CARS was not trained at it.
Both single-run rows happen to be their arm's best seed, and because AB-UPT is
the more variable of the two the single-seed comparison understates the gap.}
\label{tab:seeds300}
\footnotesize
\setlength{\tabcolsep}{3pt}
\begin{tabular}{lcccc}
\toprule
& \multicolumn{2}{c}{CARS} & \multicolumn{2}{c}{AB-UPT} \\
\cmidrule(lr){2-3}\cmidrule(lr){4-5}
Seed & Pressure & Velocity & Pressure & Velocity \\
\midrule
$42$ & $0.1168$ & $0.1473$ & $0.1690$ & $0.1843$ \\
$43$ & $0.1232$ & $0.1544$ & $0.1635$ & $0.1686$ \\
$44$ & $0.1194$ & $0.1455$ & $0.1428$ & $0.1642$ \\
\midrule
mean $\pm$ s.d. & $\mathbf{0.1198 \pm 0.0032}$ & $\mathbf{0.1491 \pm 0.0047}$
                & $0.1584 \pm 0.0138$ & $0.1723 \pm 0.0106$ \\
\midrule
\multicolumn{5}{l}{\emph{unmatched, for reference}} \\
$1234$ & n/a & n/a & $0.1300$ & $0.1624$ \\
\bottomrule
\end{tabular}
\end{table}

\subsection{Distribution of the per-case error and focal count}

Table~\ref{tab:error_tails} gives the distribution of the per-case error
behind the headline means, and Table~\ref{tab:nf_accuracy} the sensitivity of
the trained model to the focal density it is queried with.

\begin{table}[!htb]
\centering
\caption{Distribution of per-case relative $\ell_2$ error over the $300$-case
test split of each territory, in the cell-center measure of
\S\ref{sec:dataset}. The mean column is the headline figure; the remaining
columns are given because the distribution is right-skewed enough that the
mean is not a description of a typical case. Computed from the per-case
exports underlying \S\ref{sec:error_analysis}, whose means agree with the
headline to $0.002$.}
\label{tab:error_tails}
\begin{tabular}{llcccccc}
\toprule
& Field & Mean & s.d. & Median & p$90$ & p$95$ & Worst \\
\midrule
\multirow{2}{*}{LCA} & Pressure & $0.117$ & $0.138$ & $0.065$ & $0.288$ & $0.432$ & $0.950$ \\
                     & Velocity & $0.147$ & $0.072$ & $0.130$ & $0.230$ & $0.277$ & $0.520$ \\
\addlinespace[2pt]
\multirow{2}{*}{RCA} & Pressure & $0.063$ & $0.068$ & $0.041$ & $0.121$ & $0.188$ & $0.494$ \\
                     & Velocity & $0.119$ & $0.055$ & $0.105$ & $0.181$ & $0.203$ & $0.433$ \\
\bottomrule
\end{tabular}
\end{table}

\begin{table}[!htb]
\centering
\caption{Test accuracy against focal (centerline) count $n_f$ on the
$300$-case LCA test split, at fixed weights trained with $n_f = 2{,}048$. The
focal set is thinned by uniform arc-length stride from the $2{,}048$-point
resampled centerline, so the branch decomposition and the axial ordering are
preserved and only the focal density changes. Mean and median of the per-case
relative $\ell_2$ error, fp32. This measures how sensitive a trained surrogate
is to the focal density it is queried with, not the accuracy attainable if it
were retrained at each count.}
\label{tab:nf_accuracy}
\small
\setlength{\tabcolsep}{4pt}
\begin{tabular}{lcccc}
\toprule
& \multicolumn{2}{c}{Pressure} & \multicolumn{2}{c}{Velocity} \\
\cmidrule(lr){2-3}\cmidrule(lr){4-5}
$n_f$ & Mean & Median & Mean & Median \\
\midrule
$128$   & $0.1328$ & $0.0751$ & $0.1564$ & $0.1390$ \\
$256$   & $0.1187$ & $0.0646$ & $0.1489$ & $0.1298$ \\
$512$   & $0.1176$ & $0.0642$ & $0.1470$ & $0.1295$ \\
$1{,}024$ & $0.1162$ & $0.0632$ & $0.1468$ & $0.1288$ \\
$2{,}048$ (default) & $0.1163$ & $0.0631$ & $0.1468$ & $0.1292$ \\
\bottomrule
\end{tabular}
\end{table}

\subsection{Field accuracy excluding duplicate geometries}
\label{sup:dedup}

Of the $300$ test cases in each territory, $34$ (LCA) and $22$ (RCA) are
identical to a training case in both centerline trajectory and radius
profile, differing only in the inlet flow rate $Q$
(\S\ref{sec:dataset}). Because a surrogate could in
principle have memorized those geometries, Table~\ref{tab:dedup} reports the
headline field errors with them removed.

The duplicates are substantially easier than the rest: mean pressure error
$0.039$ against $0.127$ on LCA and $0.041$ against $0.065$ on RCA. Their effect
on the headline is nonetheless small, because they are a minority of the split:
removing them raises the reported LCA pressure error by $0.010$ and the RCA
figure by $0.002$. Every conclusion in the paper is unchanged under the
duplicate-free numbers, and the model-to-model comparisons are unaffected
because the partition is identical for every model.

\begin{table}[!tb]
\centering
\caption{Mean relative $\ell_2$ error on the full test split and with
identical duplicate geometries removed. Computed from the same per-case
exports as the error-structure analysis.}
\label{tab:dedup}
\small
\setlength{\tabcolsep}{4pt}
\begin{tabular}{llccc}
\toprule
Territory & Field & Full split ($300$) & Duplicates only & Duplicates removed \\
\midrule
\multirow{2}{*}{LCA} & Pressure & $0.117$ & $0.039$ ($n=34$)  & $0.127$ ($n=266$) \\
                     & Velocity & $0.147$ & $0.087$           & $0.155$ \\
\addlinespace[2pt]
\multirow{2}{*}{RCA} & Pressure & $0.063$ & $0.041$ ($n=22$)  & $0.065$ ($n=278$) \\
                     & Velocity & $0.119$ & $0.073$           & $0.122$ \\
\bottomrule
\end{tabular}
\end{table}

\section{Neural-operator baselines: configurations}\label{app:baselines}

All baselines are matched in parameter count to within $3\%$ of CARS and
trained for $100$ epochs on the same LCA dataset with identical preprocessing
($16{,}384$ volume points, $2{,}048$ centerline points, z-score
normalization); Table~\ref{tab:baseline_configs} summarizes them. Every
baseline was trained with AdamW at CARS's settings (learning rate
$5\times10^{-5}$, weight decay $0.05$, cosine schedule with $5\%$ warm-up,
batch size $1$); the AB-UPT single runs used learning rate $1\times10^{-4}$
and weight decay $0.01$, and its matched-seed runs use CARS's settings. No per-baseline learning-rate or schedule search was performed, and the
batch-size-$1$ regime may specifically disadvantage architectures designed to
train at larger batches: the $0.804/0.957$ reached by ONO is close to what
predicting the training mean would give, the signature of a baseline that has
not trained. The margins over GNOT, ONO and Transolver are therefore upper
bounds on the architectural difference. The flow rate reaches the baselines
by three routes (concatenation, sinusoidal embedding, FiLM), and the ablation
of Table~\ref{tab:ablation} prices replacing FiLM by concatenation at $0.024$
on both fields, so a quantified share of the gap to the non-FiLM baselines is
attributable to conditioning. The AB-UPT comparison, on which the paper's
claims rest, trains stably and is compared over three shared seeds.

\begin{table}[!tb]
\centering
\small
\caption{Baseline configurations. All are trained on the same splits with the
same preprocessing; the first three receive volume positions and centerline
conditioning (positions, radii, flow rate), AB-UPT receives the surface mesh
and the flow rate.}
\label{tab:baseline_configs}
\footnotesize
\setlength{\tabcolsep}{4pt}
\begin{tabular}{@{}l>{\raggedright\arraybackslash}p{0.17\linewidth}r>{\raggedright\arraybackslash}p{0.30\linewidth}>{\raggedright\arraybackslash}p{0.17\linewidth}@{}}
\toprule
Model & Width / blocks & Params & Geometric input & $Q$ pathway \\
\midrule
GNOT~\cite{hao2023gnot}          & $d{=}144$            & $3.87$M & centerline tokens (branch input) & concatenated to trunk \\
ONO~\cite{xiao2023ono}           & $d{=}276$            & $3.84$M & centerline tokens                & sinusoidal embedding \\
Transolver~\cite{wu2024transolver} & $d{=}292$          & $3.81$M & centerline tokens                & sinusoidal embedding \\
AB-UPT~\cite{alkin2024abupt}     & $d{=}132$, $h{=}2$, $5{+}5$ & $3.76$M & surface mesh, $8{,}192$ points pooled to $1{,}024$ supernodes & FiLM (as CARS) \\
CARS                             & $d{=}192$, $h{=}3$, $4{+}4$ & $3.86$M & centerline $(x,y,z,r)$           & FiLM \\
\bottomrule
\end{tabular}
\end{table}

\paragraph{GNOT} Heterogeneous cross-attention with centerline tokens as
branch inputs and mixture-of-experts feed-forward networks with geometric
gating.

\paragraph{ONO} Orthogonal neural operator built on orthonormalized neural
eigenfunctions of the kernel integral operator, with an
orthogonalization-regularized attention block.

\paragraph{Transolver} Physics-informed slice attention projecting tokens
onto a small set of learned slices, at linear complexity.

\paragraph{AB-UPT} The architecture that motivates CARS, trained from
scratch; it combines anchor attention, 3D rotary embeddings and FiLM in a
branched structure with separate surface and volume decoders. The
implementation is adapted to make the comparison like-for-like: width and
block counts bring the parameter count to $97.5\%$ of CARS; the data path is
extended to this cohort's volumetric fields and evaluation-point sets;
flow-rate conditioning is added on the same scalar through the same
feature-wise modulation; and the loss is restricted to the volume heads. The
surface pressure and vorticity heads carry zero loss weight and remain
untrained (relative error ${\approx}1.05$ throughout), so the parameter match
is nominal and AB-UPT's capacity usable for the supervised objective is below
CARS's at equal total parameters. Its geometric input is the surface mesh,
which on this cohort is a deterministic function of the centerline and radius
profile (\S\ref{sec:problem}); the data loader carries centerline coordinates
and radii into the forward signature, but they are discarded before the model
body, verified by tracing the forward pass. AB-UPT trains under mixed
precision, which its per-epoch cost requires, while CARS trains in fp32; both
are \emph{evaluated} in fp32, and evaluating AB-UPT under mixed precision
instead moves its numbers by $10^{-5}$ to $10^{-4}$, which settles the scoring
question and leaves the training question open, since this architecture
family is precision-sensitive in inference (Table~\ref{tab:density_precision}).
Both arms report the best-validation checkpoint.

Figure~\ref{fig:pareto} places every model on the latency--accuracy plane.

\begin{figure}[!htb]
\centering
\begin{tikzpicture}
\begin{axis}[
    width=0.92\linewidth, height=6.0cm,
    xmode=log, log basis x=10,
    xlabel={Per-case inference latency (ms, log scale)},
    ylabel={Test pressure rel-$\ell_2$ (lower is better)},
    xmin=6, xmax=420,
    ymin=0.05, ymax=0.92,
    xtick={10, 20, 50, 100, 200},
    xticklabels={$10$, $20$, $50$, $100$, $200$},
    grid=major, grid style={gray!20},
    label style={font=\small},
    tick label style={font=\scriptsize},
]
\addplot[draw=none, fill=red!4, forget plot] coordinates
    {(6, 0.92) (8.7, 0.92) (8.7, 0.514) (21.4, 0.514) (21.4, 0.444)
     (87, 0.444) (87, 0.116) (420, 0.116) (420, 0.92)} \closedcycle;
\addplot[thick, dashed, color=red!75!black, forget plot] coordinates
    {(8.7, 0.514) (21.4, 0.514) (21.4, 0.444) (87, 0.444) (87, 0.116) (420, 0.116)};
\draw[->, thin, gray!50, >=stealth] (axis cs:90, 0.150) -- (axis cs:90, 0.119);
\draw[->, thin, gray!50, >=stealth] (axis cs:170, 0.211) -- (axis cs:170, 0.133);
\addplot[only marks, mark=diamond*, mark size=3.0pt, color=orange!80!black,
         forget plot] coordinates {(8.7, 0.514)};
\addplot[only marks, mark=diamond*, mark size=3.0pt, color=violet,
         forget plot] coordinates {(21.4, 0.444)};
\addplot[only marks, mark=diamond*, mark size=3.0pt, color=teal!70!black,
         forget plot] coordinates {(22.5, 0.804)};
\addplot[only marks, mark=triangle, mark size=3.5pt, color=blue!70!black,
         line width=0.8pt, forget plot] coordinates {(166, 0.211)};
\addplot[only marks, mark=triangle*, mark size=3.5pt, color=blue!70!black,
         forget plot] coordinates {(166, 0.130)};
\addplot[only marks, mark=square, mark size=3.0pt, color=red!50!black,
         line width=0.8pt, forget plot] coordinates {(87, 0.201)};
\addplot[only marks, mark=*, mark size=3.0pt, color=red!75!black,
         forget plot] coordinates {(87, 0.150)};
\addplot[only marks, mark=*, mark size=4.6pt, color=red!75!black,
         draw=black, line width=0.5pt, forget plot] coordinates {(87, 0.116)};
\node[font=\scriptsize, anchor=north]       at (axis cs:8.7, 0.508)  {Transolver};
\node[font=\scriptsize, anchor=north west]  at (axis cs:22.5, 0.79)  {ONO};
\node[font=\scriptsize, anchor=north east]  at (axis cs:20.5, 0.43)  {GNOT};
\node[font=\scriptsize, anchor=south west]  at (axis cs:175, 0.213)  {AB-UPT-100$^{\dagger}$};
\node[font=\scriptsize, anchor=south west]  at (axis cs:175, 0.135)  {AB-UPT-300$^{\dagger}$};
\node[font=\scriptsize, anchor=east]        at (axis cs:80, 0.201)   {CARS-RoPE3D};
\node[font=\scriptsize, anchor=east]        at (axis cs:80, 0.150)   {CARS-100};
\node[font=\scriptsize\bfseries, anchor=east, color=red!75!black]
                                             at (axis cs:80, 0.092)   {CARS-300 (ours)};
\node[font=\scriptsize\itshape, color=red!75!black, anchor=west]
    at (axis cs:8.7, 0.74) {lower-left envelope};
\draw[->, thin, red!75!black, >=stealth]
    (axis cs:11, 0.71) -- (axis cs:13.5, 0.515);
\node[font=\scriptsize\itshape, color=gray!60!black, anchor=west, align=left]
    at (axis cs:115, 0.78) {grey arrows: $100\!\to\!300$\\\phantom{xx}epochs};
\end{axis}
\end{tikzpicture}
\caption{Latency--accuracy scatter on the LCA test set (test pressure
rel-$\ell_2$ vs.\ per-case fp32 inference latency on a single A100, both
lower-is-better). Filled markers: $300$-epoch configurations; hollow markers:
matched $100$-epoch budget; grey arrows show the within-architecture
training-budget improvement. The red dashed line is the stepwise lower-left
envelope of the models evaluated here, a property of this particular set of
runs rather than a Pareto frontier over architectures: only CARS and AB-UPT
were trained at $300$ epochs and the baselines were not tuned individually. $^{\dagger}$~AB-UPT additionally consumes the
surface mesh. The reference pipeline is off-scale to the right.}
\label{fig:pareto}
\end{figure}
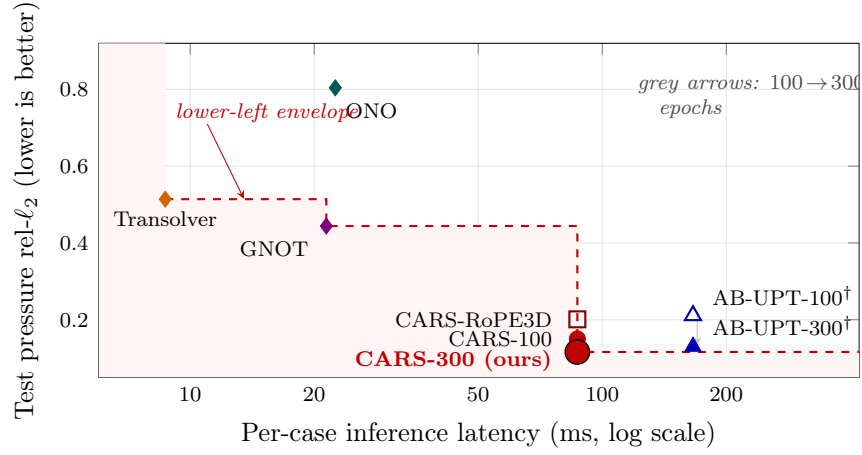

\clearpage
\section{FFR endpoint: protocol, agreement, baselines and sensitivity}\label{app:ffr_protocol}

\subsection{Lesion localization and measurement station}

The FFR endpoint is evaluated at a lesion-specific measurement point,
localized from the geometry alone so that the identical procedure applies to
the predicted and the CFD-derived field. The resampled centerline
concatenates the branches of the vessel tree, so a step exceeding ten times
the median point spacing marks a branch boundary, and all arc-length
arithmetic is confined to a single branch. A stenosis is a \emph{focal}
narrowing, distinguished from natural distal tapering by a healthy shoulder on
both sides: for every centerline point $i$ within a branch, with $s$ the arc
length along that branch,
\begin{align}
    r_{\text{prox}}(i) &= \max \{ r(k) : s_i - w_{\text{far}} \le s_k \le s_i - w_{\text{near}} \}, \nonumber\\
    r_{\text{dist}}(i) &= \max \{ r(k) : s_i + w_{\text{near}} \le s_k \le s_i + w_{\text{far}} \}, \nonumber\\
    r_{\text{ref}}(i)  &= \min\big( r_{\text{prox}}(i),\, r_{\text{dist}}(i) \big),
    \label{eq:ref_radius}
\end{align}
and the severity is
\begin{equation}
    \sigma(i) \;=\; 1 - \frac{r(i)}{r_{\text{ref}}(i)},
    \label{eq:severity}
\end{equation}
with $w_{\text{near}} = 2$~mm and $w_{\text{far}} = 10$~mm. Requiring both
windows to lie inside the same branch excludes distal branch tips by
construction. A point is eligible if in addition $r_{\text{ref}}(i) \ge 1$~mm,
a caliber gate reflecting that invasive FFR is not used below about $2$~mm
diameter, and at least $5$~mm of branch remains distal to it. The lesion is
the eligible point of maximum severity over the whole tree; one is found in
every case of both test sets.

Pressure is sampled $20$~mm of arc length distal to the lesion within the same
branch, matching the clinical convention of placing the sensor at least $2$~cm
beyond the lesion, past the pressure-recovery zone; where less vessel is
available the point is taken at the distal end of the lesion's branch, which
applies to $9\%$ (LCA) and $15\%$ (RCA) of cases. Table~\ref{tab:ffr_offset}
reports sensitivity to the offset under the nearest-point protocol
(\appref{app:ffr_exact}): the
$20$~mm station is the most favorable of the four on LCA and the least
favorable on RCA, which is what one expects of a station fixed on clinical
grounds before any test-set evaluation. FFR is computed as
\begin{equation}
    \mathrm{FFR} \;=\; P_{\text{distal}}/P_{\text{aortic}},
    \label{eq:ffr_formula}
\end{equation}
with $P_{\text{distal}} = P_{\text{aortic}} + c_\rho\, p$ from the solver's
kinematic pressure at the measurement point, the unit conversion $c_\rho$ of
Eq.~\eqref{eq:ffr_functional} and $P_{\text{aortic}} = 13{,}650$~Pa
($102$~mmHg). Since $p$ is referred to the inlet it is negative downstream of a
lesion, so $\mathrm{FFR} < 1$; the conversion is applied identically to
predicted and reference pressure.

\begin{table}[!tb]
\centering
\caption{Sensitivity of the FFR endpoint to the measurement offset distal to
the lesion, on the filtered $\mathrm{FFR}_{\mathrm{gt}}\in[0.3,1.0]$ subset,
nearest-point protocol. All offsets are evaluated from the same forward pass.
``Clamped'' is the fraction of cases in which less than the requested offset
of vessel remains and the measurement is taken at the branch end. $n$ varies
because the physiologic-range filter is applied at the measurement location.}
\label{tab:ffr_offset}
\begin{tabular}{llrrrrr}
\toprule
 & Offset & Clamped & $n$ & Pearson $r$ & MAE & Acc.\ @$0.8$ \\
\midrule
\multirow{4}{*}{LCA} & $5$~mm  & $0\%$  & 275 & 0.917 & 0.019 & 96.4\% \\
                     & $10$~mm & $2\%$  & 284 & 0.895 & 0.020 & 96.1\% \\
                     & $20$~mm & $9\%$  & 286 & \textbf{0.953} & \textbf{0.019} & \textbf{97.6\%} \\
                     & $30$~mm & $21\%$ & 283 & 0.945 & 0.024 & 96.1\% \\
\addlinespace[2pt]
\multirow{4}{*}{RCA} & $5$~mm  & $0\%$  & 294 & 0.994 & 0.008 & 97.6\% \\
                     & $10$~mm & $4\%$  & 296 & 0.993 & 0.008 & 97.0\% \\
                     & $20$~mm & $15\%$ & 293 & \textbf{0.994} & \textbf{0.008} & \textbf{96.2\%} \\
                     & $30$~mm & $27\%$ & 293 & 0.994 & 0.009 & 96.9\% \\
\bottomrule
\end{tabular}
\end{table}

\subsection{Exact-coordinate against nearest-point reading}
\label{app:ffr_exact}

Under the nearest-point protocol the
predicted pressure is read at the nearest of the $16{,}384$ sampled query
points and the reference at the nearest production cell center, so the
predicted value inherits the randomness of the evaluation draw; the primary
analysis of \S\ref{sec:clinical_ffr} appends the measurement point to the
query set instead, and the residual differences between nominally identical
baseline figures across the sensitivity tables arise in the analyses that retain
the nearest-point protocol: $97.6$/$97.9$/$97.6$ on LCA, and $96.6\%$ against
the $96.2\%$ that the exact-coordinate protocol gives at the same $20$~mm
station on RCA. The spread is the sampling term the primary protocol removes.

Table~\ref{tab:ffr_protocol} gives the two protocols of
\S\ref{sec:clinical_ffr} side by side. Exact querying leaves the threshold
decision essentially untouched, changing no case on LCA and one on RCA, and
improves the full-cohort correlation on LCA from $0.921$ to $0.941$ while
leaving the physiologic-range subset and both RCA scopes unchanged to three
decimals. The improvement is concentrated where it should be: the
nearest-point offset perturbs a small number of cases with steep local
pressure gradients, which are exactly the near-occlusive cases the subset
excludes, so the defect was invisible on the filtered numbers and visible on
the full cohort. The sampling step also contributes about $0.01$ of run-to-run
variation in $r$ on its own, since two draws of the query set place different
points near $\mathbf{x}_{\mathrm{m}}$.

\begin{table}[!htb]
\centering
\caption{Effect of reading the predicted pressure at the exact measurement
coordinate rather than at the nearest of the $16{,}384$ sampled query points,
at the primary $20$~mm station. The reference is the nearest production cell
center in both columns. The measurement station, the lesion localization and
the case counts are identical between the two, so the difference is
attributable to the lookup alone.}
\label{tab:ffr_protocol}
\small
\setlength{\tabcolsep}{4pt}
\begin{tabular}{llcccccc}
\toprule
& & \multicolumn{3}{c}{Nearest point} & \multicolumn{3}{c}{Exact coordinate} \\
\cmidrule(lr){3-5}\cmidrule(lr){6-8}
Territory & Scope ($n$) & $r$ & MAE & Acc. & $r$ & MAE & Acc. \\
\midrule
\multirow{2}{*}{LCA} & full ($300$)     & $0.921$ & $0.028$ & $97.7\%$ & $\mathbf{0.941}$ & $\mathbf{0.026}$ & $97.7\%$ \\
                     & subset ($286$)   & $0.953$ & $0.020$ & $97.6\%$ & $0.953$ & $0.019$ & $97.6\%$ \\
\addlinespace[2pt]
\multirow{2}{*}{RCA} & full ($300$)     & $0.993$ & $0.010$ & $96.3\%$ & $0.993$ & $0.010$ & $96.3\%$ \\
                     & subset ($293$)   & $0.994$ & $0.008$ & $96.2\%$ & $0.994$ & $0.008$ & $96.2\%$ \\
\bottomrule
\end{tabular}
\end{table}

\subsection{Physiologic-range subset and agreement analyses}
\label{app:ffr_agreement}

The physiologic-range subset of \S\ref{sec:clinical_ffr} keeps the cases
whose reference FFR lies in $\mathrm{FFR}_{\mathrm{gt}}\in[0.3, 1.0]$, $286$
of $300$ on LCA and $293$ of $300$ on RCA, where the reference value is itself
a threshold comparison. Excluding the tail raises $r$ and lowers MAE, because
the excluded cases carry the largest absolute reference values and the largest
absolute errors; no conclusion in the paper differs between the two scopes.
Figures~\ref{fig:roc_curves} and~\ref{fig:bland_altman} show the ROC curves
and Bland--Altman plots behind the agreement figures quoted there; the ROC AUC
carries patient-clustered bootstrap $95\%$ confidence intervals
$[0.989,0.999]$ (LCA) and $[0.990,0.999]$ (RCA). Every interval in
Table~\ref{tab:ffr_ci} is clustered on the patient, because the $300$ test
cases of each territory are variants of $96$/$111$ segmentations and are not
mutually independent. Clustering widens the intervals only marginally, by a
factor of $1.0$--$1.3$ across the six metrics and both territories, and
narrows marginally on two of them: what drives an FFR value is the lesion, and
lesion severity varies \emph{within} a patient as much as between patients, so
little of the endpoint variance is absorbed by the patient grouping. The
threshold-level conclusions are therefore not an artifact of counting
correlated cases as independent.

The two summaries answer different questions. The LCA limits of agreement
span ${\approx}0.17$ across the $\mathrm{FFR}=0.8$ level, so a case whose
reference value lies close to that level can be predicted on the other side of
it under prediction spread alone; this single-case spread is not visible in
the cohort-level threshold agreement of $97.6\%$. The RCA limits are roughly
three times tighter, consistent with the lower field errors on that territory.

\begin{figure}[!htb]
    \centering
    \includegraphics[width=\linewidth]{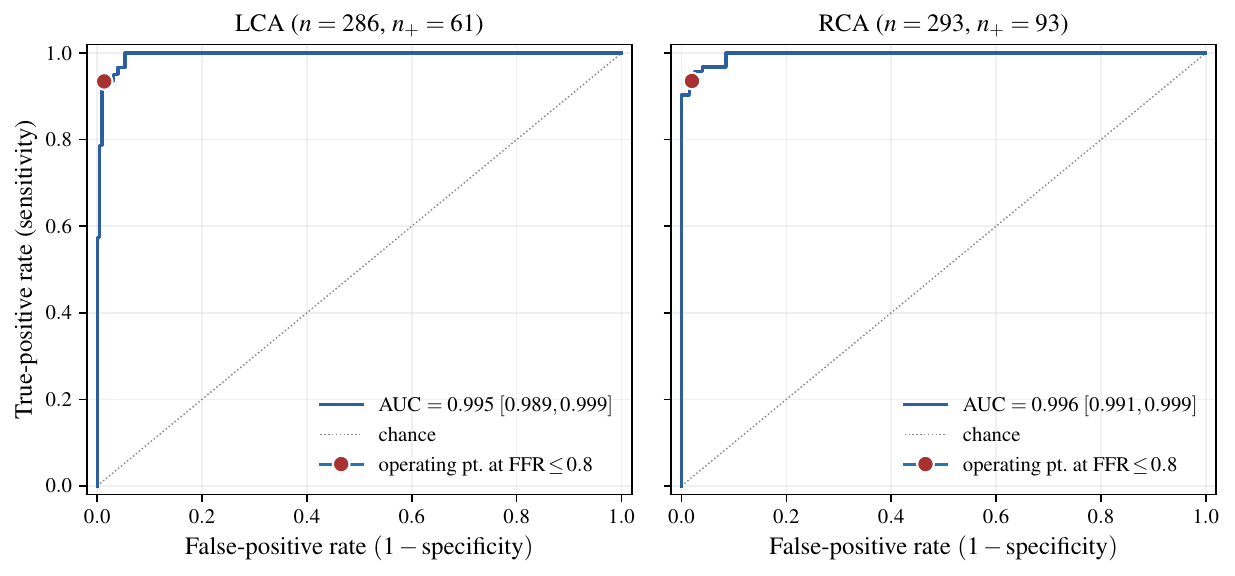}
    \caption{ROC curves for the binary $\mathrm{FFR}\le 0.8$ decision on LCA
    (left) and RCA (right). Red marker: operating point at the clinical
    cutoff. Dotted line: chance.}
    \label{fig:roc_curves}
\end{figure}

\begin{figure}[!htb]
    \centering
    \includegraphics[width=\linewidth]{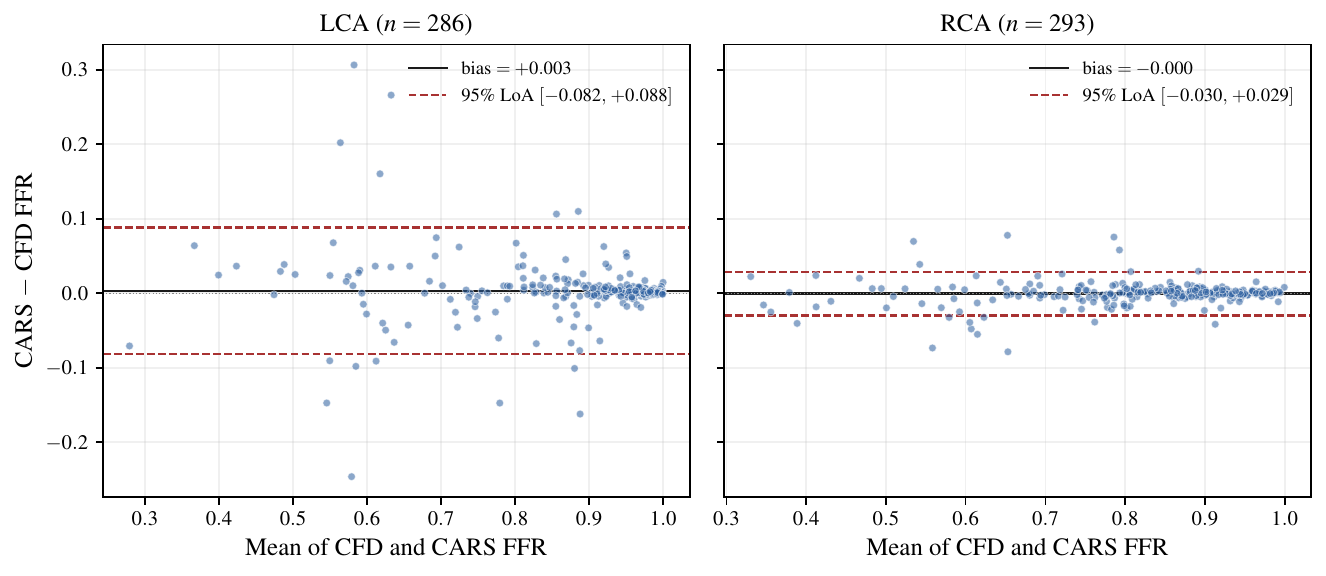}
    \caption{Bland--Altman plots of predicted vs.\ CFD-derived FFR for LCA
    (left) and RCA (right). Solid line: bias; dashed lines: $95\%$ limits of
    agreement.}
    \label{fig:bland_altman}
\end{figure}

\subsection{One-dimensional blood-flow reduced-order model}
\label{supp:rom}

The 1D ROM combines the viscous Hagen--Poiseuille pressure drop with the
convective (Bernoulli) acceleration term, integrated along the anatomical path
from the inlet to the same measurement point used for CARS:
\begin{equation}
    \Delta P \;=\; \int_{0}^{s_{\text{meas}}}
    \frac{8\,\mu\,Q(s)}{\pi\,r(s)^{4}}\,\mathrm{d}s
    \;+\; \tfrac{1}{2}\rho\,\alpha\!\left(U_{\text{meas}}^{2} - U_{\text{in}}^{2}\right),
    \label{eq:1drom}
\end{equation}
with $\alpha = 4/3$ (parabolic-profile momentum-flux correction, placed on the
convective term where it belongs; the viscous gradient is already exact for
that profile), $\mu = 4.00\times 10^{-3}$~Pa\,s, $\rho = 1{,}060$~kg/m$^3$ and
$U(s) = Q(s) / (\pi r(s)^{2})$. Placing $\alpha$ on the viscous term instead
would inflate the viscous loss by $33\%$ and move the baseline very little:
threshold agreement $49.0\%$ against $47.2\%$ on LCA and unchanged at $77.1\%$
on RCA, MAE $0.376$ against $0.381$ (LCA) and $0.099$ against $0.096$ (RCA),
Pearson $r$ $0.191$ against $0.193$ (LCA) and $0.559$ against $0.620$ (RCA).

Three points make the comparison a fair one. First, the integration path
follows the vessel tree: branch connectivity is reconstructed by attaching each
branch to the nearest point of an earlier branch, and the path is the resulting
chain from the inlet through the parent junctions to the measurement point.
Second, $Q(s)$ is the flow carried by the branch containing $s$, not the total
inlet flow: since the upstream pipeline prescribes each terminal flux in
proportion to the square of the terminal luminal radius
(Eq.~\eqref{eq:outlet_split}), a branch carries the summed terminal flux of its
own subtree,
\begin{equation}
    Q(s) \;=\; Q \,
    \frac{\sum_{t \in \mathcal{T}(b(s))} r_t^{2}}{\sum_{t \in \mathcal{T}} r_t^{2}},
    \label{eq:rom_flowsplit}
\end{equation}
with $\mathcal{T}(b)$ the terminals downstream of branch $b$, so the ROM is
driven by the same outflow distribution as the 3D reference. Third, all
quantities are converted once from the pipeline's
$(\mathrm{mm}, \mathrm{s}, \mathrm{kg})$ system to SI, so
Eq.~\eqref{eq:1drom} is evaluated with physical constants and \emph{no}
empirical flow-rate rescaling is fitted. FFR follows as
$\mathrm{FFR}_{\mathrm{ROM}} = 1 - \Delta P / P_{\mathrm{aortic}}$ with
$P_{\mathrm{aortic}} = 13{,}650$~Pa, matching Eq.~\eqref{eq:ffr_formula}, and
is clipped to $[0,1]$.

The resulting baseline reaches Pearson $r = 0.19/0.56$ and threshold accuracy
$49.0\%/77.1\%$ on LCA/RCA against majority-class baselines of
$78.7\%/68.3\%$ (Table~\ref{tab:ffr_baselines}), and is markedly weaker on LCA
than on RCA. This is a property of the model class: in a tapering tube both
terms of Eq.~\eqref{eq:1drom} scale as $r^{-4}$, and both overestimate the
pressure drop where the true loss is governed by flow separation and partial
pressure recovery rather than by fully developed viscous shear or by a
lossless Bernoulli exchange. On LCA the two terms are of the same order
(medians $1{,}939$~Pa viscous and $1{,}436$~Pa convective), and the median
per-case convective share of $|\Delta P|$ is $53\%$ against $20\%$ on RCA
(medians $1{,}132$ and $307$~Pa), so on LCA the convective term is the larger
one in more than half of cases; the mechanism is geometric, since the
measurement station sits in a distal vessel whose median radius is $0.89$~mm
against an inlet $2.05$~mm, and on $85$ of $286$ LCA cases the total predicted
drop exceeds aortic pressure and the FFR estimate clips at zero. The LCA
cohort also carries more severe lesions (median severity $0.46$ versus
$0.24$). Fitting a single global coefficient on the viscous term using
training cases only does not rescue it: the fitted coefficient collapses
towards zero on LCA, trading the over-prediction for a near-constant
$\mathrm{FFR}\approx 1$, so the uncalibrated, dimensionally consistent model is
reported. It is a physics sanity check rather than a head-to-head against
modern 1D solvers (SimVascular's 1D solver, openBF), which include vessel
compliance and proper outlet impedances and would be expected to perform
substantially better; the reduced-order comparator a computational-mechanics
reader would expect is a one-dimensional solver carrying an explicit
stenosis-loss closure (of Young--Tsai type, as in the reduced-order model
of~\cite{itu2016machine}), which is the single largest gap in the
comparison set of the paper.

\subsection{Same-data scalar-regression baseline}

An Itu-style scalar regressor~\cite{itu2016machine} and an XGBoost upper bound
were trained on the same splits and the same CFD-derived FFR labels as CARS,
using $15$ hand-engineered per-vessel geometric features (min/mean/proximal/
distal radius, stenosis severity, total and to-stenosis arc length,
tortuosity, mean and max curvature, stenosis length, $5$~mm-window mean
radius, inlet flow rate), with hyperparameters tuned on the validation split.
Table~\ref{tab:ffr_baselines} reports them alongside the ROM and CARS. Their
AUCs are not comparable to the ${\sim}0.84$--$0.93$ band reported for
machine-learning ct-FFR in the prior
literature~\cite{coenen2018ml,tesche2018ml,kumamaru2020deep}, which is measured
against \emph{invasive} FFR on clinical cohorts, whereas these are measured
against a CFD label on a synthetic product distribution of geometries and flow
rates; numerical proximity between the two carries no information about
calibration.

\begin{table}[!htb]
\centering
\caption{FFR endpoint: CARS versus the uncalibrated, dimensionally consistent
1D blood-flow ROM and Itu-style per-vessel scalar regressors on the same
CFD-derived FFR labels and the same train/val/test splits
($|\mathrm{train}|=2{,}700$), physiologic-range subset, nearest-point
protocol. ``Scalar MLP (Itu)'' is the two-layer feed-forward network of Itu
et al.~\cite{itu2016machine}; ``Scalar XGB'' is a gradient-boosted upper bound
on what scalar feature regression can extract from the same $15$ features.
AUC and accuracy carry patient-clustered bootstrap $95\%$ CIs. Always
predicting the majority class gives $78.7\%$ (LCA) and $68.3\%$ (RCA).}
\label{tab:ffr_baselines}
\setlength{\tabcolsep}{4pt}\small
\begin{tabular}{lcccc}
\toprule
Method & $r$ & MAE & Acc.\,@\,$0.8$ & AUC \\
\midrule
\multicolumn{5}{l}{\emph{LCA, $n=286$}} \\
1D ROM             & 0.191 & 0.376 & 49.0\%                          & 0.54                          \\
Scalar MLP (Itu)   & 0.730 & 0.070 & 89.2\%\,[85.0,\,92.3]           & 0.92\,[0.87,\,0.96]           \\
Scalar XGB         & 0.810 & 0.050 & 91.3\%\,[87.4,\,94.0]           & 0.96\,[0.93,\,0.98]           \\
CARS               & \textbf{0.953} & \textbf{0.019} & \textbf{97.6\%} & \textbf{0.995\,[0.989,\,0.999]} \\
\addlinespace[2pt]
\multicolumn{5}{l}{\emph{RCA, $n=293$}} \\
1D ROM             & 0.559 & 0.099 & 77.1\%                          & 0.84                          \\
Scalar MLP (Itu)   & 0.760 & 0.070 & 78.8\%\,[73.8,\,83.1]           & 0.88\,[0.83,\,0.91]           \\
Scalar XGB         & 0.840 & 0.040 & 89.8\%\,[85.8,\,92.7]           & 0.93\,[0.90,\,0.96]           \\
CARS               & \textbf{0.994} & \textbf{0.008} & \textbf{96.2\%} & \textbf{0.995\,[0.990,\,0.999]} \\
\bottomrule
\end{tabular}
\end{table}

\subsection{Sensitivity and triage tables}
\label{supp:sensitivity}

Tables~\ref{tab:triage_val_tuned}--\ref{tab:robustness_segment} are the
numerical values behind Figure~\ref{fig:robustness_triage} and
\S\ref{sec:robustness_uq}. The MC-dropout predicted standard deviation is
overconfident as an absolute interval, with $1\sigma$ coverage of $52\%$ on
LCA and $44\%$ on RCA against a calibrated $68\%$ and $2\sigma$ coverage of
$0.68$/$0.69$ against a nominal $0.95$; turning it into a calibrated interval
would take a different uncertainty formulation rather than a rescaling of this
one. The cutoff transfers less cleanly on RCA, whose validation-side $\sigma$
distribution has a heavier tail, so the realized fraction departs from its
target.

\begin{table}[!htb]
\centering
\caption{Val-tuned MC-dropout triage on the filtered
$\mathrm{FFR}_{\mathrm{gt}}\in[0.3, 1.0]$ subset ($n_{\mathrm{LCA}}=286$,
$n_{\mathrm{RCA}}=293$). The $\sigma$-percentile cutoff is set on val and
applied unchanged to test; ``realized'' is the resulting test routing
fraction. Auto-rep.\ accuracy is the FFR-$\le 0.8$ threshold accuracy on the
auto-reported subset; the MC-mean baseline (no routing) on test is $97.9\%$
LCA / $96.2\%$ RCA.}
\label{tab:triage_val_tuned}
\small
\setlength{\tabcolsep}{4pt}
\begin{tabular}{llcccc}
\toprule
Territory & Target $f$ & $\sigma_{\mathrm{cut}}$ (val) & Realized & Err.\ absorbed & Auto-rep.\ acc.\ \\
\midrule
LCA & $10\%$ & $0.0103$ & $11.2\%$ & $37\%$ & $98.0\%$ \\
LCA & $20\%$ & $0.0072$ & $23.8\%$ & $61\%$ & $98.6\%$ \\
\addlinespace[2pt]
RCA & $10\%$ & $0.0060$ & $12.3\%$ & $31\%$ & $96.1\%$ \\
RCA & $20\%$ & $0.0046$ & $25.9\%$ & $56\%$ & $96.8\%$ \\
\bottomrule
\end{tabular}
\end{table}

\begin{table}[!htb]
\centering
\caption{Input-noise robustness on the same filtered subset and the same
model checkpoint as \S\ref{sec:clinical_ffr}. The $\sigma=0$ row is the
unperturbed baseline; it differs from Table~\ref{tab:ffr_ci} because the
evaluation points are resampled per run, by up to $0.008$ in $r$, so the
perturbation rows are to be read against the $\sigma=0$ row of this table.
Centerline noise is additive isotropic Gaussian (mm) applied independently to
each of the $2{,}048$ points; radius noise is multiplicative fractional
Gaussian on raw radii.}
\label{tab:robustness_noise}
\begin{tabular}{llcccc}
\toprule
Territory & Noise type & $\sigma$ & $r$ & MAE & Acc.\,@\,$0.8$ \\
\midrule
LCA & Baseline        & $0$    & 0.949 & 0.021 & 97.9\% \\
LCA & Centerline (mm) & $0.1$  & 0.948 & 0.021 & 97.2\% \\
LCA &                 & $0.5$  & 0.949 & 0.021 & 97.6\% \\
LCA &                 & $1.0$  & 0.945 & 0.022 & 97.2\% \\
LCA & Radius (frac)   & $0.05$ & 0.946 & 0.021 & 96.9\% \\
LCA &                 & $0.10$ & 0.910 & 0.036 & 93.4\% \\
LCA &                 & $0.20$ & 0.653 & 0.127 & 62.2\% \\
\addlinespace[2pt]
RCA & Baseline        & $0$    & 0.991 & 0.011 & 96.6\% \\
RCA & Centerline (mm) & $0.1$  & 0.993 & 0.010 & 96.2\% \\
RCA &                 & $0.5$  & 0.992 & 0.017 & 93.5\% \\
RCA &                 & $1.0$  & 0.970 & 0.080 & 84.0\% \\
RCA & Radius (frac)   & $0.05$ & 0.990 & 0.012 & 96.6\% \\
RCA &                 & $0.10$ & 0.991 & 0.014 & 96.9\% \\
RCA &                 & $0.20$ & 0.912 & 0.047 & 86.0\% \\
\bottomrule
\end{tabular}
\end{table}

\begin{table}[!htb]
\centering
\caption{Segment-level structured perturbations on the same filtered subset.
A contiguous $10$-mm arc-length window is selected uniformly at random per
case and a single bias is applied to every centerline point in it:
``segment-position'' a 3D Gaussian displacement of magnitude $\sigma$~mm,
``segment-radius'' a multiplicative factor $(1+\sigma\,\varepsilon)$,
$\varepsilon\sim\mathcal{N}(0,1)$. Seed fixed across $\sigma$ values. The
$\sigma=0$ row is the unperturbed baseline for this table.}
\label{tab:robustness_segment}
\begin{tabular}{llcccc}
\toprule
Territory & Perturbation & $\sigma$ & $r$ & MAE & Acc.\,@\,$0.8$ \\
\midrule
LCA & Baseline                       & $0$    & 0.946 & 0.021 & 97.6\% \\
LCA & Segment-position (mm)          & $0.5$  & 0.949 & 0.021 & 97.9\% \\
LCA &                                & $1.0$  & 0.948 & 0.021 & 97.6\% \\
LCA &                                & $2.0$  & 0.945 & 0.022 & 97.6\% \\
LCA & Segment-radius (frac)          & $0.05$ & 0.949 & 0.021 & 97.9\% \\
LCA &                                & $0.10$ & 0.947 & 0.021 & 97.6\% \\
LCA &                                & $0.20$ & 0.947 & 0.022 & 97.2\% \\
\addlinespace[2pt]
RCA & Baseline                       & $0$    & 0.992 & 0.010 & 96.2\% \\
RCA & Segment-position (mm)          & $0.5$  & 0.991 & 0.011 & 96.6\% \\
RCA &                                & $1.0$  & 0.992 & 0.011 & 96.6\% \\
RCA &                                & $2.0$  & 0.991 & 0.012 & 95.9\% \\
RCA & Segment-radius (frac)          & $0.05$ & 0.991 & 0.011 & 96.9\% \\
RCA &                                & $0.10$ & 0.993 & 0.010 & 96.9\% \\
RCA &                                & $0.20$ & 0.965 & 0.016 & 94.9\% \\
\bottomrule
\end{tabular}
\end{table}

\clearpage
\section{Error structure, invariances and transfer}
\label{supp:limits}

Table~\ref{tab:error_covariates} is the full binned table behind
Figure~\ref{fig:error_covariates}, Table~\ref{tab:frame_sensitivity} the
frame-sensitivity measurement of \S\ref{sec:frame_sensitivity},
Tables~\ref{tab:pde_consistency} and~\ref{tab:pde_clearance} the
consistency with the governing equations, and Table~\ref{tab:xterritory} the
cross-territory transfer of \S\ref{sec:cross_territory};
\appref{supp:anatomy_disjoint} reports an anatomy-disjoint partition.

\begin{table}[p]
\centering
\footnotesize
\caption{Per-case relative $\ell_2$ error against the parameters of the
solution map, on the full $300$-case test split of each territory
(Figure~\ref{fig:error_covariates} plots the pressure column). Entries are
the mean over the bin; $n$ is the number of cases. Severity bands and branch
counts are absolute; flow-rate and radius bins are per-territory quartiles,
with edges LCA $5{,}045 / 5{,}745 / 6{,}448 / 7{,}318 / 7{,}997$ and RCA
$2{,}515 / 2{,}988 / 3{,}574 / 3{,}985 / 4{,}486$~mm$^3$/s for $Q$, LCA
$1.32 / 1.88 / 2.06 / 2.24 / 2.50$ and RCA $1.51 / 1.81 / 1.98 / 2.12 /
2.29$~mm for inlet radius, and LCA $0.26 / 0.57 / 0.72 / 0.90 / 1.50$ and RCA
$0.44 / 0.75 / 0.93 / 1.12 / 1.92$~mm for throat radius. The Spearman row is
the rank correlation of the covariate with each error over all $300$ cases.
n/a marks bins the territory does not populate.}
\label{tab:error_covariates}
\begin{tabular}{lcccccc}
\toprule
 & \multicolumn{3}{c}{LCA} & \multicolumn{3}{c}{RCA} \\
\cmidrule(lr){2-4}\cmidrule(lr){5-7}
Bin & $n$ & Pressure & Velocity & $n$ & Pressure & Velocity \\
\midrule
\multicolumn{7}{l}{\textit{Lesion severity} $1-r/r_{\text{ref}}$, fixed bands} \\
\quad $[0.00, 0.20)$ & $5$ & $0.073$ & $0.136$ & $133$ & $0.070$ & $0.118$ \\
\quad $[0.20, 0.35)$ & $42$ & $0.069$ & $0.137$ & $51$ & $0.063$ & $0.119$ \\
\quad $[0.35, 0.50)$ & $165$ & $0.121$ & $0.148$ & $80$ & $0.052$ & $0.115$ \\
\quad $[0.50, 0.65)$ & $69$ & $0.123$ & $0.141$ & $34$ & $0.059$ & $0.130$ \\
\quad $[0.65, 1.00]$ & $19$ & $0.190$ & $0.191$ & $2$ & $0.173$ & $0.118$ \\
\quad Spearman $\rho$ & $300$ & $+0.166$ & $+0.160$ & $300$ & $-0.048$ & $+0.006$ \\
\addlinespace[2pt]
\multicolumn{7}{l}{\textit{Branch count} $n_{\text{br}}$} \\
\quad $1$ & $0$ & n/a & n/a & $39$ & $0.075$ & $0.119$ \\
\quad $2$ & $3$ & $0.047$ & $0.103$ & $197$ & $0.048$ & $0.111$ \\
\quad $3$ & $79$ & $0.115$ & $0.129$ & $64$ & $0.103$ & $0.141$ \\
\quad $4$ & $122$ & $0.111$ & $0.144$ & $0$ & n/a & n/a \\
\quad $5$--$6$ & $89$ & $0.134$ & $0.173$ & $0$ & n/a & n/a \\
\quad $\ge 7$ & $7$ & $0.074$ & $0.097$ & $0$ & n/a & n/a \\
\quad Spearman $\rho$ & $300$ & $+0.175$ & $+0.239$ & $300$ & $+0.250$ & $+0.254$ \\
\addlinespace[2pt]
\multicolumn{7}{l}{\textit{Inlet flow rate} $Q$ (mm$^3$/s), per-territory quartiles} \\
\quad Q1 & $74$ & $0.129$ & $0.149$ & $75$ & $0.072$ & $0.117$ \\
\quad Q2 & $75$ & $0.126$ & $0.158$ & $74$ & $0.060$ & $0.115$ \\
\quad Q3 & $76$ & $0.096$ & $0.130$ & $76$ & $0.062$ & $0.119$ \\
\quad Q4 & $75$ & $0.120$ & $0.152$ & $75$ & $0.060$ & $0.124$ \\
\quad Spearman $\rho$ & $300$ & $-0.064$ & $-0.035$ & $300$ & $-0.090$ & $+0.124$ \\
\addlinespace[2pt]
\multicolumn{7}{l}{\textit{Inlet radius} (mm), per-territory quartiles} \\
\quad Q1 & $73$ & $0.119$ & $0.161$ & $75$ & $0.072$ & $0.124$ \\
\quad Q2 & $74$ & $0.118$ & $0.142$ & $73$ & $0.066$ & $0.122$ \\
\quad Q3 & $75$ & $0.124$ & $0.140$ & $77$ & $0.048$ & $0.113$ \\
\quad Q4 & $78$ & $0.109$ & $0.147$ & $75$ & $0.068$ & $0.117$ \\
\quad Spearman $\rho$ & $300$ & $-0.091$ & $-0.077$ & $300$ & $+0.022$ & $+0.026$ \\
\addlinespace[2pt]
\multicolumn{7}{l}{\textit{Lesion throat radius} (mm), per-territory quartiles} \\
\quad Q1 & $72$ & $0.155$ & $0.152$ & $75$ & $0.064$ & $0.132$ \\
\quad Q2 & $77$ & $0.108$ & $0.160$ & $75$ & $0.052$ & $0.109$ \\
\quad Q3 & $76$ & $0.088$ & $0.135$ & $75$ & $0.061$ & $0.116$ \\
\quad Q4 & $75$ & $0.120$ & $0.142$ & $75$ & $0.077$ & $0.119$ \\
\quad Spearman $\rho$ & $300$ & $-0.117$ & $-0.119$ & $300$ & $+0.032$ & $-0.064$ \\
\bottomrule
\end{tabular}
\end{table}

\begin{table}[!htb]
\centering
\caption{Sensitivity to the frame in which the inputs are expressed. LCA,
$100$-case subset of the test split, fp32, mean relative $\ell_2$ against the
unperturbed reference fields with predicted velocity counter-rotated. The
rotation is about the case centroid through the stated angle about a randomly
drawn axis, applied jointly to the centerline and the evaluation points; the
translation is of the stated magnitude in a random direction. The zero rows
reproduce the single-run headline errors and are the protocol control. An
equivariant surrogate would return the zero row at every perturbation.}
\label{tab:frame_sensitivity}
\begin{tabular}{lcc|lcc}
\toprule
Rotation & Pressure & Velocity & Translation & Pressure & Velocity \\
\midrule
$0^\circ$  & $0.117$  & $0.146$  & $0$~mm   & $0.117$  & $0.146$ \\
$2^\circ$  & $0.121$  & $0.150$  & $1$~mm   & $0.120$  & $0.150$ \\
$5^\circ$  & $0.134$  & $0.175$  & $2$~mm   & $0.126$  & $0.160$ \\
$10^\circ$ & $0.178$  & $0.260$  & $5$~mm   & $0.160$  & $0.232$ \\
$20^\circ$ & $0.340$  & $0.483$  & $10$~mm  & $0.414$  & $0.469$ \\
$45^\circ$ & $2.25$   & $1.40$   & $20$~mm  & $39.3$   & $21.2$ \\
$90^\circ$ & $15.1$   & $11.9$   &          &          &        \\
\bottomrule
\end{tabular}
\end{table}

\paragraph{Consistency with the governing equations: protocol}
Derivatives are taken by automatic differentiation of the network with respect
to its evaluation coordinates, with the chain rule applied through the
coordinate normalization, so all quantities are in the pipeline's physical
units. First derivatives use a single reverse pass. The Laplacian uses a
symmetric finite-difference stencil of width $0.05$~mm, because the fused
attention kernels provide no second-order backward and the exact-arithmetic
fallback does not fit in memory at the trained anchor count. Quantities are
made dimensionless with the per-case mean speed $U$ and median lumen diameter
$D$: the divergence as $|\nabla\!\cdot\!\mathbf{u}|\,D/U$ and the momentum
residual by the convective scale $U^{2}/D$; fifty test cases per territory,
$4{,}096$ interior points each (Table~\ref{tab:pde_consistency}). The stencil
width is larger than the near-wall cell height of the reference mesh, which is
$0.75\%$ of the local radius and therefore of order $0.005$--$0.015$~mm, and
the interior points are drawn from the set of mesh cell centers with no
wall-clearance criterion, so for points within a stencil width of the wall
some stencil arms fall outside the lumen, where the network was never
supervised. Measuring the distance from each sampled point to the nearest
point of the production lumen surface, $39\%$ of sampled points lie within one
stencil width of the wall and the fifth percentile of that distance is zero.
The affected quantity is the Laplacian, and hence the momentum residual; the
divergence uses first derivatives and is exact.
Table~\ref{tab:pde_clearance} recomputes the medians on subsets progressively
clear of the wall: the momentum residual falls from $2.66$ over all points to
$1.57$ at one stencil width of clearance and $1.31$ at two, while the
divergence moves comparatively little, exactly the asymmetry the mechanism
predicts. The all-points momentum residual is therefore inflated by roughly a
factor of $1.7$ by points whose stencils leave the lumen; the wall-clear value
is taken as representative, and at $1.57$ the residual is still of the order
of the convective term itself. The inlet flux diagnostic integrates over a
cross-section by disc quadrature and does not depend on the stencil.

\begin{table}[!htb]
\centering
\caption{Consistency with the governing equations, evaluated by automatic
differentiation of the trained network. Medians over $50$ test cases per
territory; ``p90 over points'' is the median across cases of the
$90$th-percentile value within a case. Dimensionless by the per-case mean
speed $U$ and median diameter $D$.}
\label{tab:pde_consistency}
\begin{tabular}{lcc}
\toprule
Diagnostic & LCA & RCA \\
\midrule
$|\nabla\!\cdot\!\mathbf{u}|\,D/U$, median            & $0.26$  & $0.30$ \\
\quad p90 over points                                  & $1.38$  & $1.18$ \\
\addlinespace[2pt]
Momentum residual $\times\,D/U^{2}$, median            & $2.56$  & $2.86$ \\
\quad p90 over points                                  & $17.2$  & $13.1$ \\
\addlinespace[2pt]
Inlet flux error $|Q_{\text{pred}}-Q|/Q$               & $12.2\%$ & $5.5\%$ \\
\bottomrule
\end{tabular}
\end{table}

\begin{table}[!htb]
\centering
\caption{PDE-consistency medians restricted to points at least the stated
distance from the lumen wall, on $50$ LCA test cases with $2{,}048$ interior
points each and the $h = 0.05$~mm stencil. Wall distance is measured to the
nearest point of the production surface mesh. This is an independent draw at half
the point count of Table~\ref{tab:pde_consistency}, so the zero-clearance
column agrees with that table to within the draw ($0.251$ against $0.26$,
$2.66$ against $2.56$); the comparison to read is across the columns of this
table, which share a draw.}
\label{tab:pde_clearance}
\begin{tabular}{lcccc}
\toprule
Wall clearance & $\ge 0$ & $\ge h$ & $\ge 2h$ & $\ge 4h$ \\
\midrule
Points retained                            & $100\%$ & $61\%$  & $51\%$  & $35\%$ \\
$|\nabla\!\cdot\!\mathbf{u}|\,D/U$, median & $0.251$ & $0.226$ & $0.207$ & $0.168$ \\
Momentum residual $\times D/U^{2}$, median & $2.656$ & $1.574$ & $1.311$ & $1.038$ \\
\bottomrule
\end{tabular}
\end{table}

\begin{table}[!htb]
\centering
\small
\caption{Zero-shot LCA$\to$RCA transfer of the $300$-epoch LCA checkpoint on
the $300$-case RCA test split, under four choices of \emph{input}
normalization. Output is always denormalized with the LCA training statistics
and compared with RCA ground truth in physical units. Errors are pooled
relative $\ell_2$. ``Best scale'' is the single global factor minimizing the
pressure error; the last two columns give the error after applying it. A best
scale of zero with a scaled error of one means the prediction carries no
usable signal. Predicted norms are $9.7$ and $3.6$ times the reference in the
first and third rows, with a predicted mean pressure of the wrong sign.}
\label{tab:xterritory}
\footnotesize
\setlength{\tabcolsep}{3.5pt}
\begin{tabular}{lccccc}
\toprule
Input normalization & Pressure & Velocity & Best scale & Pressure$^{*}$ & Velocity$^{*}$ \\
\midrule
positions, radius, $Q$ all LCA        & $15.71$ & $27.55$ & $-0.002$ & $1.000$ & $1.000$ \\
positions RCA; radius, $Q$ LCA        & $10.18$ & $22.72$ & $+0.001$ & $1.000$ & $1.000$ \\
positions, radius RCA; $Q$ LCA        & $7.01$  & $14.82$ & $+0.005$ & $0.999$ & $1.000$ \\
positions, radius, $Q$ all RCA        & $22.52$ & $31.93$ & $+0.001$ & $0.999$ & $1.000$ \\
\bottomrule
\end{tabular}

{\footnotesize $^{*}$after the best global rescaling.}
\end{table}

\subsection{An anatomy-disjoint partition of the same cases}
\label{supp:anatomy_disjoint}

The partition used throughout the paper is over samples of the parameter set
$(x, y, z, r, Q)$, which is what a surrogate of a parametric solution map is
asked to interpolate and which is applied identically to every model compared.
A different question can be asked of the same data: how well does a model
trained on one set of \emph{segmentations} predict flow in a vessel tree it
has never seen? This subsection reports that experiment. It is not the paper's
claim: this cohort cannot answer the question with useful precision, and the
interval below shows by how much.

\paragraph{Construction}
The same $3{,}300$ solved LCA cases were re-partitioned so that every case
derived from one ImageCAS segmentation falls in exactly one split. No CFD was
re-run. Training was held at $2{,}686$ cases, close to the $2{,}700$ of the main split so the
model is not handicapped, leaving $83$ anatomies for training, $7$ for
validation and $11$ for test ($241$ and $373$ cases). Because a test set of a
dozen anatomies can easily be harder or easier than the training set by
chance, the allocation was chosen by search to balance train against test on
the five covariates of Table~\ref{tab:error_covariates}: the largest
standardized mean difference is $0.107$. Normalization statistics were
recomputed from this training split alone: the coordinate bounding box used
elsewhere is computed over all three splits, and on an anatomy-disjoint
partition it could otherwise carry
the test anatomies' spatial extent into training. CARS and the strongest baseline,
AB-UPT, were then trained on this partition under the configurations of
\S\ref{sec:methodology} and scored with each model's own evaluation code.

\paragraph{Result}
Table~\ref{tab:anatomy_disjoint} reports it. Both constructions degrade
sharply: predicting flow in an unseen vessel tree is a different and much
harder problem than interpolating lesion and flow parameters on trees the
model has seen, and neither architecture does it well when trained on $83$
anatomies. CARS has the lower error of the two on both fields, on $8$ of $11$
anatomies for pressure and $10$ of $11$ for velocity.

\paragraph{Why this cohort cannot settle the question}
The $373$ test cases are not $373$ independent observations. They come from
$11$ segmentations, and $71\%$ of the variance in per-case pressure error lies
between anatomies rather than within one; per-anatomy mean pressure error
ranges from $0.26$ to $1.69$. Resampling anatomies rather than cases, the
$95\%$ interval on CARS's pressure error is $[0.44, 0.93]$, spanning a factor
of $2.1$. Treating the cases as independent would give $[0.61, 0.71]$, an
interval five times narrower and wrong. The paired margin over AB-UPT is
$+0.109\;[+0.055, +0.168]$ on velocity, which excludes zero, and
$+0.150\;[-0.017, +0.350]$ on pressure, which does not. On this partition CARS
is the more accurate of the two on both fields and on the large majority of
anatomies, the velocity margin is resolved and the pressure margin is not, and
the absolute level is determined only to within a factor of two. A cohort with
enough segmentations to resolve it is a data-collection problem, not an
architecture problem.

\begin{table}[!htb]
\centering
\caption{Anatomy-disjoint partition of the same $3{,}300$ LCA cases: $83$
training anatomies, $11$ test anatomies, $373$ test cases, covariate-balanced
against training to a largest standardized mean difference of $0.107$.
Intervals are patient-clustered bootstrap $95\%$ CIs ($n_{\mathrm{boot}} =
10{,}000$) resampling the $11$ test anatomies. The parametric column is the
$300$-epoch three-seed result of Table~\ref{tab:seeds300} for reference. Both
anatomy-disjoint runs use best-validation checkpoints; both had plateaued well
before selection, and two CARS checkpoints $60$ epochs apart agree to
$2.5\%$.}
\label{tab:anatomy_disjoint}
\begin{tabular}{llcc}
\toprule
& & Parametric split & Anatomy-disjoint \\
Field & Model & (Table~\ref{tab:seeds300}) & mean $[95\%$ CI$]$ \\
\midrule
\multirow{3}{*}{Pressure}
 & CARS    & $\mathbf{0.120}$ & $\mathbf{0.662}\;[0.445, 0.934]$ \\
 & AB-UPT  & $0.158$          & $0.812\;[0.553, 1.235]$ \\
 & \quad margin & $+0.039$    & $+0.150\;[-0.017, +0.350]$ \\
\addlinespace[2pt]
\multirow{3}{*}{Velocity}
 & CARS    & $\mathbf{0.149}$ & $\mathbf{0.681}\;[0.594, 0.773]$ \\
 & AB-UPT  & $0.172$          & $0.789\;[0.686, 0.908]$ \\
 & \quad margin & $+0.023$    & $+0.109\;[+0.055, +0.168]$ \\
\bottomrule
\end{tabular}
\end{table}

\clearpage

\section{Upstream CFD pipeline}\label{app:cfd_pipeline}

The supervisory fields are produced by an OpenFOAM pipeline applied to the
ImageCAS-derived geometries.

\paragraph{Unit system}
The pipeline is formulated throughout in a self-consistent
$(\mathrm{mm}, \mathrm{s}, \mathrm{kg})$ system: lengths in mm, kinematic
viscosity in mm$^2$/s, density in kg/mm$^3$, volumetric flow rate in
mm$^3$/s, and kinematic pressure in mm$^2$/s$^2$. The OpenFOAM
\texttt{dimensions} entries are nominally SI-labeled, but OpenFOAM does not
verify them and every quantity is consistently interpreted in the system
above, so the reported pressure drops and derived quantities are unaffected.

\paragraph{Solver and discretization}
OpenFOAM Foundation v11~\cite{weller1998openfoam}, build
\texttt{11-e1fc8c682ae6}, invoked as \texttt{foamRun -solver
incompressibleFluid} through the \texttt{simpleFoam} deprecation shim; the
solver reports operation in SIMPLE mode, so the runs are SIMPLE and not
SIMPLEC. Convection, gradient and Laplacian terms use formally second-order
schemes, and the pressure system is solved with a geometric-algebraic
multigrid. The number of non-orthogonal correctors differs between the
territory pipelines ($20$ on LCA against $2$ on RCA) and makes no difference,
since on a structured hexahedral mesh two passes already resolve the
non-orthogonality: re-solving one case with both settings to a $10^{-7}$ residual floor gives
fields that agree to $6.0 \times 10^{-9}$ relative $\ell_2$ in pressure and
$4.4 \times 10^{-8}$ in velocity, with an identical FFR to every printed
digit; both settings reach the floor in the same number of iterations
($1{,}378$ against $1{,}377$), and the higher LCA setting costs about
$1.4\times$ in core-seconds. Blood is Newtonian, $\nu = 3.77$~mm$^2$/s and
$\rho = 1.06 \times 10^{-6}$~kg/mm$^3$ ($\mu = 4.00 \times 10^{-3}$~Pa\,s,
$\rho = 1{,}060$~kg/m$^3$, physiologic values at the upper end of the usual
$3.5$--$4.0 \times 10^{-3}$~Pa\,s range and the values used in the 1D ROM of
\appref{supp:rom}), the wall rigid and no-slip, and the solve steady:
pulsatile and transient phenomena, among them wave reflection and the
instantaneous wave-free ratio~\cite{davies2012ifr,cook2017cfr}, are absent
from the supervision.

\paragraph{Boundary conditions}
The inlet carries a fixed-value kinematic pressure with a
\texttt{pressureInletOutletVelocity} velocity condition. Each terminal branch
carries a prescribed volumetric flux (\texttt{flowRateOutletVelocity}), with
the total $Q$ distributed by Eq.~\eqref{eq:outlet_split}; $Q$ is the quantity
sampled per case and consumed by CARS through the FiLM pathway. Per-case $Q$
is drawn uniformly over a per-territory range spanning the published
physiologic range for resting and hyperemic coronary flow, and
\emph{independently of the geometry}: statistically an advantage for
identifying the flow-conditioning pathway, since the model cannot recover $Q$
from geometric cues, and physiologically the reason the prevalence of
$\mathrm{FFR}\le 0.8$ in this dataset is a property of the sampling design
and not of a clinical population.

\paragraph{Mesh}
Volume meshes are generated with the structured hexahedral vessel mesher of
Decroocq et al.~\cite{decroocq2023mesh}, which sweeps a circular
cross-section along the centerline as a block-structured O-grid, so every
cross-section is circular by construction and the lumen is exactly the tubular
image of Eq.~\eqref{eq:tube_map}. Every cross-section carries $324$ cells
($26$--$30$ cells across the diameter at every station, so a $0.18$~mm throat
radius still carries the full cross-section), the near-wall cell height is
$0.75\%$ of the local radius, and the representative case of
Figure~\ref{fig:field_compare} has $292{,}572 = 324 \times 903$ cells. Axial
spacing is ${\approx}0.2$~mm and does not contract at a throat, so a focal
lesion a few millimetres long is spanned by ten to twenty axial cells; the
grid-convergence study below, whose refinement family is axial as well as
circumferential, bounds the combined effect.

\paragraph{Reynolds number}
With $\mathrm{Re} = 2Q/(\pi \bar r \nu)$ and $\bar r$ the proximal-segment mean
radius (median $1.91$~mm on LCA), the inlet Reynolds number over the test
splits has median $571$ (IQR $509$--$667$, max $1{,}045$) on LCA and $316$
(IQR $272$--$349$, max $519$) on RCA; the sampled flow rates correspond to
$300$--$480$~mL/min on LCA and $150$--$270$~mL/min on RCA, against a resting
total coronary flow of ${\approx}250$~mL/min and a coronary flow reserve of
$2.5$--$4$ under adenosine~\cite{taylor2013ffrcfd,tonino2009ffr}. At the
located lesion, with the flux the subtree's $r^2$ share, the throat Reynolds
number has median $1{,}056$ (p95 $2{,}143$) on LCA and $647$ (p95 $1{,}203$)
on RCA; only $4.3\%$ of LCA cases and no RCA case exceed the classical
transition value of ${\approx}2{,}300$, so the laminar reference is well
supported over the bulk of the cohort, with a small upper tail on the left
tree where a post-stenotic jet could be transitional.

\paragraph{Iteration, convergence and the retention filter}
The pipeline runs a fixed $4{,}000$ SIMPLE iterations per case and retains a
case only if its final pressure residual falls below $10^{-4}$. Re-solving a
representative case with a residual floor of $10^{-7}$ on both pressure and
velocity met that floor after $2{,}751$ iterations, and the two solutions
agree over all $254{,}016$ cells to $5.6\times 10^{-6}$ relative $\ell_2$ in
pressure, $2.2\times 10^{-5}$ in velocity and $7.8\times 10^{-7}$ in FFR, so
iteration error is more than three orders of magnitude below the
discretization error; global mass conservation holds to machine precision by
construction of the boundary conditions and is not independent evidence of
convergence. Of the solves that run to completion, the residual filter removes
$9.9\%$ (RCA) and $13.3\%$ (LCA), and counted against all attempted solves
$24$ of $133$ and $37$ of $102$, because early terminations are also absent.
Removal tracks mesh quality (among removed cases $46\%$ (RCA) and $70\%$ (LCA)
fail \texttt{checkMesh}, against $1\%$ and $17\%$ of retained cases), on RCA
the removed set carries higher flow (median $Q$ $4{,}271$ against
$3{,}331$~mm$^3$/s), and lesion severity was not measured on the removed set.
The filter is applied before the splits are constructed.

\subsection{Grid convergence of the reference solutions}
\label{app:gci}

A three-level Grid Convergence Index
study~\cite{roache1994perspective,roache1997quantification,celik2008procedure}
on sixteen cases, eight per territory (selection below), refines the
production meshes. Table~\ref{tab:gci} reports the result; the paragraphs below
give the construction and what the figure establishes.

\begin{table}[!t]
\centering
\caption{Grid Convergence Index at the production ($1\times$) resolution, from
three levels at $160$/$324$/$544$ cells per cross-section with realized ratios
$r_{32}=1.447$, $r_{21}\approx 1.31$, over sixteen cases. Field norms are
volume-weighted; median and interquartile range over cases. Two cases have not
reached the asymptotic range (observed orders $0.48$ and $0.47$) and are
retained; the bracketed figure is the pressure-norm median without them.}
\label{tab:gci}
\small
\begin{tabular}{lccc}
\toprule
Quantity & LCA ($n{=}8$) & RCA ($n{=}8$) & All ($n{=}16$) \\
\midrule
Pressure, vol.-weighted $\ell_2$ & $1.99\%$ & $1.36\%$ & $1.61\%$ $[{\approx}1.5\%]$ \\
\quad IQR                        & $1.14$--$5.86\%$ & $0.73$--$2.37\%$ & $0.93$--$3.56\%$ \\
Velocity, vol.-weighted $\ell_2$ & $0.27\%$ & $0.29\%$ & $0.29\%$ \\
\quad IQR                        & $0.06$--$0.47\%$ & $0.13$--$0.35\%$ & $0.12$--$0.46\%$ \\
FFR, minimum-pressure proxy      & $0.83\%$ & $0.26\%$ & $0.41\%$ \\
\quad IQR                        & $0.33$--$3.38\%$ & $0.15$--$0.55\%$ & $0.25$--$0.99\%$ \\
\midrule
Observed order $p$, pressure     & \multicolumn{3}{c}{median $2.51$} \\
Observed order $p$, velocity     & \multicolumn{3}{c}{median $3.28$} \\
\bottomrule
\end{tabular}
\vspace{-2mm}
\end{table}

\paragraph{Holding the geometry fixed across refinement levels}
A refinement study requires the same domain at every level, which is not
trivial here because the production mesher injects each stenosis through an
unseeded random draw and records neither the draw nor its result: regenerating
a case from its source centerline yields a \emph{different} lesion. Each
case's lesion parameters were therefore recovered from its production mesh.
Generating a stenosis-free mesh at production resolution gives point-for-point
index correspondence with the production mesh, so differencing the two point sets
isolates the displaced points, and mapping those onto the vessel graph
recovers the carrying edge and the lesion's start and end positions along it.
For the validation case all $71{,}904$ displaced points of $304{,}954$ lie on
a single edge, so the recovery is unambiguous, and the recovered extent is
interior to the production sampling bounds. Regenerating at production
resolution from the recovered parameters reproduces the production mesh to six
significant figures in cell count ($292{,}572$), maximum non-orthogonality
($52.3779^\circ$), maximum skewness ($2.74534$) and maximum aspect ratio
($52.7179$), with a maximum point deviation of $0.0195$~mm against a lesion
deformation amplitude of $0.578$~mm. The study therefore refines the
production geometries themselves.

\paragraph{Refinement levels and ratios}
The mesher's circumferential division count $N$ must be a multiple of eight,
which fixes the available levels at $N = 16, 24, 32$, carrying $160$, $324$
and $544$ cells per cross-section. Refinement is simultaneously axial, so the
three levels carry $96{,}320$, $292{,}572$ and ${\approx}655{,}000$ cells on
the validation case. Realized refinement ratios from cell counts are
$r_{32} = 1.447$ and $r_{21} = 1.307$; these are not equal, so the
closed-form GCI does not apply and the iterative non-constant-ratio
formulation is used, solving for the observed order $p$ from the realized
ratios of each case. Mesh quality does not degrade: maximum skewness falls
from $2.97$ to $2.70$ and maximum non-orthogonality stays within
$52$--$57^\circ$ across levels, every study mesh passes \texttt{checkMesh}, and
topology and patch ordering are identical at all three levels. The domain
converges with the mesh rather than being exactly fixed, because $N$
circumferential divisions inscribe an $N$-gon in each circular cross-section;
the resulting volume deficit is $O(N^{-2})$, giving $2.55\%$, $1.14\%$ and
$0.64\%$ at the three levels, confirmed against measured mesh volumes, so the
geometric error converges at the same second order as the schemes and enters
the GCI as part of one second-order error term. Boundary conditions are held
identical across levels by copying each production per-outlet flux verbatim,
and all levels are run to the same $10^{-7}$ residual floor.

\paragraph{Case selection}
A case qualifies when all three levels are solved and readable, all three
converged (residual floor met or final pressure residual below $10^{-5}$),
\texttt{checkMesh} is clean at all three study levels, and \texttt{checkMesh}
is clean on the production mesh (a production mesh that fails carries
inverted cells, and such cases terminate on a floating-point exception at the
finer levels). Among qualifying cases eight per territory were drawn at evenly
spaced quantiles of the production mesh's maximum non-orthogonality, so that
each cohort's runnable mesh-quality range is spanned. Of twenty-one completed
triples, eighteen qualified (nine in each territory) and sixteen were
selected. Nothing is filtered or ranked on observed order, on GCI, on lesion
presence, or on reference FFR. The single LCA case at the worst production
mesh quality degraded \emph{under} refinement, its fine level failing
\texttt{checkMesh}, so mesh quality is stable under refinement across the
range the study covers but not beyond it.

\paragraph{The quantity the GCI is computed on}
Each level's norm is the global volume-weighted root mean square on its own
mesh, $\big(\sum_i f_i^2 V_i / \sum_i V_i\big)^{1/2}$, and the three
per-level scalars enter the Richardson extrapolation as an ordinary scalar
quantity of interest; no coarse or medium field is ever interpolated onto the
fine mesh. The implemented expression is
$\mathrm{GCI} = F_s\,|(f_1-f_2)/f_1| / (r_{21}^{\,p}-1)$ with $F_s = 1.25$
applied to the relative error, so the bare Richardson estimates are the
reported values divided by $1.25$. A global root mean square is subject to
cancellation, so the GCI plausibly \emph{understates} the pointwise
discretization error and says nothing about where that error sits;
peri-stenotic restricted norms were not computed.

\paragraph{Observed orders}
The pressure median of $2.51$ is close to the formal second order and is the
principal evidence that the triples are in the regime a refinement study
requires. The velocity median of $3.28$ is not: an observed order
substantially above the formal order indicates a triple outside the asymptotic
range, since in the asymptotic regime the observed order should approach the
formal one rather than exceed it by $60\%$, so the velocity GCI is an
order-of-magnitude indication. Two of the sixteen cases sit at
$p \approx 0.48$ and are retained rather than dropped, since removing them
would reintroduce selection on the outcome. Restricting the index to triples
with observed order in a band around $2$ and reporting the remainder
separately is the analysis that would separate these populations; it has not
been done.

\paragraph{How the figure bounds the surrogate error}
First, the norms differ: the GCI is volume-weighted, whereas the surrogate
errors are unweighted over a point set drawn uniformly over cell centers,
which with wall-graded cells over-weights the near-wall region. Computing
$\|p_{\text{fine}} - p_{\text{prod}}\| / \|p_{\text{fine}}\|$ on a common
point set, and reporting the surrogate error in the volume-weighted norm as
well, is what would make the two commensurable. Second, the comparison is only
meaningful median against median: the surrogate's LCA pressure error has
median $0.065$ and mean $0.116$, so median against median the ratio is about
three and mean against median nearly six; on RCA the mean ratio is about five,
and on velocity the ratios are an order of magnitude larger ($0.146$ against
$0.27\%$ and $0.119$ against $0.29\%$). At the GCI upper quartile ($5.86\%$ on
LCA, estimated from few points) the reference error is comparable to the
surrogate's median. Third, part of the reference error is a
boundary-approximation bias: the $N$-gon inscription leaves a $1.14\%$ volume
deficit at production resolution, hence a radius deficit of ${\approx}0.57\%$,
which with $\Delta P \propto r^{-4}$ accounts for a pressure error of order
$2.3\%$, of the same size as the entire reported LCA pressure GCI. Because
this deficit is a deterministic function of the resolution applied identically
to every case, it shifts all labels in the same direction and a surrogate
trained on them reproduces the shift; to that extent the reference error
places no floor on measured agreement. Separating the two components requires
refining circumferentially with an area-preserving radius correction, or
projecting vertices onto the exact circle at every level. The territory
difference is the substantive finding: LCA carries $1.5$ times RCA's
pressure-norm error and $3.2$ times its FFR error, with a heavier upper tail,
consistent with LCA being the harder cohort throughout.

\paragraph{The endpoint station is better converged than the domain minimum}
The minimum-pressure proxy evaluates $p_{\min}/p_{\text{in}}$ over the whole
domain; the endpoint is read $20$~mm distal to the lesion within its own
branch. Because every terminal flux is prescribed, static pressure falls
monotonically along each branch, so the domain minimum sits at the end of
whichever terminal branch carries the largest cumulative resistance: on twenty
LCA test cases it lies at a station of local radius $0.40$~mm against a cohort
median of $1.12$~mm, at $88\%$ of the way along the centerline, in the
smallest cells in the domain and not at the stenosis throat. Recomputing the
FFR row at the recorded $20$~mm coordinates on all three levels gives a median
GCI of $0.055\%$ (IQR $0.007$--$0.338\%$) against $0.41\%$ for the proxy, and
a median $|\Delta \mathrm{FFR}|$ between the production level and the
extrapolate of $0.0010$ against $0.0054$; the $30$~mm station agrees
($0.068\%$; $0.0011$). The station is geometrically unavailable in three of
the sixteen cases ($3.74$, $4.27$ and $19.05$~mm of vessel distal to the
lesion) and one further case yields no usable triple, so $n = 12$; five of
those twelve are stenosis-free cases in which the station is $20$~mm from the
inlet, leaving $n = 7$ lesioned cases; one case (GCI $9.78\%$ at an observed
order of $0.94$) is not in the asymptotic range and is retained, excluding it
the maximum falls to $0.71\%$; and the station is sampled by nearest-cell
lookup, with the snap distance recorded per probe.

\paragraph{A systematic FFR bias, not noise}
FFR increases monotonically with refinement in fifteen of the sixteen cases:
coarser meshes over-predict the trans-lesional pressure drop, through
numerical dissipation and through the inscribed-polygon deficit that renders
the lumen slightly too narrow. Both mechanisms act in one direction, so the
effect is a bias. Against the Richardson extrapolate the production-resolution
FFR is low by $0.005$ at the median case; the worst case is low by $0.114$,
and it is one of the two non-asymptotic cases, so that figure is an upper
envelope on the tail. A surrogate trained on these labels inherits the bias
rather than being penalized by it, because the bias is common to training and
evaluation labels. At the endpoint the surrogate error ($0.008$ MAE on RCA,
$0.019$ on LCA) is roughly eight and nineteen times the median label bias of
$0.0010$, so the labels are not the binding constraint on either territory;
the minimum-pressure proxy would have put the RCA figures within a factor of
two of each other, which is why the station matters, and the $0.114$ tail
means the endpoint comparison should be read as a statement about medians.

\section{Further limitations of the evaluation}
\label{supp:limitations}

Table~\ref{tab:limitations} collects the limitations listed
by name in \S\ref{sec:limitations}, with what each bounds and the measurement
that would remove it.

\begin{table}[p]
\centering
\footnotesize
\caption{Further limitations of the evaluation
(\S\ref{sec:limitations}), each bounding what the reported numbers establish,
with the measurement that would remove it.}
\label{tab:limitations}
\setlength{\tabcolsep}{5pt}
\begin{tabular}{@{}>{\raggedright\arraybackslash}p{0.23\linewidth}>{\raggedright\arraybackslash}p{0.41\linewidth}>{\raggedright\arraybackslash}p{0.29\linewidth}@{}}
\toprule
Limitation & What it bounds & What would remove it \\
\midrule
Convergence filter precedes the splits (\appref{app:cfd_pipeline})
& $9.9\%$ (RCA) and $13.3\%$ (LCA) of completed solves are dropped on
residual. Removal tracks mesh quality, so the retained cohort favors
geometries the mesher handled cleanly; whether it also selects on stenosis
severity is unmeasured
& Re-meshing the failed geometries, and the severity distribution of the
removed set \\
\addlinespace[3pt]
Flow is developing over much of the tree
& $L_e \approx 130$~mm against a median LCA centerline of $182$~mm, so the
nearly-developed-profile argument of \S\ref{sec:problem} is weakest
proximally, which is also where the error field is largest; we claim no
causal link
& Inlet extensions, or a spatially resolved residual diagnostic \\
\addlinespace[3pt]
Part of the reference error is geometric, not truncation error
& The inscribed-$N$-gon bias is of the same order as the whole LCA pressure
GCI and is inherited by any surrogate, so the GCI is not a floor on measured
agreement (\appref{app:gci})
& A refinement family with an area-preserving radius correction, or vertices
projected onto the exact circle at each level \\
\addlinespace[3pt]
The cohort spans non-physiological corners
& $Q$ is drawn independently of caliber, so the smallest LCA inlet
($1.32$~mm) reaches ${\approx}1.46$~m/s at the top of the sampled range,
three to five times physiological
& The joint $(Q, \bar r)$ distribution, and endpoint metrics on the
physiological subset \\
\addlinespace[3pt]
Outcome-adjacent filtering of the endpoint subset
& Sensitivity analyses use reference $\mathrm{FFR} \in [0.3, 1.0]$, excluding
$14$ LCA and $7$ RCA of the most severe lesions
& Metrics separately on the clamped subsets \\
\addlinespace[3pt]
One acquisition source
& All anatomies come from one tertiary center via ImageCAS, so cross-vendor
and cross-protocol shift is untested
& Held-out evaluation on a second CTA+CFD cohort, which needs only
centerlines, radii and a flow rate \\
\addlinespace[3pt]
RoPE3D under length extrapolation is untested
& If RoPE3D transferred better across territories it would partly reverse the
recommendation of \S\ref{sec:ablations}
& One evaluation of the existing RoPE3D checkpoint on the other territory \\
\addlinespace[3pt]
Sensitivity analysis covers smooth perturbations only
& \S\ref{sec:robustness_uq} establishes local Lipschitz behavior in the
geometric arguments; topology-changing perturbations lie outside that
parameter set
& A segmentation-error model \\
\bottomrule
\end{tabular}
\end{table}

\fi

\end{document}